\documentclass[english,11pt,a4paper]{article}
\usepackage{babel}
\usepackage{makeidx}
\usepackage{layout}
\usepackage[latin1]{inputenc}
\usepackage{latexsym,amssymb,amsmath}
\usepackage{amsfonts,amsbsy}
\usepackage{eufrak}
\usepackage[mathscr]{eucal}
\usepackage{verbatim}
\usepackage{amscd}
\usepackage{textcomp}
\usepackage{indentfirst}
\usepackage[all]{xy}

\begin{document}
\date{}
\title{\bf {\large{INTRODUCTION  TO  SPORADIC  GROUPS\\
for physicists}}}
\author{\normalsize{Luis J. Boya\footnote{\texttt{luisjo@unizar.es}}} \\
\normalsize{Departamento de F\'{\i}sica Te\'{o}rica} \\
\normalsize{Universidad de Zaragoza} \\
\normalsize{E-50009 Zaragoza, SPAIN}} \maketitle

\newtheorem{defn}{Definition}[section]
\newtheorem{thm}{Theorem}[section]
\newtheorem{prp}{Proposition}[section]
\newtheorem{cor}{Corollary}[section]
\newtheorem{lem}{lemma}[section]




\quad \quad \quad \quad \quad \quad \quad \quad \quad \quad \quad
\quad \quad \quad \quad  \quad \quad \quad \quad Juan SANCHO
GUIMERÁ\

\quad \quad \quad \quad \quad \quad \quad \quad \quad \quad \quad
\quad \quad\quad \quad \quad\quad \quad \quad \quad\quad\quad  \emph{In Memoriam}\\

\begin{abstract}

We describe the collection of finite simple groups, with a view on
physical applications. We recall first the prime cyclic groups
$Z_p$, and the alternating groups $Alt_{n>4}$. After a quick
revision of finite fields $\mathbb{F}_q$, $q = p^f$, with $p$ prime,
we consider the 16 families of finite simple groups of Lie type.
There are also 26 \emph{extra} ``sporadic'' groups, which gather in
three interconnected ``generations'' (with 5+7+8 groups) plus the
Pariah groups (6). We point out a couple of physical applications,
including constructing the biggest sporadic group, the ``Monster''
group, with close to $10^{54}$ elements from arguments of physics,
and also the relation of some Mathieu groups with compactification
in string and M-theory.

\end{abstract}
\pagebreak

\tableofcontents

\pagebreak

\section{Introduction}

\subsection{General description of the work}

1.1.1 \emph{Motivation.} The algebraic structures called groups first appeared
distinctly in mathematics in the first third of the 19th century,
after the work of Gauss (on congruences), and Lagrange, Ruffini,
Abel and above all Galois, in relation to the solvability of
polynomial equations by radicals: the groups appearing did permute
the roots of the equation. In physics also \emph{finite groups} were
tacitly used, e.g. by Bravais, to classify some crystal lattices
around 1850, and better
and more directly by Fedorov, at the turn of that century.\\

\emph{Lie groups}, i.e. infinite continuous groups with some
differentiability properties, were first considered by Lie
around 1870 in relation to solutions of differential equations,
trying to imitate what Galois had done with algebraic equations.
Cayley had provided in 1854 the first formal definition of a group,
the same we use today. Klein (since 1872) identified many
continuous groups acting by transformations on the diverse
geometries, and even characterized these \underline{Geometries}
(euclidean, metric, affine, projective, conformal, etc.) by the
group of allowed transformations, in the so-called
``Erlangen-Program'' [1]. By the turn of the 19th/20th centuries,
groups were taking a prominent role in mathematics (Poincaré,
Hilbert, etc.), consolidated around 1920 as an indispensable part of
Modern Algebra, indeed the core of it.\\

With the advent of Quantum Mechanics (QM) in Physics in 1925, as
evolution of the Old Quantum Theory (of Planck, Einstein and Bohr,
1900-1925), some Lie groups and their representations were to be
used in physics, starting with the three dimensional rotation group
$SO(3)$. Besides some original books in group theory at the very
beginning [2-4], the three standard books for physics at the
time were by Weyl, van der Waerden and Wigner ([5, 6], and ([7]).
The Pauli exclusion principle led also to some properties of the
representations of the symmetric group $Sym_n \equiv S_n$, with $n!$
elements (Heisenberg, and Dirac, independently, in 1926).\\

Soon the interest in atomic and nuclear physics and also in particle
physics shifted wholly to general, continuous, Lie groups: isospin
group: Heisenberg (1932); Poincaré group, Wigner (1939); Gell-Mann's
$SU(3)$ flavor group (1962), etc, i.e. finite-dimensional
manifolds which support a compatible group structure; as symmetry
operations enter QM through \emph{linear} (and
\emph{projective}) representations, this formalism, including
decomposition of products, traces, covering groups, etc., was more
and more used in that period, as evidence the books mentioned above.
Some people even talked about the \emph{Gruppenpest}, [8]; so it was
not before around 1962 that quantum physicists realized Group Theory
was an unavoidable part of their
mathematical toolkit.\\

While continuous (\emph{simple Lie}) groups were classified already
by Killing and Cartan around 1887/95, similar work on \emph{finite}
groups was very much delayed. After the foundational period with
Jordan, Mathieu, Klein, etc, simple groups of Lie type (over finite
\emph{fields}) were studied by Dickson and Miller. Since around 1920
group theory took a preeminent role in mathematics, as said. By
1960, mathematicians were busy completing the list of
\emph{families} of finite simple groups (FSG),
continuing the work started by Dickson (as mentioned) at the
turn of the 20th century, and the later one by Chevalley, Borel
etc. about 1950 on some finite groups of Lie type.  These
FSG groups occur in several families (2+16, as we shall
see), plus some isolated, \emph{sporadic} groups (26 of them),
initially discovered (the first five) by Mathieu in 1860:
these first sporadic groups are still called \emph{Mathieu groups} today. After a full century with no new sporadic groups found, the
list was completed by the biggest of all, the \emph{Monster} group
$\mathbb{M}$, conceived around 1973 by Fischer and Griess and
definitively established in 1980 by Griess; see e.g. his [14]; it
was given an alternative built-up form, inspired by string theory, a
physical construct, by Frenkel \emph{et al}. in 1984 [9], completed
by Borcherds ([10], see [11, 12]).
This was important also for physics, and indeed soon after other
``sporadic'' finite groups (e.g. $M_{24}$, the largest Mathieu
group) were used, in relation with the compactifying space $K3$ [13].\\

It is the purpose of this review to introduce the theory of
 FSG, in particular the sporadic groups, to a physics
audience. We feel the time is ripe for that, as in one hand
theoretical physics is in a standstill in microphysics, with no real
progress made since around 1976, when the so-called Standard Model
was completely established, so any new avenue is worth exploring,
and on the other hand many physical clues point to concrete, finite
symmetries, mostly realized as finite groups (examples will be
discussed later). We recognize there was the construction of the
Monster group via the Vertex Operators appearing in superstring
theory [9], which triggered the interest of physicists in this
somewhat exotic branch of pure mathematics.\\

At the same time, we realized that while the practice of Lie groups
today makes use of the tools known to the average theoretical
physicist (e.g. differential geometry, representation theory), this
is not the same for the finite groups; we want to supply a decent
mathematical basis for individuals to engage in actual research in
\emph{physics} dealing with finite groups.\\

So we have aimed to expose the theory from the beginning, although
the very first definitions and concepts are glossed over rather
quickly; it is our idea this review will serve as reference, so
we shall start by recalling even the simplest notions of modern
mathematics, e.g. sets and maps. In that spirit, we have strived to
present things in a modern mathematical language.\\

Although intended mainly for physicists, we have aimed at presenting
the material from a modern point of view, also from the mathematical
side: for us, mathematics is not only an instrument to describe the
physical world, say helping to solve equations, but it also
represents a conceptual frame, a building, in which physical events
take place and develop. So we hope some mathematicians looking at
this review can benefit, too.\\

1.1.2 \emph{What we cover}. As we want to present the pure mathematical doctrine in actual
terms, and often the mathematical instruction of physicists does not
use several of these modern concepts, we have tried to show a modern
unified treatment of algebra in general and group theory in
particular starting from scratch, although naturally many results,
theorems etc. are given only a cursory treatment. Our idea is that
all concepts, definitions and results will be clearly if briefly
stated, while many results are taken for granted and several others
presented with proofs, or just hints of the proof, and rather
briefly.\\

Thus we start by some definitions and results in pure set theory,
emphasizing notions as subsets, maps or functions, inverses,
equivalence relations, sequences and commutative diagrams, etc.
Next, the two big branches of mathematics are presented, namely
algebra and topology, with the fundamental definitions on
substructures, equivalences, natural maps, morphisms and
automorphisms, etc. From abstract topology we descend to geometry,
conceptually perhaps the most important branch of mathematics for
physics.\\

In modern times axioms are subsumed in category
theory, to which we also refer briefly, mainly for notational and descriptive purposes.\\

For the study of finite groups proper we consider, first, groups in
general, and study substructures (subgroups), superstructures
(extensions, e.g. direct products) and morphisms (in particular
endomorphisms/automophisms and isomorphisms). Then we specialize in
finite groups, with partition by classes, chain compositions,
subgroup structure, etc. Next we included detailed studies of
some families of finite groups, etc; here, we first notice the
appearance of FSG, in particular
primarily the cyclic groups of prime order, $Z_p$, and the
alternating groups $Alt_n$ (for $n > 4$): the first two (infinite)
families of FSG. But we consider also other families as well, such as
the symmetric groups $S_n$, dihedral $D_n$, dicyclic $Q_n$, Clifford
groups $\Gamma_n$, etc.\\

This leads us to study concretely the other known families of
(finite) simple groups, which we shall first try to describe. The study
will require brief excursions on finite groups of Lie type, a large
subject, which we shall resume rather than develop in much detail:
we begin by recalling the finite fields $\mathbb{F}_q$ (Galois
fields; $q$ is a power of a prime, $q = p^f$), with vector spaces
and groups of matrices over them, starting with the most general
group $GL_n(q)$: they provide eventually the rest of the
\emph{families} (16 out of
18) FSG.\\

Next we present the case of the 26 sporadic groups, that
is, FSG not in the above 18 families, which is
really the main novelty content (for physicists) of this review. We
shall see that they appear in \emph{three} well-defined and
inter-related blocks, plus some (six) unconnected exceptions, the
so-called \emph{pariah} groups [14]. We shall see
the special role of the number 24, for some mysterious reasons:
Mathieu's $M_{24}$ is the most important group of the first sporadic
series, and the Leech lattice, operating on a 24-dimensional space,
is the starting point for the second series of sporadic groups (e.g.
the Conway's $Co_{{1,2,3}}$ groups). The third series runs around
the Monster group $\mathbb{M}$, the biggest of the
sporadic groups, of order $\approx 10^{54}$: in fact all sporadic
groups but a few are contained in the Monster as
subgroups or subquotients (= quotients of subgroups).\\

As noticed above, we believe mathematicians around 1980 have agreed
that \emph{all} sporadic groups were already known, so it is a good
moment to present their list to a non-specialized audience, such as
physicists (see e.g. [15]).\\

We are not interested in studying all possible applications in
physics, mainly because we feel many new ones are to be expected and
are to be developed in the future, but our review will be rather
incomplete if we do not refer to some of the most recent
applications in physics, like the string-theoretic construction of
the Monster group and also the Mathieu group $M_{24}$ in relation to
the K3 manifold, a favourite space for essays on superstring \&
M-theory compactification.\\

1.1.3 \emph{Detailed plan of the work}. We proceed now to a brief section-by-section description of our
work. Subsection 1.2 is devoted to
presenting the very first concepts in mathematics; we recall first the
simple notions of sets and maps in the following section, with some
considerations about subsets, inverse maps, equivalence relations,
sequences and diagrams, and the like: the most fundamental and
elemental mathematical
concepts.\\

The two fundamental branches of (modern) mathematics,
algebra and topology,  (Weyl (1939)
speaks of the ``angel'' of topology and the ``demon'' of algebra,
[16]) are recalled on Sect. 1.2.2. We present briefly geometry as
evolving from topology, and mention analysis, which originally
sprung before modern algebra. We emphasize first the conceptually
different definitions: topology (and geometry) deal essentially with
structureless elements, or \emph{points}, while in algebra the
elements might be distinguished or related (like the unit $e$,
$a^{-1}$ inverse of $a$, powers $a^m$, etc), and are to combine
with each other (for $a$, $b$, the composition $ab$ is defined, for
example). We emphasize the crucial role of geometry in modern
physics, as presenting the ``frame'' or a ``model'' in which
physical events take their place.\\

We concentrate then on algebra henceforth, starting by the
definition and classification of the \emph{six} more common types of
algebraic structures, from groups (with rings, fields, modules and
vector spaces) to algebras proper, as the six will arise along our
work. General properties, like substructures, products and
extensions, natural maps (morphisms and automorphisms) are recalled
also in our Sect. 1.2.3. Finally, as we eventually shall use the
convenient notion of \emph{categories}, we introduce also them
briefly. (See e.g. [17] for the final section 1.2.4 of this Introductory part).\\

Section 2 deals with finite groups proper. After review of the
elementary notions of general groups in section 2.1, the five-parts
\emph{cadre} or box is set up, with the concepts of subgroups,
morphisms and extensions in section 2.2. Subgroup structure, chains of
quotients, and classes of groups (simple, solvable, etc.) are
considered in 2.3; the studies of properties of morphisms of groups are
collected in 2.4. Direct and semidirect products, as examples of
extensions of groups, including the holomorph of a group, are dealt
with in section 2.5. Common families of finite groups are recalled in section
2.6. Abelian groups are considered in section 2.7, and particulars of the
symmetric groups $S_n$ are in section 2.8. Two elementary guides on
finite groups
are the books by Lederman [18] and Carmichael [19].\\

Section 3 looks at more specialized items of group theory necessary
for our work, including: action of groups on manifolds (section 3.1),
representation theory (sections 3.2 and 3.3), and introduction to the extension
theory from the point of view of homological algebra (section 3.4): this
will also cover briefly notions and examples of the Schur's
multipliers. Section 3.5 shows systematic properties of the 20+8
(Abelian and non-Abelian) groups up to order 16, as examples
[20]. Section 3.6 explains different forms of presentation of a particular group.\\

Section 4 deals with the problematics of FSG and describes first an historical introduction to
the families of FSG (4.1); after a short review to
finite fields and vector spaces (section 4.2), we describe briefly the
sixteen families of FSG of Lie type (sections 4.3 and 4.4),
starting with the most common bi-parametric family $PSL_n(q)$. As
in the continuous case, candidates to (finite) simple groups are
among the subgroups or subquotients of $GL_n(q)$ conserving a
regular bilinear form, either symmetric or antisymmetric (sections 4.3 and 4.4);
main reference here is [21]. The exceptional simple Lie groups (e.g.
$G_2$) have counterpart \emph{families} in the finite-field case;
there are also another two families of FSG, one coming from the Lie
groups with automorphisms,
the other from some non-simply-laced groups, section 4.5.\\

Section 5 deals with description and properties of sporadic groups
(section 5.1): they constitute three series plus the ``Pariah'' groups: the
three series have respectively five, seven (sections 5.2 and 5.3) and eight (sections 5.4 and 5.5) cases, and
are related to each other (curiously, as noted, depending on the
pure number 24), plus the six unrelated Pariah groups (section 5.6) for a
total of the 26 sporadic (finite simple) groups. It is to be
expected that these unrelated \emph{Pariah} groups will be better
understood in
the future.\\

Section 6 deals with some physical applications: here we have to
limit ourselves to the most important ones, leaving
for the future perhaps new ones. We indulge in the ``vertex
operator'' construction of the Monster group [9], and include some
modern applications of the Mathieu's $M_{24}$ group.

\subsection{Initial mathematics}

\emph{Set and maps.} To start with, it is advisable to recall some notions in \emph{set theory.} Concepts like \emph{sets}  $X$, $Y$; \emph{elements} $x, y
\in X$; \emph{subsets} $Y \subset X$; and \emph{maps} or
\emph{functions} $f: X\longrightarrow Y$ are supposed to be known to
the reader. Unions $Y\cup W$ and intersections $Y\cap W$ are also
well-defined operations. The number of elements in a set $X$, if
finite, is called the cardinal of the set and denoted  $card(X)$ or
$|X|$; given a set $X$, a \emph{subset} $Y \subset X$ defines the
complementary $Y^{\sim}$, such $Y\cap Y^{\sim} =\emptyset$ , $Y\cup
Y^{\sim} = X$. The empty set $\emptyset$ and the total space $X$ are
considered as \emph{improper} subsets of $X$. The totality of
subsets of set $X$ is noted (sometimes) $\mathcal{P}(X)$; it
includes $\emptyset$ and $X$; for example, if $X$ is \emph{finite},
with $card (X) = N$, then \emph{card} $\mathcal{P}(X) = |
\mathcal{P}(X) | = 2^{|X|} = 2^{N}$; with the symmetric union
$U+V:=U\cup V\setminus U\cap V$, the set $\mathcal{P}(X)$ forms
an Abelian group, with $\emptyset$ as unit.\\

A \emph{Map} or application or \emph{function} $f: X\longrightarrow
Y$ implies for any $x \in X$, $f(x) = y \in Y$ is well defined and
unique. It defines the subset image $f(X) \equiv Y'\subset  Y$; the
set of maps from set $X$ to set $Y$ is named $Map(X, Y)$; for
example, if $card(X, Y) = (n, m)$, then \emph{card} $Map(X, Y) =
nm$. Maps $f: X\longrightarrow Y$ and $g: Y\longrightarrow Z$
\emph{compose} to a map $g\circ f: X\longrightarrow Z$, given as
$(gf)(x) := g(f(x))$; composition of maps is automatically
\emph{associative}: $f$, $g$, $h$ in $Map(X, Y)$, resp. $Map(Y, Z)$,
$Map(X, Z)$ verify

\begin{equation}\label{eq:1}
(hg)(f) = (h(gf))
\end{equation}

The map $f: X\longrightarrow Y$ is \emph{injective} if $x\neq y$
implies $f(x)\neq f(y)$; it is \emph{surjective} if $f(X)=Y$; it is
\emph{bijective} when it is both injective and surjective; in this case,
for finite sets, $|X| = |Y|$. So in this case also the inverse map
$f^{-1}$ can be defined, by $f^{-1}(y) = x$, when $x$ is the
\emph{unique} element in $X$ with $f(x) = y$, and there is a
bijection $X\longleftrightarrow Y$.\\

A map $f: X\longrightarrow   Y$ defines another one $F$, \emph{among
subsets} of $X$ into subsets of $Y$; although $f$ might not have an
inverse, $F$ \emph{always has}: indeed, $F^{-1} (V)$, where
$V\subset Y$, $V$ is a subset of $Y$, is the set $U\subset X$ of
elements in $X$ whose image spans $V$, so $F^{-1}(V) = U$;  if there
is none, we write $F^{-1}(V) = \emptyset $, still a subset of $X$.\\

We use the notation $X \backslash U$ to mean the set $X$ without the
subset $U$: if $U$ contains only an element, say $e$, we write $X
\backslash \{e\}$ .\\

General references here are e.g. the books by Lang [22] and
Birkhoff-MacLane [23].\\

A \emph{diagram}, in general, is a collection of sets and arrows
(maps); it is \emph{commutative} if the final result does not depend
on the path taken.  A diagram with a single line is called a
\emph{sequence} (\emph{suite}, fr.).\\

Example of sequence

\begin{equation}\label{eq:2}
A\longrightarrow  B\longrightarrow  C
\end{equation}

Example of diagram

\small{\begin{equation}\label{eq:3}
\begin{CD}
A @>>>B @>>> C\\
@VVV @VVV @VVV\\
D @>>> E @>>> F\\
\end{CD}
\end{equation}

It is supposed to be commutative, so the route $ABE$ is the same as
$ADE$, etc.\\

We write always $\mathbb{N}$ for the natural numbers 1, 2, 3,
\ldots, $n$, and $\mathbb{Z}$ for the integers, 0, $\pm1$, $\pm2$,
etc., forming an Abelian \emph{infinite} group and $\mathbb{Z}^+
\equiv \{0\}\cup \mathbb{N}$ (called non-negative integers); of
course, $\mathbb{N}\subset \mathbb{Z}$, $\mathbb{Z}^+\subset
\mathbb{Z}$ and $|\mathbb{N} | = |\mathbb{Z} | = \infty$.
$\mathbb{Q}$ means the \emph{field} of rational numbers
($\exists$ $m/n$, $n\neq0$), with $\mathbb{R}$ the real field, and
$\mathbb{C}$ the field of complex numbers. Recall the definition of
$\mathbb{R}$ (and hence that of $\mathbb{C}$ also) requires some
kind of \emph{transfinite}
induction [24].\\

The (Cartesian) product set $X\times Y$ is the set of
ordered pairs $(x, y)$. The \emph{graph} of a map $f:
X\longrightarrow   Y$ is the subset $(x, f(x))$ in $X\times Y$.\\

An equivalence relation in a set $X$, named
$\mathcal{R}$ or $x\mathcal{R}y$, is a relation between two elements
in $X$ which is \emph{reflexive}, $x\mathcal{R}x$, \emph{symmetric}
$x\mathcal{R}y \Longrightarrow y\mathcal{R}x$, and
\emph{transitive}, $x\mathcal{R}y$ and $y\mathcal{R}z$
$\Longrightarrow x\mathcal{R}z$. It partitions the elements of $X$
into \emph{disjoint classes}, so $X =\cup\{classes\}$. For example,
in the natural numbers $\mathbb{N}$ the relation $x\mathcal{R}y$
given by $x - y$ even is of equivalence, and divides the set
$\mathbb{N}$ in two classes: even and odd numbers. Conversely, to
define a \emph{partition} in a set $X$ means to express $X$ as union
of disjoint subsets, $X =\cup H_i$, with $H_i\cap H_j =\emptyset$.
\emph{Warning}: for any natural number $n\in \mathbb{N}$, a
partition means expressing $n$ as sum of natural numbers; if
$Part(n)$ is the number of possible partitions of $n$ , we have
$Part(3) =3$,
$Part(4) = 5$, $Part(5) = 7$, etc.\\

A partial ordering in a set $X$, written $x \leq y$, is
a relation which is reflexive ($x \leq x$), antisymmetric ( $x \leq
y \Longrightarrow y \geq x$) and transitive ($x \leq y$  and $y \leq
z \Longrightarrow x \leq z$). If the relation holds for all, i.e. if
either $x \leq y$ or $y \leq x$ for any pair $(x, y)$, the ordering
is \emph{total}. For example, we shall see that the set of subgroups
$H$ of a group $G$ is partially ordered, by inclusion; on the other
hand, the integers $\mathbb{Z}$ is a totally ordered
set.\\

1.2.2 \emph{Algebra and topology}. Historically, geometry and number theory were the first branches of
mathematics, started by the Chinese, the Indians, and the
Babylonians; one associates \emph{algebra}, as the word itself, to
the Arabs, for solving equations (algebra in Arabic means
``reparation of a broken member''). Analysis came to be the main
branch of applied (and pure) mathematics after Newton and Leibniz,
already in the 17th century; the 18th century in
mathematics is dominated by the name of Euler, while in physics
and other branches of natural science it is an impasse century, to
be much revitalized in the 19th (mathematics, chemistry) and 20th
centuries (physics): the great century in mathematics is really the
19th. Today, all branches of mathematics start with the big split
between algebra and topology, with all other branches included as
part of these, and therefore we start also by this dichotomy. It is
claimed
sometimes that [25] mathematics stem on \emph{four} concepts:  \emph{number}, \emph{set}, \emph{function}, \emph{group}. The later will be defined soon.\\

The two big branches of \emph{modern} mathematics are topology and algebra. Both start with the
concept of \emph{set} $X$, and that of \emph{map}, $\mu:
X\longrightarrow Y$, as recalled above. Geometry became
subsumed by topology already in the 20th century, and
analysis and number theory, much more older
branches, are also related to algebra. Since 1940,
category theory structures much of organizational
mathematics.\\

In topology, we select in a set $X$ a family
$\mathcal{O} = \mathcal{O}(X)$ of subsets among all of them
$\mathcal{P}(X)$: $\mathcal{O}(X)\subset \mathcal{P}(X)$, and called
this the family of ``open sets''; this family must be \emph{stable}
under arbitrary \emph{unions} of these subsets and (finite)
\emph{intersections}:

\begin{equation}\label{eq:4}
 \cup o_i\in \mathcal{O};\quad \quad  o_1\cap o_2\in\mathcal{O};\quad \emptyset\quad  \textrm{and} \quad X \quad \textrm{are  in} \quad
 \mathcal{O}
\end{equation}

So the \emph{empty} subset $\emptyset$ and the \emph{whole} space
$X$ are declared \emph{open}: Then $X$ becomes, by definition, a
\emph{topological space}. \emph{Closed sets} are the sets
complementary to open sets, hence $\emptyset$ as well as the whole
$X$ are both open and closed. Thus a \emph{topological space}
$(X,\tau )$ is a set $X$ endowed of a collection $\tau$ of subsets
(called \emph{open}), which are still open under arbitrary unions
and finite intersections, with the empty set and the whole space
included in the family.\\

Geometry comes as particular forms of topology. In
metric spaces $E$, whose topology is defined from the
metric (i.e. open sets verify distance relation $|x-y| < r$),
compact sets are the closed and \emph{bounded} ones. We take for
granted and known elementary topological concepts as
\emph{connectedness}, (the space $X$ presents itself in a single
piece), \emph{simple} connectedness (any loop (map $f$ from ($I:
0\leq x\leq 1$) to points, with $f(0) = f(1)$) is
\emph{contractible}, shrinks continuously to the constant map),
\emph{compact} etc. For a good elementary introduction see [26]; the
most natural property of a space, as we are used to consider it, is
its \emph{dimension}. There is a purely topological definition, but
we shall restrict to dimensions as defined for \emph{manifolds}
(following generally the standard book of Kobayashi-Nomizu, [27]).\\

We shall not touch many topological issues in this review, except
when we mention Lie groups. The characteristic property of
(topological) maps, i.e. maps among topological spaces, is
\emph{continuity}: a map $f$ among topological spaces $f:
X\longrightarrow   Y$ is \emph{continuous}, if the inverse map of
open sets is open: $F^{-1}$ (open set in $Y$) is an open set in $X$:
recall (just above) that for any function $f: X\longrightarrow   Y$,
the inverse function is always defined \emph{among the subsets}, as
$F^{-1}$($V$ in $Y$) = $U$ in $X$, such $f$($x$ in $U$) $\in V$. Continuous maps are the natural maps in topology. The equivalence of topological spaces is called \emph{homeomorphism}:
two topological spaces $(X,\tau  )$ and ($Y, \tau')$ are
homeomorphic, if there is a bijection $f: X\longrightarrow   Y$
which is \emph{bicontinuous}, that is, a bijection continuous from
$X$ to $Y$ with inverse continuous from $Y$ to $X$. In any set $X$
one can always define the \emph{trivial} (or \underline{discrete})
topology, $\tau_o$, in which any element
(point) is open (hence also closed).\\

Algebra is nowadays the study of algebraic structures.
An algebraic structure $\mathcal{A}$ in a set $X$ is
established by giving some \emph{composition} laws, either internal
or external.  A map $f: X\times X\longrightarrow X$ is an
\emph{internal} composition law; another map from $K\times
X\longrightarrow X$ is \emph{external}, where $K$ is another
algebraic structure, given in advance: (e.g. in a $K$-vector space
$(x+y)$ and $(kx)$ are well-defined operations). We shall need
usually one or two composition laws as given, one perhaps
external.\\

The main class of maps between algebraic structures is the natural map, called \emph{morphism}: a map between
\emph{analogous} structures $\mu: A\longrightarrow   A'$ preserving
the laws (defined precisely below in each case). For example, if
there is only an internal composition law, written, for $g$ and
$g'$, as ($gg'$), the \emph{natural} map  $\mu: A\longrightarrow A'$
verifies $\mu(gg') =\mu (g)\mu (g')$ for any pair $(g, g')$ in $A$.
Notice the word ``analogous'' above: for example, an Abelian
group $A$ might be isomorphic to the Abelian \emph{group} underlying
the sum in a vector space $V$, but it \emph{cannot} be isomorphic to
the \emph{vector space} as such: the group has only one composition
law, whereas the vector space needs two: they are \emph{not}
analogous!.\\

Certain structures and their allowed maps (morphisms) define a
\emph{category} (we elaborate a bit below); we shall loosely speak
of the category $\mathcal{T}op$ of topological spaces and continuous
functions, $\mathcal{G}$ will be the category of groups and
morphisms, or $\mathcal{A}b$ the category of Abelian groups, or even
$\mathcal{E}ns$ or \emph{Set}, the category of all sets and maps; see [28].\\

When two structures are equivalent? One introduces
\emph{different} concepts: \emph{homeomorphism} as equivalence of
topological spaces, and \emph{isomorphism} as equivalence for
algebraic structures; to repeat:

\begin{itemize}
\item \emph{Homeomophism} (as said) \emph{in topological spaces}. Two topological spaces $X$ and $Y$ are homeomorphic,
 written $X \sim Y$ or $X \approx Y$, if there exists a map $f: X\longrightarrow   Y$ which is bijective and
  bicontinuous, see above.

\item \emph{Isomorphism} \emph{in algebraic structures}. Two \emph{analogous} algebraic structures $K$ and
$H$ are isomorphic, when there is morphism $K\longrightarrow   H$
bijective, with the inverse map also a morphism: one supposes
automatically that \emph{morphism} means to preserve \emph{all}
composition laws. Then one writes, in general, $K \approx H$.
\end{itemize}

As said, according to Herman Weyl, the angel of Topology and the
demon of Algebra are always fighting each other in Mathematics
[16].\\

Notice the \emph{big conceptual difference between topological versus
algebraic structures} also referred to above: in the first
(topology), the elements are just ``points'', all structureless,
while the structure is established on collections of them, or in
maps between them. In algebra, in contrast, the elements
\emph{combine}, there is usually identity and inverses, etc.
Although the main frame for physics is the spaces (geometry),
things happening in spaces are described by action of some algebraic
agent (e.g. transformations through symmetries, etc.). Gauss
thought that algebra, i.e. numbers, existed only in our minds, while
geometry, i.e. points, exist independently of us. Today we
consider both Algebra and Topology to be a \emph{free} creation of
the human mind.\\

1.2.3 \emph{Algebraic structures.} There are SIX main algebraic structures $\mathcal{A}$ one should
consider:\\

One law, internal \quad \quad \quad               GROUPS\\

Two laws, internal $\left\lbrace
\begin{array}{l}
 \text{RINGS\quad \quad MODULES}\\
\quad\quad\quad\quad\quad\quad\quad\quad\quad\quad\quad\quad\quad\text{One internal, one external}\\
\text{FIELDS\quad VECTOR SPACES}\\
\end{array}
 \right.$\\

Two internal plus one external\quad     ALGEBRAS\\

Now we present a brief description of them:\\

1) In a $\emph{group}$ $G$, one has just an internal law, $G\times
G\longrightarrow   G\quad (g, h\longrightarrow gh)$ with unity (or
identity or neutral: $\exists e$, with $e·g=g=ge$), inverse (for all
$g$, $\exists g^{-1}$ with $g·g^{-1} = g^{-1}·g = e$) and
associative ($(gh)k = g(hk)$). The group consisting only of the
identity is named $I$: $I = \{e\}$. First example of groups are
$\mathbb{Z}_2$, with elements  $(e, a)$ with $a^2 = e$, or in
general $\mathbb{Z}_n$, the cyclic group, with a
\emph{generator} $b$ and the relation $b^n =e$. The symmetric group
$Sym_n$ or simply $S_n$ is also supposed to be known to the
reader. When the composition law is \emph{commutative}, i.e.
verifies $ab = ba$, we speak of an \emph{Abelian group}. The name
group is due to Galois (1832); the first modern definition is due
to Cayley, (1854). The structure of group is by far the most
important structure in mathematics (and in
physics!). For an introduction to Abelian groups, see [28].\\

\emph{Other structures}. It is convenient to have defined and at
hand other structures, as we shall need them also. Most of them (but
not all) were originated and named in Germany in the second half of
the 19th century.\\

2) A ring $R$ has two internal laws, sum and product:
the sum makes it an Abelian group, noted ``+'' with 0 as the neutral
element. The second law, the product, also internal, is noted
multiplicatively; it is associative $(xy)z = x(yz)$, and
distributive with respect the first law: $x(y+z) = xy + xz$. The
paradigmatic example of ring is the ring of the integers
$\mathbb{Z}$, with the usual addition and multiplication; the
notation $R^{*}$ is commonly designating the units of $R$, i.e. the
\emph{invertible} (for the product) elements; for example,
$\mathbb{Z}^{*} = \pm1$. French (Spanish) for ring is \emph{anneau
(anillo)}.\\

A more sophisticated example of a ring is: the set of endomorphisms
of an Abelian group $A$, $End(A)$, makes up a ring (the product is
the composition, and the sum is defined as $(\mu_1 + \mu_2)(a) :=
\mu_1(a) +\mu_2(a)$; one checks the distributive law. So $End(A)$ is
a ring).\\

3) A Field $\mathbb{F}$ is a ring in which any element
$k\neq0$ has an inverse under the product, so $k\cdot k^{-1} = 1$;
this multiplicative group is written then $\mathbb{F}^{*} \equiv
\mathbb{F} \backslash\{0\}$. The natural example is the field
$\mathbb{Q}$ of rational numbers $n/m$, with $n$, $m$ integers
($m\neq0$), but also the real numbers $\mathbb{R}$ and the complex
numbers $\mathbb{C}$ will be much used as fields, as supposedly
known. Notice the English concept of \underline{field} must be
translated as \emph{cuerpo} in Spanish, \emph{corps} in French and
\emph{K\"{o}rper} in German: for a generic field we shall use $K$.
The modern definition of a \emph{field} includes the
\emph{commutativity} of the product.  If only commutativity fails,
one speaks of a \emph{skew-field}; for
example, the quaternions of Hamilton are a skew-field.\\

For any field $\mathbb{K}$, the \underline{characteristic},
$Char(\mathbb{K})=\chi=\chi(\mathbb{K})$ is the minimum natural $n$
such $n\cdot e = 0$; if no finite $n$ exists, we say the
characteristic is \emph{zero}; for example, the rationals
$\mathbb{Q}$ have $\chi(\mathbb{Q})=0$, as the reals and the complex; we shall see that the characteristic $\chi$  is a prime number
$p$ or zero. We shall use also \emph{finite fields}, of order $q=
p^f$ with $p$ prime, $f$ natural number, named
$\mathbb{F}_q$, to be defined precisely later.\\

4) A module $M$ (or $R$-module) is an Abelian group
(composition law noted ``+'', unit 0) with a ring $R$ of operators,
that is, there is an external law $R\times M\longrightarrow M$ with
$m(x+y) = mx + my$, $l(m(x))= (lm)x$, $l$, $m\in R$. Our first
example will be the integers $\mathbb{Z}$ acting in any Abelian
group $A$ as $2\cdot a = a+a$, etc: any abelian group is
automatically a $\mathbb{Z}$-module (the reader should enjoy
providing a full proof by himself). Module theory is an important
branch of modern algebraic theories; we shall
say more about this later.\\

5) A vector space $V$ is a module in which the ring is a
field $K$; it is the first structure a physicist finds, so e.g.
$\mathbb{R}^3\approx V_3(\mathbb{R})$ is the usual $3D$-vector space
over the reals, with the natural operations $\mathbf{x} +
\mathbf{y}$ and $k\mathbf{x}$ defined as usual. To recall the
concept of \emph{dimension}, let us define $\{\mathbf{x}\}$ as the
ray of the vector $\mathbf{x}$, the set $\{k\mathbf{x}\}$, $k\neq0$
in $K$. Vectors $\mathbf{x}$, $\mathbf{y}$ are (linearly)
independent if one is \emph{not} in the ray of the other. The maximum
number of linearly independent vectors, if finite, is an invariant
of the vector space, called the dimension. We shall
consider mainly only finite-dimensional vector spaces. By $End(V)$
we mean the whole set of matrices $n\times n$ with entries in $K$,
if $V$ is a $n$-dim $K$-vector space; the invertible ones form a
group, denoted $GL(V)$ or
$GL_n(K)$.\\

6) In an algebra $A$ we have two \emph{internal} laws
($a+b$ and $ab$, making a ring) and an external one, with a field
$K$ operating such that $A$ is a $K$-vector space for the addition
in $A$, and also with the property that $\lambda\mu(x) =\lambda(\mu
(x))$, $\lambda$ and $\mu$ in $K$ etc. Matrix algebras, Lie algebras
and Jordan algebras are three generic examples, to be defined
precisely later, but now an approximation is made here to the first
two:\\

If $V$ is any $K$-vector space, the matrices $n\times n$ ( =
morphisms of $V$ in $V$) with entries in $K$, say $M$, are our first
example of an algebra, with the three laws $M + N$, $MN$ and $kM$.
Sometimes (e.g. in a Lie algebra) one \emph{omits} the associative
law for the product, and writes instead the Jacobi identity
(shown below).\\

It is good here to quote the first self-contained book on modern
algebra [29].\\

And, if $G$ is a Lie group, that is a finite-dimensional manifold
$\mathcal{V}$ with a compatible group structure, the space ``close''
to the identity becomes an algebra, called the Lie algebra of the
Lie group (discovered by Lie himself; the actual name is due to Weyl); the composition in a Lie algebra is written $[x, y]$, and
instead of associativity one has $[x,x]=0$ and the so-called Jacobi
identity $[x, [y, z]] = [[x, y], z] + [y, [x,
z]]$.\\

\emph{Morphisms}. The important maps among structures either
topological or algebraic are, as said, the natural maps,
i.e. these conserving the structure; for algebraic ones, recall: a
morphism $\mu$ among \emph{two analogous} (algebraic)
structures $A_1$, $A_2$ is a map preserving all the laws; for
example, if $A_1$ and $A_2$ are $K$-algebras, a morphism $\mu$  is a
map $\mu: A_1\longrightarrow A_2$ verifying three conditions:

\begin{equation}\label{eq:5}
     \mu(a+b) =\mu(a) +\mu(b);\quad \mu(ab) = \mu(a)\cdot \mu(b);\quad \mu(ka) =k\mu(a),\quad k\in \mathbb{K}
\end{equation}

(To repeat: In topological spaces $X$, $Y$, the natural maps are the
\emph{continuous functions}, $f: X\longrightarrow Y$ is continuous
if the preimage of an open set is open; notice $F(f)^{-1}$ is well
defined, as functions from subsets in $Y$ to subsets in $X$; recall
the empty set $\emptyset$ and the whole set $X$ are ``subsets'' of
the very set $X$).\\

    Books on algebra are legion; we wish just to add one,
    [30].\\

1.2.4 \emph{Category  theory}. A category $\mathcal{C}$ contains a set of \emph{objects}, $A$, $B$,
$C$,\ldots $ob(\mathcal{C})$; any two objects $A, B\in
ob(\mathcal{C})$ define a set $Mor (A, B)$, called the set of
morphisms of $A$ in $B$, which compose: for three objects $A$, $B$,
$C$, there is a composition law\\

\quad \quad \quad $Mor (A, B)\times Mor (B, C)\longrightarrow Mor
(A, C)$\\

\noindent with three conditions: (see e.g. Lang [22]).\\

CAT 1: $Mor(A, B)$ and $Mor(A', B')$ are \emph{disjoint}, unless
$A=A'$ and $B=B'$, then \emph{identical}.\\

CAT 2: $id_A\in  Mor(A, A)$ is the identity.\\

CAT 3: Composition is \emph{associative}: $f\in   Mor(A, B)$, $g\in
Mor(B, C)$ and $h\in   Mor(C, D)$ $\Longrightarrow $ $(h \circ g)
\circ f = h \circ (g \circ f)$.\\

$f\in Mor(A, B)$, is an isomorphism, if $\exists g\in
Mor(B, A)$, with $g \circ f = id_A$, $f \circ g = id_B$\\

$f\in Mor(A, A)$ is called an endomorphism. If
\emph{isomorphisms},
it becomes an automorphism.\\
\begin{lem}
For any object $A$ in category $\mathcal{C}$,
$Aut(A)$ is group.
\end{lem}

\quad \quad Some examples of categories:\\

\quad \quad Ex. 1: $\mathcal{E}ns$, the category of (all) sets and
(all) maps between them.\\

\quad \quad Ex. 2: $\mathcal{G}$, the category of (all) groups and
homomorphisms between them.\\

\quad \quad Ex. 3: $\mathcal{A}b$, the category of all Abelian groups with morphisms.\\

\quad \quad Ex. 4: The category $\mathcal{T}op$ of topological
spaces and
continuous maps.\\

\quad \quad Ex. 5: The category $\mathcal{D}iff$ of differentiable
manifolds and $C^{\infty}$ maps.\\

Categories are related by \emph{functors} in the following
(abbreviated) way: Let $\mathcal{R}$, $\mathcal{R}'$ be categories,
with objects $A$, $B$. A \emph{covariant} functor $F:
\mathcal{R}\longrightarrow \mathcal{R}'$ carries objects $A$ in
$\mathcal{R}$ to objects $FA$ in $\mathcal{R}'$, and morphisms $\mu$
in $Mor(A, B)$ to morphisms
$F\mu$ in $Mor (FA, FB)$, again with some natural conditions:\\

FUN  1: $F(Id_A ) = id_{F(A)}$\\

FUN 2: $f: A\longrightarrow   B$  and $g: B\longrightarrow   C$
$\Longrightarrow$ $F(g \circ f) = F(g) \circ F(f)$\\

The Functor is \emph{contravariant} if $F(g \circ f) = F(f) \circ
F(g)$.\\

A Functor between different (algebraic) structures may be
forgetful; two examples will suffice: Between Category
$\mathcal{G}$ and Category $\mathcal{E}ns$, the functor ``forgets''
the group structure (composition), as there is none in
$\mathcal{E}ns$. A functor between category $\mathcal{V}$, of
$\mathbb{K}$-Vector spaces and Abelian groups $\mathcal{A}b$
``forgets'' about the ($k$, $\mathbf{x}\longrightarrow
k\mathbf{x}$) operation in $\mathcal{V}$, as only the \emph{sum} is preserved.\\

For more references, see [23] or
[31].\\

\section{Generalities   about   groups}
\subsection{Elementary notions}

To repeat: a group structure in a set $X$ is defined by an inner
composition law:  $X\times   X\longrightarrow   X$ associative with
unity and inverse:
\begin{equation}
\begin{aligned}
&\quad (x, y)\longrightarrow   xy = z \quad \quad\quad\quad
\quad\quad \quad\quad \quad \textrm{Product or
Composition}\\
&\quad \exists\quad \textrm{$e$ unique, with}\quad ex = xe = x,
\forall x\quad \quad\textrm{Unity, neutral element}\\
&\quad \textrm{for any}\quad  x, \exists x^{-1}\quad \textrm{unique,
with}\quad  x x^{-1} = x^{-1} x = e\quad \quad
\textrm{ Inverse}\\
&\quad (xy)z = x(yz)\quad \textrm{for any triple}\quad x, y, z\quad
\quad\quad \quad\quad\quad \textrm{Associativity}
\end{aligned}
\end{equation}

This (modern) definition of group was first clearly stated by Cayley in 1854. \\

The ``model'' for the group structure are the \emph{bijective} maps
of a set $X$ on itself, $Map_{bij}(X, X)$: composition, unity,
inverse are natural, and associativity is automatic. When the set is
finite, $|X| = n$, the group is the symmetric group $Sym_n = S_n$,
with $n!$ elements.\\

Associativity can be extended to $a(b(cd)) = (ab)(cd)$ etc, so the
parenthesis are superfluous (but not the ordering!).  We shall call
$\mathcal{G}$ the category of all groups (and their morphisms),
$\mathcal{G}^0$ those finitely generated (e.g. $\mathbb{Z}$, the
integers) and $\mathcal{G}^{00}$ the finite-order ones, $|G| < +
\infty$. The \emph{order} of a group $G$ is the cardinal $|G|$,
supposed finite, $< + \infty$. We shall be busy with finite groups
(category $\mathcal{G}^{00}$); then each element $g$ has a
\emph{period}, that is, the smallest natural number $n$ such
$g^n = e$.
\newline
Period $g = 1\Leftrightarrow g = e$.\\

If the composition law is \emph{commutative}, i.e. if $ab= ba$ for
any pair $(a, b)$, the group is called Abelian. The
category $\mathcal{A}b$ of Abelian groups admits therefore the
subcategory $\mathcal{A}b^o$ of finitely generated Abelian groups
(e.g. integers $\mathbb{Z}$ with the addition), and the subcategory
$\mathcal{A}b^{oo}$ of finite Abelian groups (studied in detail in
section 2.7; e.g. the cyclic group of integers mod $n$: $Z/nZ =Z_n$). For
Abelian groups, we write sometimes $A = \stackrel{{\rm\,o}}A$,
meaning by $\stackrel{{\rm\,o}}A$ the ``opposed group'', with
composition law $(a, b)=ab$ given by $\{a,
b\}=ba$.\\

If $\mu$ is a morphism (or natural map) between two groups, $\mu :
G_1\longrightarrow G_2$ we have the \emph{exact sequence}\\
\begin{equation}
1\longrightarrow  \textrm{Ker}\ \mu \longrightarrow
G_1\longrightarrow \textrm{Im}\ \mu\longrightarrow 1
\end{equation}
where  $\mu^{-1}(e_2)\equiv$ Ker $\mu$, $\mu(G_1)\equiv Im
\mu\subset G_2$, and \emph{exactness} in the i-th place for a
general exact sequence means $Im(G_{i-1}\longrightarrow
G_i)=Ker(G_{i}\longrightarrow G_{i+1})$. $Ker$ stands for
Kernel (nucleus), and $Im$ for Image.
$\mu(G_1)\equiv$ Im $\mu$ is clearly a
\emph{subgroup} of $G_2$.\\

Ker $\mu$ is more: as $\mu(g^{-1}) =[\mu(g)]^{-1}$, we have, for
$g_0\in$ Ker $\mu$ and $g$
arbitrary\\

\begin{equation}
\mu(g\cdot g_0\cdot g^{-1}) =\mu(g)\cdot e\cdot(\mu(g)^{-1}) = e
\end{equation}

We shall say that the subgroup Ker $\mu$ is \emph{invariant under
conjugation}; such a subgroup is called normal or
invariant or distinguished (see section 2.3).\\

As mentioned, the $n!$ substitutions in a set of $n$ symbols compose
to make up the permutation group $Sym_n$ or $S_n$, the
prototype of finite groups, already alluded to. The first symbol can
go to any ($n$) places, then the second to $(n-1)$, the third to
($n-2$), etc, so $|S_n| = n!$. Even simpler is the notion of
cyclic group $Z_n$; if we have a regular polygon in the
plane, the rotations: any vertex to the next, generate this group,
of $n$ elements and abelian; we write this cyclic group (also
already mentioned) of $n$ elements as $Z_n =\{g; g^n=e\}$, where $g$
is a generator.\\

As ``incomplete'' known structures, we can mention: the natural
numbers $\mathbb{N}$ (1, 2, 3\ldots) have a sum without unit nor
inverse. The set of integers $\mathbb{Z}$ is an Abelian group under
the sum (as said), and a \emph{ring} considering the product. If $V$
is a $K$-vector space, the set of endomorphisms $End(V)$ are the
$n\times n$ matrices, with entries in the field $K$. The  matrix
product $(M, N\longrightarrow  MN)$ is associative, with unit but
not always with inverse. The product of octonion numbers
$\mathbb{O}$ (described later in this review) have unit and inverse
$o\longrightarrow o^{-1}$ (if $o\neq0$), but product it is
\emph{not} associative: so the octonion product does not generate a
multiplicative group. Matrices $n\times n$ with entries in
$\mathbb{K}$ with $Det\neq0$ have inverses, so they form the
(multiplicative) group $GL_n(\mathbb{K})$ studied in detail in  section 2.6.
By $\mathbb{Z}^+$ we mean the nonnegative integers, also the natural
numbers $\mathbb{N}$ plus the zero 0: they also do \emph{not} make
any of the six algebraic structures in section 1.2.3.\\

So groups are the simplest of the algebraic structures, and by far
the most important ones: the reasons will be clear along the work.\\

As symmetry of (geometric) figures, the notion of group is very
ancient, implicit even with the greeks. Lagrange, Gauss and Ruffini
are the ancestors of (abstract) group theory, as symmetry operations
in algebraic equations (and Gauss' congruences); the culmination of
the idea of group occurs with Galois (1832); for this theory,
see
e.g. [32].\\

Let $G$ be now a concrete finite group with $|G| = n$. Period 2
elements are called \emph{involutions} ($a$ involution: $a\neq e$,
$a^2 = 1$). If a group $G$ contains only involutions (besides the
identity $e$), it is abelian: $a^2 = b^2 = (ab)^2 = e$ imply $abab =
abba=e$, or $ab = ba$.\\

A finite group $G$, $|G| < +\infty$, is usually expressed by
\emph{generators and relations}. For example, $Z_n$, the cyclic
group of order $n$, can be specified (as said) as $\{g, g^n=e\}$: a
single generator and a single relation. The symmetric group $S_3$
can be defined by  $\{g^3 = a^2=e; a\cdot g\cdot a = g^2\}$,  two
generators and an extra relation. We shall see many more examples.\\

The symmetric group, $S_n$, of order $n!$, is non-abelian for $n >
2$. It has the subgroup of even permutations, called the alternating
group, $Alt_n$, of order $n!/2$; it is abelian for $n = 3$ (in fact,
$Alt_3 = \mathbb{Z}_3$).
We have $S_1 = I$, $Sym_2 = Z_2$; $Alt_1 = Alt_2 = I$.\\

The literature on groups is very extensive. We quote just here [33]
as a modern and complete textbook, and [34] as the most
 complete reference for \emph{finite} groups.\\

\subsection{The framework or box}

A group $G$ is the simplest algebraic structure, that is, there is a
single composition law $G\times G\longrightarrow G$, with identity,
inverse and asociativity. As for \emph{any} algebraic structure
$\mathcal{A}$ one considers, in principle, FOUR general situations
with homologous laws: \emph{substructures}, $B\subset \mathcal{A}$,
\emph{superstructures} ( or extensions) $\mathcal{A}\subset\hat{A}$;
natural maps or morphisms $\mu: \mathcal{A}_1\longrightarrow
\mathcal{A}_2$ or endomorphisms, $\mu: \mathcal{A}\longrightarrow
\mathcal{A}$. We would like to exhibit the four items in a box or
\emph{cadre}:\\

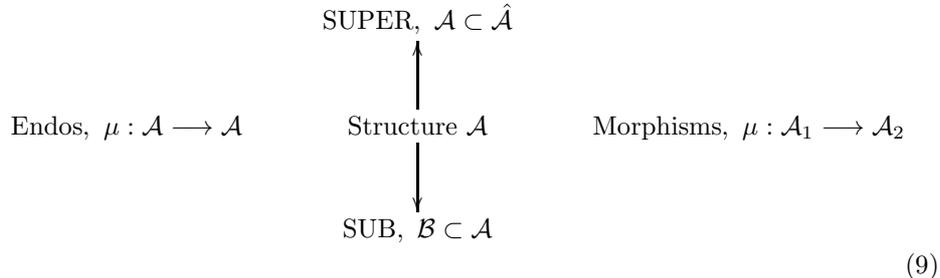
\begin{equation}
\xymatrix{& \textrm{SUPER},\ \mathcal{A}\subset
\hat{\mathcal{A}}\\ \textrm{Endos},\ \mu:\mathcal{A}\longrightarrow\mathcal{A}&
\textrm{Structure}\ \mathcal{A}\ar[d] \ar[u] & \textrm{Morphisms},\ \mu:\mathcal{A}_1\longrightarrow\mathcal{A}_2\\
& \textrm{SUB},\ \mathcal{B}\subset\mathcal{A}}
\end{equation}

Now we concentrate on groups $G$, and talk of subgroups $H \subset
G$, extensions $G\subset\hat{G}$, morphisms  $\mu:G\longrightarrow
K$,
etc.\\

Particular classes of morphisms are the invertible ones:
isomorphisms between two groups $\iota : G_1\Longleftrightarrow G_2$
and automorphisms (\emph{autos}) among the very same object $G$;
$\alpha : G\Longleftrightarrow G$. The set of \emph{autos} of a
given group $G$ makes up a \emph{very important} group (under
composition), as it contains identity and inverse, called $Aut(G)$.\\

Under \emph{autos} $\alpha$, the identity $e$ goes to itself; more
generally, the order is maintained: $g^n= e \Longrightarrow
\alpha(g)^n = e$ (proof is elementary, as $\alpha(g\cdot g) =
\alpha(g)\cdot\alpha(g))$ ).\\

For any $G$, $Aut(G)$ is an outstanding \underline{group}; for
example, $Aut(\mathbb{Z}_2) = I$ (the $a\neq e$ element has to go to
itself), $Aut(\mathbb{Z}_3) = \mathbb{Z}_2$ (interchange generator
$a$ with $a^2$), etc. In principle, there is NO relation between the
group $G$ and the group $Aut(G)$; in particular, as we shall see in
other examples, $G$ can be abelian and $Aut(G)$ nonabelian, etc.\\

For any algebraic structure, there is an \emph{enumerative} problem:
how many structures of certain type are there up to isomorphism? For
example, how many groups are with a given order $n$? To set the
problem properly, one needs first to state clearly when two
algebraic structures are (fully) equivalent. In general one can say
that two (finite) groups, $G$ and $H$ are equivalent, if there a map
1-1 between them, preserving the product in each: we take
\emph{isomorphic} groups as
\emph{equivalent} structures.\\

For groups, the enumerative question is an open problem even today,
although the \emph{abelian} case is solved (see later 2.7). We do
not know, ``a priori'' how many different groups of a given order
there are. To gauge the complexity, there are about fifty thousand
million groups of order $2^{10}=1024$ [35]. Simpler cases are also
solved: for example, for any natural number $n$, there is a single
\emph{cyclic} group of order $n$, which we label $\mathbb{Z}_n$. The
group with just the unit $e$ is
noted $I$ in this review; so $I =\{e\}$ .\\

There are no nonabelian groups of order less than 6: if $G$ is not
abelian, it contains (at least) two generators $a$ and $b$, with
$ab\neq ba$, but then, $e$, $a$, $b$, $ab$ and $ba$ are all
different. We shall see immediately that for $|G| = 5$ there is only
the cyclic group $\mathbb{Z}_5$, abelian; the smallest nonabelian
group is $Sym_3$,
of order 6.\\

We proceed now to a systematic study of these properties, in the
case of (finite) groups.

\subsection{Subgroups}

A subset $H$ of a group $G$, $H\subset G$, is a \emph{subgroup}, if
it is a group by itself, that is, it contains $e$, the product of
any two $h$, $h'$: $h''= h h'$, and the inverses $h^{-1}$ for each
$h$ are also in $H$. The identity $e$ and the whole group $G$ are
natural (improper) subgroups of any $G$, the (possible) others are
called \emph{proper} subgroups. For example $\mathbb{Z}_4= \{ a;
a^4=1\}$ has a natural proper subgroup, $\mathbb{Z}_2= \{e, b = a^2;
b^2 =e\}$. If $g\in G$ has order $n$, it generates the (sub)group
$\mathbb{Z}_n$. The elements $z$ obeying $zg=gz$ $\forall g\in G$
form a natural subgroup, called
the \emph{center} of group $G$; see below.\\

For a subgroup $H\subset G$, and $G  \ni g\notin H$, the set $gH
(Hg)$ is called the left- (right-) \emph{coset} (of $g$); one has
$|gH| = |Hg| = |H|$, as $g$ only reshuffles the elements in $H$.
Hence, $G$ is union of (e.g. left-) cosets, $G =\bigcup_{suff\
g}\quad gH$, each with $|H|$ elements, and it follows at once the
fundamental\\

\emph{Lagrange Theorem}: For $G$ finite, and $H$ a subgroup,
$|G| : |H|$; the quotient is called the \underline{index} of $H$ in
$G$, noted $[G:H]$. Two consequences:

\begin{itemize}
\item [(1)] $\mathbb{Z}_p$, the cyclic groups of \emph{prime} order, are the
\emph{only groups} with no proper subgroups.

\item [(2)] Any element $g\in G$ and its powers $g^2,\ldots, g^m=e$ generate a $\mathbb{Z}_m$ subgroup.
\end{itemize}

Lagrange's is the first of the fundamental theorems on finite
groups. As
other consequences,\\

\begin{lem} $G$ is of \emph{even} order iff it contains
involutions (Cauchy); if so, the number of them is odd.
\end{lem}

$\mathbf{Proof}$: If $a$ in $G$, $a^2=e$, $\{e, a\}$ make up the
$\mathbb{Z}_2$ subgroup; hence, $|G|$ even from Lagrange's theorem.
If $|G|$ even, couple any $g$ with the inverse $g^{-1}\neq g$. Only
$e$ and involutions $a$ are left over, hence even number; so
 number of involutions
$a$ is odd.\\

\begin{lem}
 $|G|$ is divisible by prime $p$ if it contains
elements of order $p$.
\end{lem}

 $\mathbf{Proof}$, like above; please note the
second part of previous lemma does \emph{not} follow: for example,
the number of period-3 elements is even (if $a$ is \emph{cubic}, so
$a^3=e$, $a^2$
is also \emph{cubic}). See [33].\\

Call $H\subset G$ \emph{normal}, if it is invariant under
congugation, so $gH = Hg$ (as sets), $ghg^{-1}=h'$. In particular,
if $H$ has index 2, there is only the subgroup, $H$ and a coset, say
$gH$; hence $Hg = gH$, and
$H$ is normal in $G$:\\

\begin{lem} Any subgroup of index 2 in normal (only one
coset,
so $gH = Hg$).
\end{lem}

\begin{lem} If $\mathbb{Z}_2$ is normal, it
is central (as $g\cdot\mathbb{Z}_2\cdot g^{-1}=\mathbb{Z}_2\Longrightarrow gag^{-1}=a$).
\end{lem}

So, as a \emph{normal} subgroup $H\subset G$ is invariant under
conjugation, i.e., $g\cdot h\cdot g^{-1} = h'$, defines a product in
the cosets, as $(gH)\cdot (g'H) = (gH\cdot Hg') = (gg'H)$, and one
obtains a factor or quotient group $Q$,
noted $G/H$ for $H$ normal in $G$, as the natural composition of
cosets. So
$H\longrightarrow G\longrightarrow G/H=Q$. One also writes (see (7))\\

\begin{equation}
1\longrightarrow  H\longrightarrow   G\longrightarrow
Q\longrightarrow   1
\end{equation}
as an exact sequence, just meaning $H$ normal in $G$, and $G/H
\approx
Q$.\\

 \emph{Theorem of Cayley}: Any finite group $G$, with $|G| = n$, can be
considered as subgroup of the symmetric group $Sym_n$.\\

This is obvious, as $Sym_n=S_n$ is the maximal group permuting $n$
symbols. Therefore, in a way the symmetric group is the most general
finite group; this does not help much in finding all finite groups,
because the subgroups of $S_n$ are not yet classified!, and recall
the order: $|S_n| = n!$, growing very fast with $n$.\\

The map $\alpha_h: g\longrightarrow h g h^{-1}$ is (homo-)morphism
of $G$ in $G$, in fact an automorphism, called \emph{inner}
automorphism; $g$ is left fixed under $h$ if and only if it commutes
with it. The elements commuting with all others constitute a special
subgroup, as said, called the center of the
group, $Z_G$ or $Z(G)$; if we call $Int(G)=Inn(G)$ the set of inner
\emph{automorphism}, we have the
exact sequence\\

\begin{equation}
1\longrightarrow Z_G\longrightarrow   G\longrightarrow
Inn(G)\longrightarrow   1;
\end{equation}
$g$ and $k$ are \emph{conjugate} if $k = j\cdot g\cdot j^{-1}$ for
some $j$ in $G$. ``Conjugacy'' is a relation of equivalence (trivial
proof), so it partitions $G$ into classes (of conjugate
elements). $G$ is abelian iff each conjugacy class has only a
member. For example, in the smallest non-abelian group, which is
$S_3$, with 3! = 6 elements, there are \emph{three} classes: $e=
(1)(2)(3)$; $(12)(3)$, the three transpositions (fixing 3, 2, and
1); and the cycle (123) and its square (132). Each class $i$ has a
stabilizer subgroup $H_i$,
so $\sharp cl(i)\cdot |H_i|=|G|$.\\

A group with no proper \emph{normal} subgroups is called a
simple group. Simple groups are the \emph{atoms} in the
category of groups $\mathcal{G}$, that is, any group is either
simple or composed (in a certain sense, to be explained) of smaller
groups; for example, $\mathbb{Z}_p$ is simple for $p$ prime number:
it has no proper subgroups at all (by Lagrange's theorem). In this
review we shall be busy searching for the FSG; our first result is worth stressing\\

\begin{lem}
Let $A$ be abelian and simple; then $A
=\mathbb{Z}_p$ for any prime
number $p$.\\
\end{lem}

The \emph{commutator} of two elements $g$, $k$ $\{ g, k\}$, is
defined by $g\cdot k\cdot g^{-1}\cdot k^{-1} \equiv\{  g, k\}$, and
it is $e= Id$ iff $g$ and $k$ commute: the set of commutators of any
group $G$ generates a normal subgroup, called the
commutator subgroup, noted $G'$ or Der($G$)) or
$\Omega_G$. The quotient group $G/G'$ is obviously abelian (all
``noncommutativity'' is enclosed in the kernel, or  commutator
subgroup), it is called the \emph{abelianized} group, $Ab(G)
= G/G'$:\\

\begin{equation}
1\longrightarrow   G'\longrightarrow   G\longrightarrow   Ab(G)
\equiv G/G'\longrightarrow   1
\end{equation}
and one shows easily that

\begin{lem}The map $G\longrightarrow   Ab(G)$ is the
\emph{maximal} abelian image of
$G$ (under morphisms).
\end{lem}

For example, $Alt_n$ is the commutator subgroup of the symmetric
group $Sym_n=S_n$ (for $n > 4$), as the quotient is $\mathbb{Z}_2$
and
$Alt_{n>4}$ is simple (see section 2.8).\\

As said, an automorphism is a map $\alpha: G\longrightarrow   G$,
\emph{invertible} (and morphism, of course); their set $\{\alpha\}$
form, as said, a (new) group, called the group of
automorphism, $Aut(G)$, but the concept is more general: for any
algebraic structure $\mathcal{A}$ (or even geometric structure
$\mathcal{V}$), the set of bijective maps preserving the structure
is always a group, called $Aut(\mathcal{A})$, (or
$Aut(\mathcal{V}$)). For groups, one distinguishes \emph{inner}
automorphisms (as conjugations; see above) from general,
\emph{external} automorphisms; it is also easy to prove that inner
\emph{autos} $Int(G)=Inn(G)$ are a normal subgroup of $Aut(G)$; the
quotient is called \emph{the group of classes of}
(external or outer) \emph{automorphisms}: $ Aut(G) / Inn(G) := Out(G)$:\\

\begin{equation}
1\longrightarrow   Int(G)\longrightarrow   Aut(G )\longrightarrow
Out(G)\longrightarrow   1
\end{equation}

The following Diagram, called ``the cross'' explains, for any group
$G$ part of what we
have said\\

\begin{equation}\label{eq:58}
\begin{CD}
@. Z_G @. @. @. @. @. @. @.
@. @.\\
@. @VVV @. @.\\
G' @>>> G @>>> Ab(G)
@. @.\\
@. @VVV @. @.\\
@. Int(G) @>>> Aut(G)
@>>> Out(G)\\
\end{CD}
\end{equation}

For example, if $S_3$ is the symmetric group of three symbols, of
order 3!=6, the above structure is\\

$\xymatrix{& & 1\ar[d] & & \\1\ar[r] &\mathbb{Z}_3\ar[r]& S_3\ar@{=}[d]\ar[r]&\mathbb{Z}_2\ar[r]&1\\
& &Inn(G)=S_3=Aut(G)\ar[d] & &\\& & 1& &} $
\\

As said, a group is called \emph{simple} if it has no
\emph{proper normal} subgroups; it is called \emph{complete} if it
has neither outer automorphisms, nor center $\neq I$. It is
equivalent to say: the conjugations, $k\longrightarrow gkg^{-1}$
exhaust all automorphisms and $G\approx Aut(G)$.  It is called
\emph{perfect}, if $Ab=I$. $Z_p$ ($p$ prime) is an (abelian and)
simple group. $S_3$
is complete, and $SL_2(\mathcal{R})$ is perfect. More examples later.\\

A subgroup $H\subset   G$ is called \emph{characteristic}, if it is
invariant under all automorphisms. Let us prove: the center $Z(G)$
is characteristic: in $g\cdot z\cdot g^{-1} = z$ apply a generic
\emph{auto} $\alpha$ : $\alpha(g\cdot z\cdot g^{-1})
=\alpha(g)\alpha(z)\alpha(g)^{-1}$: when $z$ runs through the
center, and $g$ runs over all $g's$, so $z'=\alpha(z)$ is still
central. The same argument applies to the commutator or derived
subgroup
 $G'$: because $gkg^{-1}k^{-1}$ remains a commutator under any \emph
 {auto} $\alpha$. So both $Z_G$ and $G'$ are characteristic.\\

\emph{Partition by classes}. Let $G$ be a finite group; as
said, the class of the element $g$ is the set $cl(g): =
\{ kgk^{-1}\}$ for all $k$ in $G$. The identity $e$ is class by
itself, and so are the elements in the center $Z_G$; if $G = A$
abelian, there are one class per element, and viceversa: if all
elements are a class by themselves, the group is abelian.
``Belonging to a class'' in a group $G$ is an equivalence relation
(proof very easy), so it partitions $G$ into disjoint subsets. For
example, for $G = \mathbb{Z}_2, \mathbb{Z}_3$ we write

\begin{equation}
|\mathbb{Z}_2| = 2 = 1\cdot1_1(e) + 1\cdot1_2(a),\quad |\mathbb{Z}_3| = 3 = 1\cdot1_1(e) + 2\cdot1_3(a, a^2)\\
\end{equation}
meaning e.g. for $Z_2$: there is one class of 1 element of order 1
(the identity, $e$) and another class of 1 element, of period 2
($a$), etc. Let us
prove: \emph{elements in the same class have equal order}. Proof:
$a^n=e\Longrightarrow (g\cdot a\cdot g^{-1})^n =(ga g^{-1})\cdot(ga
g^{-1})\ldots = (g\cdot a^n\cdot g^{-1}) = e$.\\

For the smallest nonabelian group, $S_3$, we have\\

 $|S_3| = 3! = 6 =1\cdot1_1(e) + 1\cdot3_2(12) + 1\cdot2_3 (123)$
 or:   $IA_1(e)$, $IIA_3$(12), $IIIA_2$ (123). For each class we write a representative
element; e.g. $1\cdot3_2(12)$ means: one class, with three elements,
of order two, for example (12), meaning: there is unity ($e$), and
three elements of order two (e.g.
the transposition (12)), and two elements of order three (123).\\

Also $ab$ and $ba$ are in the same class: $a^{-1}(ab)a=ba$

Write the order ($n$) of a finite group $G$ with $r$ classes as

\begin{equation}
n=c+h+h'+h''\ldots
\end{equation}
where $|G|=n$, $|Z_G|=c$, $h=[G:N]$, $h'=[G:N']$, $h''=[G:N'']$,
etc., where $c\geq1$ is the order of the center, $N'$, $N''$,
$N'''$, \ldots are centralizers(stabilizers) of the non-central
classes of $G$ (so $h(i)\geq2)$, etc; there are $r-c$ summands
$\{h\}$ in (16), as the number of classes in $G$ is $r$. The above
equation is called the \underline{class equation} and it is very
useful; of course, if $G$ abelian, $n=c=r$; if $G$ non-abelian and
simple, $c=1$.

\subsection{Morphisms}

Two groups (as types of algebraic structures) are \emph{isomorphic},
if there is an allowed invertible map between them (\emph{allowed}:
morphism; \emph{invertible}: one-to-one). For example, in the
abelian category $\mathcal{A}^{00}$, we have $\mathbb{Z}_6$ and
$\mathbb{Z}_2\times \mathbb{Z}_3$ isomorphic: if $a^6=e$, $a$
generates $\mathbb{Z}_6$; but $b=a^3$ and $c = a^2$
generate $\mathbb{Z}_2\times \mathbb{Z}_3$, and $bc$ is of order 6.\\

For an abelian group $A$, the set of endomorphisms, $End(A)$ or
\emph{endos}, makes up a \emph{ring}. $(\alpha  +\beta )(a)\equiv
\alpha (a) +\beta (a)$;  $(\alpha\beta )(a) = \alpha  ( \beta(a))$
for
$\alpha$, $\beta$ \emph{endos} and $a$ in $A$.\\

In particular, modules as algebraic structures (section 1.2.3)
are generated from abelian groups with \underline{a} ring of
\emph{endos} (not
necessarily THE ring of \emph{endos}). Let us prove:\\

\begin{lem}. The category of abelian groups $A$ and the
category of $\mathbb{Z}$-modules coincide.
\end{lem}

Define $2\cdot a = a+a$ for $a \in A$ abelian, and 2 in
$\mathbb{Z}$: that makes any abelian group a $\mathbb{Z}$-module;
but also, if $\mu: A\longrightarrow A'$ is a morphism between
abelian groups, $\mu (2a) =\mu(a+a) =\mu (a) +\mu (a) =2\mu(a)$,
hence any
morphism is still a morphism in the $\mathbb{Z}$-module category. Define $(-1)a=-a$ as the inverse.\\

To find $Aut(G)$ for an arbitrary group $G$ is important. As
automorphisms keep order ($a^n=e\Longrightarrow  \alpha(a)^n = e$),
one should select a set of generators, and see how they combine with
each other in search of automorphisms; for example, for $G =\mathbb{
Z}_4$, the only possible non-trivial automorphisms is the map
$a\longrightarrow a^3$, where $\mathbb{Z}_4$: $\{a; a^4=e\}$. For
non-abelian groups, one should search for classes of external
automorphisms directly: for these to exist, there must be more than
one class with the same number of elements of same order: For
example, let $Q$ be the so-called quaternion group, $Q$
= $\pm\{ 1, i, j, k\}$ with 8 elements, where $i^2 = -1$, etc. The
\emph{quaternion} numbers (W.R. Hamilton, 1842) form a skew field in
$\mathbb{R}^4$; if $q=u+ix+jy+kz$ with $u,x,y,z\in\mathbb{R}$, we
define $k=ij$, $i^2=j^2=k^2=-1$ and $ij+ji=0$; the skew field of the
quaternion numbers is called $\mathbb{H}$. One can also define
$\mathbb{H}$ by the set $q = (u, \mathbf{x})$ with $u\in\mathbb{R}$
and $\mathbf{x}\in\mathbb{R}^3$; then $q = u + \mathbf{x}$. One
defines the product $q q'$ as $qq'= (uu'- \mathbf{x}\cdot\mathbf{x}'
+ u\mathbf{x}'+ u'\mathbf{x} + \mathbf{x}\wedge \mathbf{x}')$, and
then the conjugate as $\bar{q} = (u - \mathbf{x})$,  and the norm is
$\mathcal{N}(q) =\bar{q}q\in\mathbb{R}$, $>0$, so
the inverse is $q^{-1} =  \bar{q} / \mathcal{N}(q)$, ($q\neq0$).\\

The class equation (16) is now\\

\begin{equation}
|Q| = 8 = 1\cdot1_1(e) + 1\cdot1_2(-1) + 3\cdot2_4 (\pm i, \pm j,
\pm ij)
\end{equation}

The three order-four classes can be permuted, and $Out(Q) = S_3.$\\

If $V = V(\mathbb{K})$ is a $n$-dim $\mathbb{K}$-vector space, the
endomorphisms are all matrices, as they verify $M(x+y) = Mx + My$
and $M(\lambda x) =\lambda Mx$, i.e., matrices keep the structure of
a vector space; so one writes $End(V_n(\mathbb{K})) =
Mat_n(\mathbb{K})$. The restriction to the invertible ones ($det M
\neq0$) makes up the $Aut$ group: $Aut(V_n(\mathbb{K})) \equiv
GL_n(\mathbb{K})$ = $\{$set of invertible matrices, under matrix
product$\}$. Recall, if $M$ and $N$ are invertible, $M+N$ needs not to be.\\

In any abelian group $A$, taking the inverse is an automorphism
(because $(gg')^{-1} = g'^{-1} g^{-1}$  and abelianess). If, in an
arbitrary group $G$, we have $\beta(gg')=\beta(g')\beta(g)$, we speak of $\beta$ as an antiautomorphism.\\

For the simple abelian groups $\mathbb{Z}_p$ we have\\

$Aut(\mathbb{Z}_p) = \mathbb{Z}_{p-1}$. \emph{Proof}: the $p-1$
elements $\neq e$ are on equal footing, so a generator $a$ ($a^p
=e$) can
go to any other power,   $\alpha(a) = a^q$ ($ q\neq 0)$ , e.g. $q=2$.\\

Let us prove:\\

\begin{lem}$G=\mathbb{Z}_2$ is the only group with
$Aut(G)=I$.
\end{lem}

 $\mathbf{Proof}$: if $G=A$ abelian, $a\longrightarrow
a^{-1}$ is automorphism; if $A$ contains only involutions, one
permutes them;
 and, if $G\neq\stackrel{{\rm\,o}}G$, conjugation is an automorphism.\\

\subsection{Extensions}
In the Cartesian product of two groups $G$ and $K$ we
establish a group \emph{law} naturally by\\

\begin{equation}
(g, k)\cdot (g', k') := (gg', kk')
\end{equation}
which is called the (group) \underline{direct product} of the groups
$G$ and $K$, $G\times K$. If both are finite, one has $|G\times K| =
|G|\times |K|$; for example, $\mathbb{Z}_2\times \mathbb{Z}_3 =
\mathbb{Z}_6$, but $\mathbb{Z}_2\times \mathbb{Z}_2\neq
\mathbb{Z}_4$ : $\mathbb{Z}_4$ has elements of order four, but
$\mathbb{Z}_2\times \mathbb{Z}_2$ has not. The smallest example is
this $V := (\mathbb{Z}_2)^2$, called F. Klein's \emph{Vierergruppe};
and $(\mathbb{Z}_3)^2$, with $\mathbb{Z}_9$ are the two possible
groups of order
9.\\

 One has also

 \begin{lem}: if $|G|$ and $|K|$ have no
common factors,
$Aut(G\times K) = Aut(G)\times Aut(K)$ (Because \emph{autos} keep order).
\end{lem}

Let now $A$ be an abelian group, and suppose there exists a map
$\mu: B\longrightarrow Aut(A)$ between another group $B$ and the
group of automorphisms of $A$; this permits a very important
construction, the \emph{semidirect} product $A\rtimes B$: there is a
new group (law) in the set ($A\times B$, still the Cartesian
product), in this
 way:\\

 \begin{equation}
 (a, b) (a', b') := (a +  \mu_b(a'), bb')
 \end{equation}
where $\mu_ b(a')$ is that element of $A$ obtained from $a'$ via the
automorphism $\mu_b$. We shall use this construction very often.
For example:\\

In any even order $2n$ there is a \emph{Dihedral group},
namely\\

\begin{equation}
D_n = \mathbb{Z}_n\rtimes \mathbb{Z}_2
\end{equation}
where the \emph{auto}morphism consists in taking the inverse, which
is \emph{auto} iff $A$ is abelian; for example, the smallest
dihedral groups are $D_3 = S_3$; $D_5$; $D_7$, etc. $\mathbb{Z}_2$
itself has no \emph{autos} $\neq e$, hence there is no $D_2$;
instead,
there are two groups of order four, as said, namely $\mathbb{Z}_4$ and $V := (Z_2)^2$. And $S_3 = D_3 = Z_3\rtimes Z_2$.\\

Let $G$ be a group and $Aut(G)$ be given. The \underline{holomorph}
$Hol(G)$ can be defined as the semidirect extension by the whole
\emph{Aut} group, so\\

\begin{equation}
Hol(G) := G\rtimes Aut(G)
\end{equation}

For example, $Hol(\mathbb{Z}_3) = \mathbb{Z}_3\rtimes \mathbb{Z}_2 =
D_3 = Sym_3=S_3$: extension by the automorphism $a\longrightarrow
a^2$ in
$\mathbb{Z}_3$; for another example ($V=(\mathbb{Z}_2)^2$):\\

$Hol(V) = V\rtimes S_3\approx S_4$: the three involutions $a$, $b$
and $ab$ in $V$ can be permuted.\\

For a nonabelian group $G$ to have external automorphisms, as said,
it must have more than one class (of conjugate elements) with the
same number of elements of the same period. For example, in $D_4$,
as $D_4= \mathbb{Z}_4\rtimes\mathbb{Z}_2$, we have:  with
$a^4=\beta^2=e$, $\beta\cdot a\cdot \beta = a^3$:

\begin{equation}
| D_4| = 8 = 1\cdot 1_1(e) + 1\cdot1_2(a^2) + 2\cdot 2_2
(\beta,\beta a^2; \beta a, \beta a^3) + 1\cdot 2_4 (a,
   a^3)
\end{equation}\\

The outer (class of) automorphism permute the \emph{two} clases (of
two elements each) of order two. One shows $Aut(D_4)\approx D_4$,
and the ``cross'' is

\begin{equation}
\begin{CD} @. \mathbb{Z}_2 @. @. @. @. @. @. @.
@. @.\\
@. @VVV @. @.\\
\mathbb{Z}_2 @>>> D_4 @>>> V
@. @.\\
@. @VVV @.\\
@. V@>>>D_4@>>>\mathbb{Z}_2 @.@.\\
\end{CD}
\end{equation}
\\

\subsection{Families of finite groups}

The following families of finite groups will be used in
the sequel:\\

Cyclic groups $\mathbb{Z}_n$, $n\in\mathbb{N}$: abelian,
order $n$; one and only one for each $n$; simple iff $n=p$ prime;
$\mathbb{Z}_1 = I$. $Aut(\mathbb{Z}_p) = \mathbb{Z}_{p-1}$, as any
$a\neq e$ can go, under \emph{autos}, to any other $a^m\neq e$.
$\mathbb{Z}_n$ it is the \emph{rotation} symmetry group of the
regular $n$-sided polygon; the alternative definition $\mathbb{Z}_n
= \mathbb{Z}/n\mathbb{Z}$ was implicitly used by Gauss in his
``congruences''.\\

Dihedral groups $D_n = \mathbb{Z}_n\rtimes\mathbb{Z}_2$;
order $2n$. Non-abelian; $D_3 = S_3$. As $\mathbb{Z}_2$ has no autos
($\neq Id$),``$D_2$'' should be the direct product,
$\mathbb{Z}_2\times \mathbb{Z}_2$. $D_n$ is the (full, orthogonal)
symmetry group of the
regular $n$-sided polygon.\\

For any abelian group $A$ one can define as said the dihedral
extension as $Dih(A) = A\rtimes \mathbb{Z}_2$ (\emph{auto} to the
inverse); for
example, $Dih(\mathbb{Z})\approx2\mathbb{Z}$ \emph{as set}.\\

\emph{Symmetric} Groups $S_n$, also $Sym_n$, also
$\Sigma_n$, sometimes $Perm_n$. Order $n!$, abelian only $S_2
=\mathbb{Z}_2$ ($S_1 = \{e\} = I$). Studied further, in section 2.8.\\

\emph{Even} permutations make up the alternating group, $Alt_n$,
also $A_n$; order $n!/2$. \underline{Simple} for $n
> 4$ (Galois, 1832); in particular $Alt_2 = I$, $Alt_3 = \mathbb{Z}_3$.
Also $Alt_4 = V\rtimes\mathbb{Z}_3$, as $Aut(V) = S_3$, and
$\mathbb{Z}_3\subset S_3$. $Alt_5$ (of order 60) turns out to be the
 smallest nonabelian \emph{simple} group. \\

One shows also: $Alt_4$ is the \emph{rotation} symmetry group of the
regular tetrahedron $T_3$, as $Sym_4$ the corresponding for the cube
$H_3$, and $Alt_5$ for the icosahedron $Y_3$; see e.g. [37].\\

$Q_n$ are called \emph{dicyclic}, order $4n$; $Q_n\equiv
\mathbb{Z}_{2n}\rtimes_{/2} \mathbb{Z}_4$ . By generators and
relations, it is $\{a^{2n}=b^4=e, a^n=b^2, b\cdot a\cdot b^{-1} =
a^{-1}\}$. For example (check!) $Q_1=V$, $Q_2 = Q$(quaternion group,
$\pm(1, i, j, ij))$, $Q_3$ (order 12)$\approx
\mathbb{Z}_3\rtimes\mathbb{Z}_4$.\\

$\Gamma_n$ are called finite \emph{Clifford} groups (group of Dirac
matrices; e.g. for $n=4$, the usual 4-dim. complex Dirac
matrices\ldots). Invent $n$ complex square matrices $\gamma_{\mu}$
satisfying
(Dirac)\\

\begin{equation}
\{\gamma_\mu,\gamma_\nu\}=-2\delta_{\mu\nu}\quad \textrm{$\mu,\nu$:
1 to $n$}
\end{equation}

Then, there is a finite group with $2^{n+1}$ elements, called the
\emph{Clifford group} $\Gamma_n$\\

\begin{equation}
\{\pm1,\pm\gamma_\mu,\pm\gamma_\mu\gamma_\nu,\ldots,\pm\gamma_5\},\quad
\textrm{where}\quad \gamma_5:=\gamma_1\gamma_2\dots\gamma_n
\end{equation}

The \emph{even} products $\pm1,\pm\gamma_\mu\gamma_\nu\ldots$ make
up the \underline{restricted} Clifford group $\Gamma_n^+$, with
order $2^n$.
See [36].\\

Given any group $G$, with $Aut(G)$ known, one forms, as said the
\underline{holomorph} as the semidirect extension with $Aut(G)$:\\

\begin{equation}
Hol(G) := G\rtimes Aut(G)
\end{equation}

As the \emph{inner autos} depend only on $G$ itself, so $Hol(G)$
would ``repeat'' something,  we shall mainly use the holomorph for
an \emph{abelian} group, that is $Hol(A) = A\rtimes Aut(A)$. Some
simple examples follow:\\

    $Hol(\mathbb{Z}_3) = \mathbb{Z}_3\rtimes\mathbb{Z}_2 = D_3 =S_3.- Hol(V) = V\rtimes S_3 =
    S_4,$
    $| Hol(\mathbb{Z}_5)| = 20$,   as   $Aut(\mathbb{Z}_5) = \mathbb{Z}_4$. Etc.\\

\emph{Groups up to two factors}. We have now enough information
to calculate the form of all groups up to order $|G| = pq$, i.e. two
(equal or unequal) \emph{prime} factors:

\begin{itemize}
  \item a) If $|G| = p$, as we argued repeatedly, there is only the group $\mathbb{Z}_p$:
  finite, abelian and simple.

  \item b) If $|G| = p^2$, there are only the two abelian groups of above for $p=2$, namely $\mathbb{Z}_{p^2}$ and
   $(\mathbb{Z}_p)^2$:
  if $a$, $b$ generate the group, $ab\neq ba$ implies there are more than $p^2$
  elements.

  \item c) If $|G| = pq$ ($p < q$), one has two cases: i) if $q-1:p$, we say $p$ and $q$ are \emph{compatible};
   then there is, besides the direct product $\mathbb{Z}_p\times \mathbb{Z}_q$, the semidirect
   product

   \begin{equation}
       \mathbb{Z}_q\rtimes \mathbb{Z}_p
    \end{equation}

\end{itemize}
because then $\mathbb{Z}_p$ can act as \emph{autos} of
$\mathbb{Z}_q$, as $Aut(\mathbb{Z}_q) = \mathbb{Z}_{q-1}$. $p=2$ is
always compatible,
as $q-1$ is even, so the dihedral groups enter here.\\

For example, for $|G| = 3\cdot7 = 21$, there is a\emph{ nonabelian}
``Frobenius group'' $G_{21} = \mathbb{Z}_7\rtimes\mathbb{Z}_3$, as
$7-1=6 = 3\cdot2$. But for $|G| = 15 = 3\cdot5$, when $p$, $q$ are
incompatible, there is only the \emph{(abelian)} direct product
$\mathbb{Z}_p\times\mathbb{Z}_q$.\\

\emph{Coxeter groups}. These are groups generated by
involutions $a$ $(a^2=e)$. They are defined once the order of the
product of two $(a_ia_j)$ is known. The book [41] is in part devoted
to them. We restrict ourselves to write simple examples:\\

$a$ alone generates $\mathbb{Z}_2$.- $a$, $b$ with $(ab)^2= e$
produces $V$.- with $(ab)^3=e$, it is $S_3$, $(ab)^m=e$ generates
$D_m$.\\

The diagram $\circ\!\!-\!\!\circ\!\!-\!\!\circ$ means
$a^2=b^2=c^2=(ab)^3=(bc)^3=(ac)^2=e$ and generates $S_4$ etc. The
finite Coxeter groups are all known: they make up the
symmetry groups of
polytopes, and the Weyl groups of simple Lie groups [41].\\

\emph{p-groups}. A group $G$ with $|G|=p^f$, power of a prime,
is called a \emph{p-group}; they
are also very important (see, e.g. [46]). For $f\leq3$, the number of possible groups is easy to count:\\

$f=1$: only $\mathbb{Z}_p$, as said. For $f=2$, only abelian, so
$\mathbb{Z}_{p^2}$ and $(\mathbb{Z}_p)^2$. For $f=3$, besides the
three abelian (see next Section), there are two non-abelian ones
(see Sect. 3.5 for $p=2$). Here we
just prove an elementary theorem on $p$-Groups:\\

\begin{lem} If $|G|=p^f$, the center is not trivial, i.e.,
for
$|G|=p^f$, $|Z_G|>1$.
\end{lem}

$\mathbf{Proof}$. Any subgroup and quotient of $G$ has order divisible
by $p$. Write the partition in classes:

\begin{equation}
p^f=c+h+h'+h''\ldots
\end{equation}
where $c=|Z_G|$, $h=[G:H]$, $h'=[G:H']$ etc., where $H$, $H'$,
$H''$\ldots are the stabilizers of the non-central classes. Now
as $p^f$ and $h$,$h'$\dots, divide $p$, \emph{also $c$ does}, as $h_i>1;$ the smallest possible center is $\mathbb{Z}_p$, \emph{qed}.\\

A (finite) $p$-group $G$ is called \underline{extra-special} if
$Z_G$ is cyclic and $Z_G=G';$ it follows that $G/Z_G$ is an
elementary abelian group [52].

\subsection{Abelian groups}

We write for $A$ abelian groups, $A =\stackrel{{\rm\,o}}A$. The
\emph{atoms} in the category $\mathcal{A}b$ of abelian groups are
the \emph{cyclic
groups of prime order}; we repeat:

\begin{thm}. $A$ abelian is simple iff $A = \mathbb{Z}_p$,
for any prime number $p = 2, 3, 5,\ldots$
\end{thm}

So now we consider the category of finite abelian groups,
$\mathcal{A}b^{00}$ (for a short introduction see [28]. See also [37]).\\

Any finite abelian group is the direct product of cyclic groups of
order power of a prime: this is the fundamental result; see e.g.
[22]. The partition by classes is also simple, e.g. for
$\mathbb{Z}_7$ we have $1\cdot1_1 + (p-1)\cdot1_p$ for     $p = 7$.
For this standard theorem, see again [22].\\

For any number $n$, it is easy to write down all abelian $A$ groups
of this order: first, write the prime factor decomposition of $|A|$,
say $\prod p_i^{n_i}$; then there are as many different abelian
groups as $Part(n_1) \cdot Part(n_2)\cdot\ldots\cdot Part
(n_{last}),$ where \emph{Part(n)} means the
partitions of the integer $n$ in natural numbers: here there are some results:\\

\begin{table}[h]
\begin{tabular}{ r r r r r r r r r r r r r r r r r r r r}
$n$& 1 & 2  & 3  & 4  & 5 & 6 & 7 & 8 & 9 & 10& 11 & 12 & 13 & 14 & 15 \\
Part($n$)&1 & 2 & 3 & 5 & 7 & 11 &15 &22 & 30 & 42 & 56 & 77 &101 & 176 &\ 231 \\
\hline
\end{tabular}
\end{table}

For example, there are 3 abelian groups of order $8 = 2^3$, namely
$\mathbb{Z}_8$, $\mathbb{Z}_4\times \mathbb{Z}_2$ and
$(\mathbb{Z}_2)^3$. For $|A| = 720 = 2^4\cdot3^2\cdot5$, there are
$5\cdot 2\cdot 1= 10$ abelian groups, etc. For $|A| = 1024 =
2^{10}$,
 there are Part(10) = 42 abelian groups.\\

For an abelian group $A$, the group $Aut(A)$, as said, does not have
much to do with $A$ itself: it could be non-abelian, of small or
bigger size, etc. The holomorph $Hol(A)$ for an abelian group $A$ is
the semidirect product $A\rtimes Aut(A)$; for example,
$Hol(\mathbb{Z}_3) = S_3 = \mathbb{Z}_3\rtimes \mathbb{Z}_2$, as
$Aut(\mathbb{Z}_3)=\mathbb{Z}_2$.
There are more examples in [20].\\

For example, $Aut(V) = S_3$: the three $\neq e$ elements $a$, $b$
and $ab$ can be arbitrarily permuted; later we shall use this
result; in particular, we shall see that $Hol(V) := V \rtimes S_3 =
S_4$. Of course, $Hol(\mathbb{Z}_3) = S_3$.\\

The abelian groups of structure $(\mathbb{Z}_p)^m$ ($p$ prime, $m$
arbitrary in $\mathbb{N}$) are called \emph{elementary abelian
groups}; we shall see later (Section 4) that they are the $m$-dim
vector spaces over the (finite) prime \emph{fields} $\mathbb{F}_p$.
Just an example: $V$ is like $\mathbb{F}_2^2$, which justifies the
notation
$GL_2(2) = Aut(V) = Sym_3$.\\

For example, there are 2 abelian groups of order $12 = 2^2\cdot3$,
namely $\mathbb{Z}_{12}=\mathbb{Z}_{4}\times \mathbb{Z}_3$ and
$V\times \mathbb{Z}_3=\mathbb{Z}^2_2\times\mathbb{Z}_3$.\\

For more of finite abelian groups see e.g. [2].\\

\subsection{Symmetric group}

\emph{Permutation groups}: We already mentioned several times
the symmetric or permutation group $S_n = Sym_n$, with $n!$
elements, and also the index-two subgroup, the alternative group,
$Alt_n$, with $n!/2$ . For small $n$ we repeat:

\begin{equation}
S_1 = I,\quad Alt_1 =I, \quad  Sym_2 = \mathbb{Z}_2,\quad  Alt_2 = I
\end{equation}

\begin{equation}
\begin{aligned}
&Sym_3 = D_3 = Hol(\mathbb{Z}_3),\quad Alt_3 = \mathbb{Z}_3\\
&\textrm{in classes}\quad 3!=6
= 1\cdot1_1+ 1\cdot3_2+1\cdot2_3;\\
\end{aligned}
\end{equation}

\begin{equation}
\begin{aligned}
&Sym_4 = V\rtimes S_3:\quad 4! = 24 = 1\cdot1_1 + 1\cdot6_2 +
1\cdot3_2 + 1\cdot8_3 + 1\cdot6_4\\
&Alt_4 = V\rtimes \mathbb{Z}_3:\quad 12 = 1\cdot1_1 + 1·\cdot3_2 +
2\cdot4_3
\end{aligned}
\end{equation}

\begin{equation}
\textrm{For $n > 4$, $Alt_n$ is \underline{simple} and}\quad
Sym_n/Alt_n = Ab(S_n) = \mathbb{Z}_2
\end{equation}

The \emph{conjugation classes} of the symmetric group are given by
the partitions of number $n$, as is well known, e.g.
[33]. The partitions can be labelled as Ferrer graphs with dots. We
specify just the $n= 4$ case: it has 5 partitions ([4],
[3,1], $[2^2]$, $[2,1^2]$ and $[1^4]$:\\

$e$ is the partition $[1^4]$.\\

(12) cycles are in $[2, 1^2]$: 6 of them.\\

(12)(34) are in $[2^2]$: 3 of them.\\

(123) are in the [3,1] class, with 8 elements.\\

 Finally, (1234) are in [4], with 6.\\

There are simple rules to compute the number of permutations in each
class $\approx$ partitions. For example, for the partition
(123)(45)(6) in $Sym_6$, the number is $6!/3\cdot2\cdot1 = 120$: the
stabilizers are the cyclic groups  $\mathbb{Z}_3\times
\mathbb{Z}_2\times \mathbb{Z}_1$; when there are repetitions, one
permutes them. For example, (12)(34)(56), still in $S_6$, has
$6!/2\cdot2\cdot2\cdot6 = 15$ elements, where the 6 is $|S_3|$, as
the three 2-cycles are to be permuted. See [33] for a detailed
explanation.\\

Any permutation is composed of \emph{cycles}, where e.g. (12), cycle
of two elements, is called a \emph{transposition}. For example,
$S_3$ or order 3! = 6, has three types of cycles, (1)(2)(3) as the
unit, (12), (23) and (13) as transpositions, and (123) and (132) as
3-cycles (as said). Any permutation can be written as product of
transpositions, and the parity of their number is an invariant:
hence even permutations, those obtained from an even number of
2-cyles, make up a subgroup, and being of  order two is normal: so
we have

\begin{equation}
Alt_n\longrightarrow   S_n\longrightarrow \mathbb{Z}_2\quad
(n>1,\quad \textrm{as}\quad S_1 =I)
\end{equation}

It turns out that, for $n\leq4$, the structure is very simple (as seen in (29) to (32)).\\

 For $n=5$ on we have the fundamental result of Galois (1832; see e.g. [32]):\\

 \quad \quad $Alt_{n>4}$    is    simple.\\

What about $Aut(Sym_n)$? We shall exhibit the case of $Sym_6$, the
only one with external automorphisms. The partition by classes is\\

$|Alt_6|: 6!/2 =
360=1\cdot1_1(e)+1\cdot45_2(12)(34)+1\cdot40_3(123)+1\cdot40_3(123)(456)
+1\cdot90_4(1234)(56) + 2\cdot72_5(12345)$\\

$|Sym_6|: 720 = 1\cdot1_1(e) + 1\cdot15_2(12)+ 1\cdot45_2(12)(34) +
1\cdot15_2(12)(34)(56)+ 1\cdot40_3(123)+1\cdot120_6(123)(45) +
1\cdot40_3(123)(456) +1\cdot90_4(1234)+ 1\cdot90_4(1234)(56) +
1\cdot144_5(12345) + 1\cdot120_6(123456)$\\

Notice $Alt_6$, besides the expected double class $2\cdot72_5$
(given rise to $S_6$), has also $2\cdot40_3$ as another potencial
\emph{outer} automorphism; later (in Sect. 5) we shall see the
relation with the smallest sporadic group, $M_{11}$; see in this
context [38].\\

Finally, the groups $Alt_6$ and $Alt_7$ have anomalous
Schur multipliers (see section 3.4).\\

For a general reference on permutation groups, see [43].\\

\pagebreak

\section{More advanced group theory}
\subsection{Groups operationg in spaces}

The normal use of groups, both in mathematics as in physics, is to
act as transformations on sets (spaces). According to Felix Klein,
geometries are characterized by the group of allowed
transformations. Modern physics abound in symmetry groups, that is,
groups of transformations leaving the physics invariant: for
example, the Lorentz group $O(3, 1)$ is the group of special
relativity; in particle physics $U(1)$ is the gauge group of
electromagnetism, $SU(3)$ is the ``gauge'' (color) group of strong
interactions, etc.\\

Here we \emph{categorize} this action, of groups $G$ acting on
spaces $X$ or $\Omega$. We use the notation $G \
\circ\!\!\longrightarrow X$, or $G \ \circ\!\!\longrightarrow
\Omega$ to distinguish from
$G\longrightarrow X$, reserved for morphisms.\\

Let a group $G$ and a space (or just a set) $\Omega$ be given. We
say that \emph{$G$ acts on $\Omega$} if there is a map
$G\times\Omega\longrightarrow\Omega$ verifying $e(x) = x$ $\forall
x$, and $(gg')(x) = g(g'(x))$, the natural ``transformation law'' in
$\Omega$ due to $G$. For example, if $\Omega$  is the 2-sphere $S^2$
and $G$ are the rotations $SO(3)$, $g\cdot x$ is the rotation of the
point on the sphere $x\in S^2$ by the rotation $g\in SO(3)$. If $X$
is a finite set, with $n$ elements, the maximal transformation group
is isomorphic to $Sym_n$ or $S_n$, as we have said; for any set $X$,
finite or not, write $Perm(X)$ the group of all permutations among
its elements. For another trivial example, a group $G$ acts on
itself at least in \emph{three} ways: on the left, as
$g:k\longrightarrow gk$; on the right, as $g:k\longrightarrow kg$;
and by conjugation, as\\

\begin{equation}
g:k\longrightarrow   g\cdot k\cdot g^{-1}.
\end{equation}

The definition $G \ \circ\!\!\longrightarrow X$ or $\Omega$ is
equivalent to the existence of a morphism  $\mu:G\longrightarrow
Perm(\Omega )$, because really $G$ does permute the elements in
$\Omega$ .  The action is called \emph{effective}, if Ker $\mu$ =
$I$; otherwise, is called ineffective. In this second case, there is
a natural action $G'(:=G/Ker\mu)\ \circ\!\!\longrightarrow X$, which
is, by construction, effective. Effective really means that no
elements in $G$, but the
identity, acts trivially (i.e., not moving any point) in the set.\\

For example, in Quantum Mechanics, it is the group $SU(2)$ which
performs rotations; it acts \emph{ineffectively}, and the effective
group is $SO(3) = SU(2)/Z_2$.\\

Consider again $G \ \circ\!\!\longrightarrow \Omega$ . Take $G(x)$
as the set of points  $\{g(x), \forall g \in G\}$: it is called the
\emph{orbit} of $x$ under $G$; it is a subset of $\Omega$ . Two
orbits either coincide or are disjoint, because ``belonging to an
orbit'' is an equivalence relation (trivial proof). Hence, under $G$
the space
$\Omega$ \emph{splits into a union of} (disjoint) \emph{orbits}; write\\

\begin{equation}
G =\bigcup_{\textrm{suff x}} G(x)
\end{equation}

Points which are orbits by themselves are called \emph{fixed
points}, for obvious reasons.  If there is only an orbit, we speak
of \emph{transitive} action (of $G$ on $X$). For each orbit $G(x)$
define the \emph{stabilizer} subgroup $G_x$ as the fixing set $\{ g;
g(x) = x\}$. It is trivial to show that points in the same orbit
have \emph{conjugate} stabilizers, so as abstract groups, stabilizers
characterize orbits, not just points; in physics stabilizer is
called, sometimes (Wigner), \emph{little group}, see [39].\\

As an example, consider the rotation group $SO(3)$ acting in the
vector space $\mathbb{R}^3$: the action is effective. The orbits
are: the origin, which is the (unique) fixed point, and the spheres
of arbitrary radius $r > 0$; the stabilizer of the fixed point is
the whole group, of course, but the stabilizers of the spheres are
$SO(2)$ (think of rotations around parallels, and the North and South
poles). If an action (of $G$ in $X$, say) is transitive with trivial
stabilizer, we say the action is \emph{free}; in the finite case one
has then $|G| = |X|$.\\

For example, in the (three) actions of a group $G$ on itself (see
above), left and right actions are free, i.e. transitive with
trivial stabilizer, while under conjugation, the orbits are the
classes of conjugate elements, the centrals $z\in Z_G$ are the fix
points, and each class has its own stabilizer, which is the whole
group for
centrals.\\

If we now suppose both $G$ and $X$ finite, for any point $x \in X$
we have:

\begin{equation}
|G| = |G_x|\cdot |G(x)|
\end{equation}

That is to say, points per orbit times order of the stabilizer
equals the order of $G$, (which is obvious).\\

Suppose now $G$ is transitive in $\Omega$ (= just an orbit), with as
stabilizer of point $x$ the subgroup $H\subset G$. It is obvious
that $H$ acts in $\Omega$  also, leaving $x$ fixed, so in
particular, in $\Omega\backslash\{x\}$ \emph{might} act transitively
also: in this case we say $G$ is \emph{doubly transitive} in
$\Omega$ . This is equivalent to taking two points $x\neq y$ to two
preestablished images, $x'\neq y'$: that is why the name. The
process can be iterated, and define the action of $G$ in $\Omega$
$k$-transtitive, if $k$ arbitrary distinct point ($x_1$,\ldots,
$x_k$) can be taken to $k$
preestablished distinct images $x_1'$,\ldots $x_k'$.\\

For example, $S_n$ acts naturally $n$-transitively in the set of $n$
points; it is easy to see that $Alt_n$ is only ($n-2$) transitive in
the same set, as $Alt_3 = \mathbb{Z}_3$, abelian with 3 elements,
acts \emph{free} in the 3-element set.\\

If $G$ is $k$-transitive in $\Omega$, we say it is \emph{sharp} or
strictly $k$-transitive if the last action leaves no
little group $> e$ (i.e. it is $I$). In this sense $S_n$ acting in
$n$ symbols is sharp $n$-transitive. We shall see that, besides
$Sym_n$ and $Alt_n$, actions more than 3-transitive are very rare:
that was the argument leading to the discovery of the first
\emph{sporadic groups}, the Mathieu groups (section 5). Also the free
 action of $G$ on set $X$ means the same thing as sharp 1-transitive action.\\

For example, let $Aff_1(\mathbb{R})$ be the affine group in the real
line, taking the point $x\in \mathbb{R}$ to $ax+b$, $a\neq0$: the
action is transitive, with stabilizer of 0 the dilations $a$: call
it $\mathbb{R}^*$: this acts in the complement  $\mathbb{R}^* =
\mathbb{R}\backslash\{0\}$ transitively, with the identity as
stabilizer: in other words, the action of this affine group in the
line is sharp 2-transitive. More examples later\ldots\\

Suppose $Pol_m$ is a regular polygon with $m$ sides (lying in a
plane): the cyclic group $\mathbb{Z}_m$ rotating orderly the
vertices is a symmetry group, as it is also the reflection in the
line through the center and vertices: the whole $2m$ operations make
up the
dihedral group\\

\begin{equation}
D_m=\mathbb{Z}_m\rtimes\mathbb{Z}_2
\end{equation}

For another example, we repeat Wigner's 1939 [39] analysis of
\emph{elementary quantum systems}; let $L$ be the (homogeneous)
Lorentz group acting in the $\mathbb{R}^4$ space of four momenta
$p_{\mu}$: the action is effective, with many orbits: any
hyperboloid $p_{0}^2 - \mathbf{p}^2 = m^2 > 0$ is an orbit, as well
as the origin $p_{\mu} \equiv 0$, the light cone $V_0$ ($m=0$) and
the ``spacelike'' hyperboloids ($m^2 < 0$). The little group is
$O(3)$ for $m^2 > 0$, the full $L$ for the origin (only fixed
point), the euclidean plane group $E(2)$ for the light cone, and
$O(2, 1)$ for the $m^2 < 0$
hyperboloids.\\

Wigner characterizes the elementary particles as mass, spin
(helicity) and sign of energy $[m, s,\varepsilon ]$ or $[0,
h,\varepsilon]$; he considers the ``covering group''
 $SL_2(\mathbb{C})$ of the Lorentz group: then the \emph{physical} little
groups for $m>0$ are $SU(2)$, with representations $s$ of dimension
$2s+1$ ($s$ = 0, 1/2, 1, \ldots) or $U(1)$ in the massless case,
with representation label $h$, the helicity; both the time-like
hyperboloid and the lightcone sets split into positive and negative
energy, which is the label $\varepsilon$. For example, the graviton
is $[m=0, h=2,
\varepsilon=+1]$.\\

For another example, take $\mathbb{C}^n$ as the $n$-dim. vector
space over the complex field; the set of complex invertible $n\times
n$ matrices makes up the group noted $GL_n(\mathbb{C})$: the action
on $\mathbb{C}^n$ is effective, with the origin 0 as the unique fix
point, transitive in the rest, $\mathbb{C}^n\backslash \{0\}$, with
stabilizer the affine group $Aff_{n-1}(\mathbb{C})$. See [40].\\

\subsection{Representations}

In mathematics it is very usual, when dealing with some objects, to
look for a ``visual'' characterization of them, making them
analogous (isomorphic) with some already known structure. E.g. for
real vectors in three-space one imagines lines drawn from a point.\\

For groups, the best image is perhaps to ``realize'' the group by
groups of matrices (under product); that started very early in group
theory [40]. This leads to the following
definitions:\\

A (linear) \emph{representation} of group $G$ in the vector space
$V$ (over some given field $\mathbb{K}$) is a realization of the
group as matrices (endomorphisms) in $V$, or more precisely, a
representation is a homomorphism $D$ (initial of the german
\emph{Darstellung}) into the group of invertible matrices:\\

\begin{equation}
D: G\longrightarrow   Aut(V) = GL_n(\mathbb{K})
\end{equation}
between our abstract group $G$ and the invertible matrices in the
$\mathbb{K}$-vector space $V$; the dimension of the representation
is that of the vector space. In physics the field $\mathbb{K}$ is
invariably $\mathbb{R}$ or $\mathbb{C}$, but the dimension could be
$\infty$; very often the unitary restriction $D(G)\subset U(n)$ is
enforced. We shall consider finite dimensional representations over
\emph{arbitrary} fields. As $D(eg) = D(e) D(g) = D(g)$, the identity
is
\emph{always} represented by the unit matrix.\\

We know that this is a fundamental tool when dealing with groups in
physics (in part because the physical space in Quantum Mechanics,
for example, is a (Hilbert, complex) vector space, and symmetries of
our physical systems must be realized as unitary transformation in
that space).\\

A representation $D: G\longrightarrow   Aut(V)$ is \emph{faithful}
if $Ker D = I$; that is, if it is effective, as action in the vector
space; otherwise it is unfaithful. It is \emph{reducible}, if there
is a closed subspace $W$ of $V$ such $D(G)W\subset   W$, that is,
$W$ is an \emph{invariant} subspace. If there is no such, the $D$ is
called irreducible. A reducible representation $D$ is
called \emph{completely reducible}, if it can be expressed as direct
sum of irreducible ones. For \emph{compact} groups, in particular
for discrete groups, all representations are completely reducible.
Two representations $D$, $D'$ of the same group $G$ in spaces $V$,
$V'$ are called \emph{equivalent}, if conjugate: there exists
an invertible map $f: V\longrightarrow V'$, with $D'(g) = f\cdot
D(g)\cdot f^{-1}$.\\

The search for irreducible inequivalent representations ($\equiv$
\emph{irreps}) is a formidable industry, developed during more than
a century ago (Frobenius, Schur), with plenty of applications in
mathematics and physics. For any group $G$ the \emph{identical}
representation $D_{id}(g)= e$ exists always and it is trivially
irreducible; as the set of \emph{irreps} is a well-defined one, one
has always to include the identical
\emph{irrep} in this family.\\

For example, for the simplest (cyclic) group $\mathbb{Z}_2=\{ a,
a^2=e\}$ , there are two \emph{irreps}, called $D_0$ and $D^-_0$,
with $D_0$ the identical $D_0(a) = +1$, and $D^-_0 (a) = -1$. For
the above groups $SU(2)$ and $SO(3)$, we have $D_j$, with dimension
$2j+1$, $2j$ integer, and the restriction to $j = l$= integer for $SO(3)$.\\

The sum $D\oplus D'$ and the product $D\otimes D'$ of
representations correspond to the same operations with
representative matrices (direct sum and tensor product of matrices).
An important problem is to decompose the product of two
\emph{irreps} $D_1$ and $D_2$ in a sum of \emph{irreps}: in quantum
physics this problem arises for the group $SO(3)$, where it is
called the \emph{Clebsch-Gordan Problem};
for example, if ``\emph{l}'' labels the \emph{irreps} of $SO(3)$, we have\\

\begin{equation}
D_l\otimes D_{l'} = \sum ^{l + l'}_{|l-l'|} D_k
\end{equation}

We include here an important result without complete demonstration:
Let $G$ be a finite group, of order $n$, with $r$ classes (of
conjugate
elements). Then

\begin{thm}. The number of \emph{irreps} for a finite group
$G$ coincides with the number of classes. The order of the group is
the sum of the squares of the dimension of the \emph{irreps}:
\end{thm}

\begin{equation}
|G| (=n) =\sum_{1=i}^r d_i^2
\end{equation}

\emph{Hint of the Proof}. (See e.g.[16]). We pass from the finite
group $G$ (order $n$) to the \underline{group algebra}
$\mathcal{A}_K(G)$, by multiplying formally the groups elements
$g_i$ by arbitrary numbers $k_i\in \mathbb{K}$:\\

\begin{equation}
\mathcal{A}_K(G) :=\{  x, x = \sum  k_i g_i\}
\end{equation}\\
which becomes a finite dimensional associative algebra in virtue of
the group law, when $g\cdot g' = g''$ generates $x\cdot x'= x''$. As
$\mathbb{K}$-algebra, dim $\mathcal{A} = |G| = n$, of course.\\

The center of this ``Group Algebra'' consists of all the elements of
the form \quad $\sum_g g\cdot k\cdot g^{-1}$ for any $k$, that is,
the conjugate
\emph{class} of the group element $k$. So\\

\begin{equation}
\textrm{dim (Center of $\mathcal{A}$) = number of classes of $G$,
say $r\leq n$ = Ord $G$}
\end{equation}
with equality ($r=n$) iff $G$ abelian. Now (this is the hard part of
the result!) it is a well-known fact in algebras that any matrix
algebra splits through the center in \emph{simple} matrix algebras,
of square dimension, as many as the dimension of the center; so in
our case, in addition to the theorem we have that $\mathcal{A}$
splits in $r$
simple algebras, each a square:\\

\begin{equation}
n := |G| =\sum   d_i^2=1^2+\ldots
\end{equation}

We shall often call (40) the \emph{Burnside relation}. Each simple
algebra supports an irreducible
representation of $G$, and any \emph{irrep} is so included!\\

As corollaries, we have

\begin{lem}: There are always 1-dim \emph{irrep} (because so
is the identical \emph{irrep}). For $G = A$ abelian (and only then),
all \emph{irreps} are
unidimensional ( as then $r = n$). For example\\

\quad \quad \quad $8 = 8\cdot1^2$\\
is the relation (43) for the three abelian groups of order 8.
\end{lem}

\begin{lem}: If $H$ normal in $G$, the \emph{irreps} of $G$
includes those of $G/H$, as the map $G\longrightarrow  G/H$ extends
to
$G\longrightarrow$ \emph{irreps} of $G/H$. In particular
\end{lem}

\begin{lem}: The number of 1-dim \emph{irreps} is the order
of the abelianized,
$Ab(G) = G/G'$. e.g. 2 for $Sym_n$, as $Ab=\mathbb{Z}_2=Sym_n/Alt_n$.
\end{lem}

Another result, not easy to prove (Simon) is this: the dim's of
the \emph{irreps} divide the order of $G$, $|G|: d_i$; see [48].\\

The simplest non-abelian case is the symmetric group $S_3$, with
order 6 and number of classes 3: so the only solution (for (40)) is
$6 = 2\cdot1^2 + 1\cdot2^2$: two \emph{irreps} are one-dimensional,
and the other one is bidimensional. Even for $|G| = 8$, the unique
solution for the nonabelian case is $8 = 4\cdot1^2 + 1\cdot2^2$, so
the two nonabelian order 8 groups (namely, $D_4$ and the quaternion
group
$Q$) have five classes, and a single matrix \emph{irrep}.\\

For $|G| = 12$ we have the first case of two Burnside relations,
both fulfilled:\\

\begin{equation}
\begin{aligned}
&\textrm{For}\quad Alt_4 = V\rtimes \mathbb{Z}_3,\quad \textrm{it
is}\quad 12 =
3\cdot1^2 + 1\cdot3^2\\
&\textrm{For}\quad  D_6 = \mathbb{Z}_2\times S_3,\quad  12 =
4\cdot1^2 + 2\cdot2^2
\end{aligned}
\end{equation}

The traces of the matrices of the \emph{irreps} define the \emph{character} of
the representation, $\chi_i(g) = Tr D_i(g)$, so  $\chi_i$ maps $G$
into $\mathbb{C}$. In particular, $Tr_i(e)$ = dim $D_i$. As
$Tr(ABC)=Tr(CAB)$, the trace is a class function: elements
in the same class have the same characters, and equivalent representations also.\\

Representations for direct and semidirect products: It is fairly
obvious that $D( G_1\times G_2) = D(G_1)\otimes D(G_2)$.

\begin{equation}
\begin{aligned}
&\textrm{e.g.}\quad D(Dih_6) = D(\mathbb{Z}_2\times S_3) =
2\cdot(2\cdot1^2 + 1\cdot2^2) = 4\cdot1^2 + 2\cdot2^2
\end{aligned}
\end{equation}

The semidirect product occurs so often that is worth to compute
\emph{irreps} given those of the factors (Wigner): we exemplify this
by the non-trivial case $S_4 = V\rtimes S_3 = Hol(V)$: we take the
four 1-dim \emph{irreps} of $V$ first; then let $S_3$ acts on them: the
$Id$ \emph{irrep} is fixed, so we are free to represent $S_3$
($2\cdot1^2 + 1\cdot2^2$); the other three are permuted under $S_3$,
with $\mathbb{Z}_2$ as stabilizer: the result is 2 \emph{irreps} of
dim 3: in total

\begin{equation}
|S_4| = 4! = 24 = 2\cdot1^2+ 1\cdot2^2 + 2\cdot3^2
\end{equation}

The group $Aut(G)$ operates in the set of \emph{irreps} of $G$: if
$=\alpha\in Aut(G),$ $D^{\alpha}(g):=D(\alpha(g))$; if
$\alpha$ internal, $D$ is equivalent to $D^{\alpha}$.\\

\subsection{Characters. Fourier series}

If $A$ is an abelian group, its \emph{irreps} are one-dimensional,
as said. Hence, the very \emph{irreps} coincide with their
\emph{trace} or character. For example, for the Vierergruppe $V =
\mathbb{Z}_2\times \mathbb{Z}_2$ the full
character table is ($a^2=e$ etc., so any number has to be $\pm1$):\\

\pagebreak

\begin{table}[h]
\begin{center}
\begin{tabular}{ r r r r r}
 & e & a  & b &ab \\
 \hline
$\chi_{0}$ & 1 & 1 &1 &1 \\
$\chi_{1}$ & 1 & -1 & 1 &-1 \\
$\chi_{2}$ & 1 & 1 & -1 &-1 \\
$\chi_{3}$ & 1 & -1 & -1 &1 \\
\end{tabular}
\end{center}
\end{table}

The set of characters fulfils a \emph{completeness relation} that we
are to exhibit in the context of Fourier series, which is no doubt
known to the reader. Consider the \emph{infinite} abelian group
$U(1) = SO(2)$ of rotations on the circle $S^1$. The \emph{irreps}
of $U(1)$ convert the additive group of angles $\phi$  (on the
circle) into
multiplication, so define (with the $\frac{1}{\sqrt{2\pi}}$ as the normalization) the 1d \emph{irreps} as\\

\begin{equation}
 \chi_n(\phi) := \frac{1}{\sqrt{2\pi}}e^{(in \phi)},\quad \textrm{for any}\quad  n\in
 \mathbb{Z}
\end{equation}

\emph{Completeness} of the characters $\chi_n(\phi)$ is shown in
that any complex function $f: S^1\longrightarrow  \mathbb{C}$ can be
expressed as expansion in the characters:\\

\begin{equation}
f(\phi ) =\sum_{n\in \mathbb{Z}} c_n  \chi_n(\phi )
\end{equation}

where\\

\begin{equation}
c_n  =\frac{1}{\sqrt{2\pi}}\int_{-\pi}^{+\pi} f(\phi )e^
{(-in\phi)}d\phi
\end{equation}

The set of characters $\chi$ of an abelian group $A$ forms the dual
group $\hat{A}$ under composition; in our case, we have $\hat{U}(1)=
\mathbb{Z}$ (the integers). Fourier analysis is just to express any
complex function from the group $A$ in terms of the ``basic''
functions, namely the \emph{irreps} of the dual group; the general
theory is due to Pontriagin (1940): the duality holds for all
locally compact abelian groups (LCA groups) [45].\\

From that one sees reasonable the orthogonality relations among the
characters; as this industry is well-known (e.g. see the books of
Weyl [5], Van der Waerden [6], or Wigner [7]), we just consider the
character table for $S_3$ ($3\times 3$, as there are three classes
$\equiv$3
\emph{irreps}):\\

\begin{equation}
6 = IA(e) + IIA((12)\quad\textrm{etc}) +
IIIA((123)\quad\textrm{etc}) = 2\cdot1^2 + 1\cdot2^2
\end{equation}

With the semidirect-product structure $S_3 = D_3 =
\mathbb{Z}_3(a)\rtimes \mathbb{Z}_2(\alpha)$ the character table is
inmediate:
we just write it\\

\pagebreak

\begin{table}[h]
\begin{center}
\begin{tabular}{ r r r r }
 & IA & IIA  & IIIA \\
 \hline
$D_0$ & 1 & 1 & 1 \\
$D_0'$ & 1 & -1 & 1 \\
$D_{2}$ & 2 & 0 & -1 \\
\end{tabular}
\end{center}
\end{table}

For the third, $D_2$, 2-dim \emph{irrep}, the $\alpha$  in
$\mathbb{Z}_2$ is antidiagonal; the $\mathbb{Z}_3$ normal subgroup
is diagonal, with entries  $\{\omega, \omega^2\}$, where  $\omega
=e^{(2\pi i/3)}$: so
the traces are 0 and -1, as $1+\omega +\omega^2 = 0$.\\

The unitarity relations are: let cl(1,2,3) be the numbers (1,3,2) of
elements per class; then\\

\begin{equation}
 \bar{c}_i c_j = (3!/cl(i))\cdot \delta_{ij}
\end{equation}

\begin{equation}
\begin{aligned}
&\textrm{For example,}\quad \bar{1}\cdot 1 = 1^2 + 1^2 + 2^2
=6,\quad \textrm{as}\quad  cl(1) = 1 = \{e\}.\\
& \bar{1}\cdot 2 = 0,\quad  \bar{3}\cdot3 = 3 = 3!/2,\quad \textrm{
as}\quad  cl(3) = 2
\end{aligned}
\end{equation}

The reader can verify (51) in the  $\chi$-Table for the
Vierergruppe, see previous page.\\

The mathematical reason for these orthogonality relations is that
finite groups are particular case of compact ones, and the biggest
compact complex group is the unitary group $U = U(n)$; a unitary matrix $u$ verifies $u^{\dag}=u^{-1}$. \\

Another property is full reducibility: if $D=D(G)$ is an arbitrary
representation, $D(G)\subset U$ $\Longrightarrow D^{\perp}(G)$ also
$\subset U$, so any $D(G)$ splits in sum of irreducible ones. For finite groups this was first stated by Maschke (1898).\\

\subsection{Homological algebra and extension theory}

If $K$(for kernel) and $Q$ (for quotient) are arbitrary, an
\emph{extension} $E = E(Q, K)$ of $K$ by $Q$ is roughly a group $E$
in which $K$ is a normal subgroup and $E/K=Q$. We have the exact
sequence

\begin{equation}
1\longrightarrow   K\longrightarrow   E\longrightarrow
Q\longrightarrow   1
\end{equation}

An extension is named \emph{split}, if $E$ is semidirect product,
$E\approx K\rtimes Q:$ that means there is
a map $Q\longrightarrow   Aut(K)$, as defined before.\\

As $K$ is normal in $E$, the conjugation in $E$ ammounts to a map
$E\longrightarrow    Aut(K)$; completing
 the diagram, we have\\

 \small{\begin{equation}\label{eq:3}
\begin{CD}
Z_K @. @. \\
@VVV @. @.\\
K @>>>E @>>> Q\\
@VVV @VVV @VVV\\
Inn(K)@>>>   Aut(K)@>>>Out(K)
\end{CD}
\end{equation}
\\

So any extension $E(Q, K)$ induces a map  $\mu: Q\longrightarrow
Out(K)$. $\mu$ is called the \emph{coupling} between $Q$ and $K$. On
the other hand, $Out(K)$ acts naturally in the center $Z_K$ because
conjugation is trivial in the center, so \emph{autos} mod internal
ones act identically, so they are classes of outer
automorphisms. By the coupling $\mu$, this generates a $G$-module
structure in $Z_K$; this gives rise naturally to
\underline{cohomology}, which
indeed is the right tool to deal with extension problems.\\

These two things (the coupling $\mu$ and the $G$-module structure in
$Z_K$ via $\mu$ ) are the
essentials for the extension theory. We shall express the theory very succintly.\\

Notice given $K$ and $Q$, there are \emph{always} extensions, as
$K\times Q$ is one. Indeed,
    the set $Hom(Q, Out K)$ contains always the ``zero'' homomorphism.\\

In \emph{extension theory}, there are three general questions:\\

The first question is: given a coupling $\mu : Q\longrightarrow  Out(K)$, does it produce extensions?\\

The second question is: if $\mu$  is ``good'', i.e, generates extensions, how many?\\

The third question is: when two extensions can be considered to be ``equivalent''?\\

In the following we shall give partial answers to these questions,
hinging more in the answers than in the arguments for them. We rely
 heavily in [33, chapter 11] and in [44, section 4].\\

The answer to the first question is: $\mu$ does not always generate
extensions. The precise cohomological answer will
be given later; we shall remark here two positive cases:\\

1) If $\mu$ is the zero homomorphism, there are always extensions ( $K\times Q$ is one)\\

2) If $K = A$ is abelian, any coupling $\mu$  (now  $\mu:
Q\longrightarrow   Aut(A)$)
does generate extensions, as the semidirect product $A\rtimes_{\mu}  Q$ always exists.\\

Let us consider in some detail the abelian case, $K = A$. A
\emph{section} $s: Q\longrightarrow   E$ will
 be a function such that $\pi\cdot s = Id_Q$, where $\pi$ is the projection $E\longrightarrow   Q$.
 To have a group structure, i.e. to form $E$, we ``compare'' $s$ in two
 points: $s(q)$ and $s(q')$, for $q, q'\in   Q$, with $s(qq')$: define  $\omega(q, q')$
 by the shift: $s(q) s(q') :=  \omega(q, q') s(qq')$. The functions $\omega$ live in $K$;
 they are called \emph{factor sets}. Associativity in $E$ makes a restriction in $\omega$,
 and changing the section (for the \emph{same} extension) $s$ to $s'$
 defines an equivalence relation: the factor
 sets $\{\omega\} $ with these two restrictions
 is written $H = H^2_{\mu} (Q, K)$ and named the \emph{second cohomology group}
 of the $Q$-module $A=K$; we cannot elaborate, unless extending this section very much.
 One shows, as conclusion\\

 \begin{equation}
 \textrm{Extensions $E$ with abelian kernel with respect to the coupling $\mu\approx H^2_{\mu} (Q, A)$}
 \end{equation}

The answer to the first question, namely when a morphism  $\mu:
Q\longrightarrow    Out(K)$ will generate extensions, and how many
there are is this: first, any $\mu$ endows $Z_K$ with a $Q$-module
structure, as said. Then, it is shown that the same $\mu$
``percolates'' to the third cohomology group $H^3_{\mu} (Q, Z_K)$:\\

\begin{equation}
\tau(\mu)\in H^3_{\mu} (Q, Z_K)
\end{equation}

This  $\tau(\mu )$ is called the \emph{obstruction} to $\mu$. Then,
one answers completely the first question: any  $\mu$  in $Hom(Q,
Out(K))$ generates
 extensions \emph{if and only if} the obstruction  $\tau(\mu )$ is zero (of $H^3$, of course); see [44].\\

If $\mu$ is obstruction-less, or $\tau(\mu)=0$ how many extensions does it produce? \emph{Answer}: the second cohomology group:\\

\begin{equation}
\textrm{For $\mu$  obstruction-less, extensions
$\Longleftrightarrow H^2_{\mu} (Q, Z_K)$}
\end{equation}

We do not eleborate in the third question (equivalences) except for
mentioning: that an extension of $K$ by $Q$ is an exact \emph{suite}
 $1\longrightarrow  K\longrightarrow   E\longrightarrow   Q\longrightarrow  1$: it is more
 restrictive that finding the middle group $E$; in other words, it might be that
 \emph{different} extensions would generate the \emph{same} extension group $E$.\\

\underline{Schur multipliers.} Suppose you try to extend the $Z_2$ group by some group $Q$:\\

\begin{equation}
Z_2\longrightarrow   E\longrightarrow   Q
\end{equation}

As we know $Aut(Z_2) = I$, any possible extension has to use the
trivial morphism $Q \longrightarrow  I$.   This problem occurs e.g.
in quantum physics for the following reason: the state space is a
projective Hilbert space, as vectors in the same ray represent the
same physical state ; so one has to find \emph{projective}
representations of the pertinent symmetry groups (e.g. $SO(3)$);
this theory was started by I. Schur around 1900, and it turns out
that projective representations of a group G can be usually obtained
from linear ones from an extension $\hat{G}\longrightarrow   G$: if
the
 kernel is $Z_2$, we have the case for $SO(n)$ and $Spin(n)$ groups (see e.g. [42]):\\

 \begin{equation}
Z_2\longrightarrow   Spin(n)\longrightarrow  SO(n)
\end{equation}

For example, $Spin(3) = SU(2)$, which is understood here as a
\emph{central} extension of $SO(3)$ (central, as $Z_2$ is injected
in the centre of $SU(2)$). Schur mutiplicator
 or multiplier $M(G)$ is precisely the \emph{homology}
 group (which we do not describe in detail)

\begin{equation}
M(G) := H_2(G, Z)
\end{equation}

This is important in at least three contexts: for projective
representations (the original purpose of Schur), for central
extensions (where, in  $K\longrightarrow   E\longrightarrow   Q$,
$K$ abelian enters in the centre of $E$, which needs not even be
abelian), and for topological reasons (as e.g. $SO(n)$ is not simple
connected, but $Spin (n)$, $n  > 2$, is).
See [43], [45].\\

As an example, let us note that $Alt_n$ always admits a 2-extension:
If $T_2$ is the regular triangle, and $T_n$ the $n$-dim ``hyper''
tetrahedron, the rotation symmetry group is $Alt_{n+1}$
(e.g. $Alt_4$, of order 12, for the ordinary tetrahedron $T_3$), Now we have the diagram\\

$Z_2\longrightarrow 2\cdot Alt_{n+1}-\!\!\!-\!\!\!\longrightarrow Alt_{n+1}$\\

$\parallel$\quad \quad \quad\quad \quad $\cap$\quad \quad \quad \quad\quad $\cap$\\

$Z_2\longrightarrow Spin(n)-\!\!\!-\!\!\!-\!\!\!\longrightarrow SO(n)$\\

where the ``2'' in $2\cdot Alt_{n+1}$ is called also a Schur multiplier.\\

In crystallography, $2\cdot Alt_{4}$ is called ``binary
tetrahedral'' group.\\

For $n>3$, $Alt_n$ admits multipliers. Indeed [47]
$Alt_{6,7}$ admit $6\cdot Alt$, the others only $2\cdot Alt$.\\

\subsection{Groups up to order 16.}

To have a taste of the smallest groups, a brief study is made here
of all finite groups up to order 16, $|G| < 16$.  See e.g.
Thomas-Wood [20]
or Coxeter [41].\\

If Ord $G \equiv |G| = p$ is a prime, there is only the cyclic group
$\mathbb{Z}_p =\{ g; g^p=e \}$. So for primes 2, 3, 5, 7, 11 and 13
the problem is solved at once. We know also the automorphism
group, $Aut(\mathbb{Z}_p ) = \mathbb{Z}_{p-1}$.\\

For $|G| = 4$, if there is a 4-th order element, the group is
generated by it, and it is $\mathbb{Z}_4$; if there are no 4-th, the
group has three involutions (plus $e$), hence they commute and the
group must be $\mathbb{Z}_2\times \mathbb{Z}_2$, called the
``Vierergruppe'' by F. Klein.\\

The reader will convince himself easily that for $|G| = 6$, there
are \emph{only} the two (known) solutions, namely $\mathbb{Z}_6 =
\mathbb{Z}_2\times \mathbb{Z}_3$ and $D_3 = S_3$. As said, the non
abelian group $S_3$ has to have 3 classes, as the Burnside relation
is uniquely $6 = 2\cdot1^2 + 1\cdot2^2$. As automorphisms keep
order,
we have\\

\begin{equation}
Aut(\mathbb{Z}_6) = Aut(\mathbb{Z}_2\times \mathbb{Z}_3) =
Aut(\mathbb{Z}_2)\times   Aut(\mathbb{Z}_3) = I\times \mathbb{Z}_2 =
\mathbb{Z}_2
\end{equation}

 Also $Out(S_3) = I$. For order 8, we know already there are three abelian groups, namely
$\mathbb{Z}_8$, $\mathbb{Z}_4\times   \mathbb{Z}_2$ and
$\mathbb{Z}_2^3$, the last is an \emph{elementary abelian group}. If
$|G|=8$ and $G$ nonabelian, it cannot contain an 8-order element
(because then it will be $\mathbb{Z}_8$), and if all elements $\neq
e$ are involutions, we have ($\mathbb{Z}_2^3$). So there must be
order four elements; suppose we have one, $a$, $a^4=e$; if $b$ is
another element, and $ab\neq ba$, thus $bab^{-1}\neq a$, so it can
only be $a^3$, as is an automorphism and $a^2$ has order two;
$b^2=e$: the elements then are  $\{e, a, a^2, a^3, b, ba^2, ab,
ba\}$.
  The group is then the \emph{dihedral group}
$D_4 = \mathbb{Z}_4\rtimes \mathbb{Z}_2$, also called \emph{octic
group}.\\

 If there are at least two different elements $a$, $b$ of order 4, $ab\neq ba$, one
shows $ab$ is also of order four, and the group becomes the
quaternion group $Q:=\{ a^4 = b^4, a^2=b^2 = -1, ab = -ba\}$, or (as
said) $Q = \pm(1,i, j, ij)$. Class partitions are\\

\begin{equation}
D_4: 8= 1\cdot1_1 + 1\cdot1_2 + 2\cdot2_2 + 1\cdot2_4,\quad \quad
Q:  8 = 1\cdot1_1 + 1\cdot1_2 + 3\cdot2_4
\end{equation}

Both $D_4$ and $Q$ have repeated analogous classes, hence there are
outer automorphisms.
Indeed, one shows\\

\begin{equation}
Out(D_4) = \mathbb{Z}_2,\quad \quad          Out(Q) = S_3
\end{equation}

Both have the (unique) Burnside relation $|Q| = 8 = 4\cdot1^2
+1\cdot2^2$.\\

For order ten we have just two groups, the expected $\mathbb{Z}_{10}
= \mathbb{Z}_5\times \mathbb{Z}_2$ and the dihedral, or extension of
$\mathbb{Z}_5$ by the inverse, a cyclic automorphism (as
$\mathbb{Z}_5$ is abelian):\\

\begin{equation}
\mathbb{Z}_{10}= \mathbb{Z}_5\times \mathbb{Z}_2,\quad \quad
D_5 = \mathbb{Z}_5\rtimes \mathbb{Z}_2
\end{equation}

The partitions by classes and by \emph{irreps} are clearly\\

\quad \quad $\mathbb{Z}_{10}$: $1\cdot1_1 + 1\cdot1_2 + 4\cdot1_5 +
4\cdot1_{10}$, and, as it is abelian, Burnside relation is $10 = 10\cdot1^2$\\

\quad \quad $D_5$: $1\cdot1_1 + 1\cdot5_2 + 2\cdot2_5$,   and $10 =
2\cdot1^2 + 2\cdot2^2$\\

For order 14, again, there are the cyclic $\mathbb{Z}_{14} =
\mathbb{Z}_7\times \mathbb{Z}_2$ and the dihedric $D_7$. Now for
order 15 there is only a group, the cyclic, as $15 = 3\cdot5$ and 3
and 5 are \emph{incompatible} primes (simplest proof is by
Burnside relation: If $15 = s^2\cdot1^2 + m\cdot3^2$, uniquely $m=0$ as $s\geq1$).\\

So there is only order 12 which requires some attention; first,
there are \emph{two} abelian groups, as $12 = 2^2\cdot3$ and
$Part(2) = 2$, namely $\mathbb{Z}_{12} = \mathbb{Z}_4\times
\mathbb{Z}_3$ and
$V\times   \mathbb{Z}_3$\\

We just list the three non-abelian groups with some properties:\\

$Dih_6 = \mathbb{Z}_6\rtimes \mathbb{Z}_2  = \mathbb{Z}_2\times
S_3$; class split: $1\cdot1_1 + 1\cdot1_2 + 1\cdot2_2 +1\cdot4_2 + 1\cdot2_3 + 1\cdot2_6$ Burnside relation:  $12 = 4\cdot1^2 + 2\cdot2^2$\\

$Alt_4 = V\rtimes \mathbb{Z}_3$  ; $1\cdot1_1 + 1\cdot3_2 +
2\cdot4_3$; Burnside relation is $12 = 3\cdot1^2 + 1\cdot3^2$\\

$Q_3 = \mathbb{Z}_3\rtimes \mathbb{Z}_4$ The action is defined as
$\mathbb{Z}_4\longrightarrow \mathbb{Z}_2 = Aut(\mathbb{Z}_3)$,
therefore $12
= 4\cdot1^2 + 2\cdot2^2$\\

The subgroup structure is clear in most cases.\\

The following expresses the whole results. For groups
$G$, $|G| < 16$, there are  \emph{five} types:\\

1) $|G| = p$ prime.- $p$= 2, 3, 5, 7, 11, 13 and $G=I$.\\

 class:
$1\cdot1_1(e) +(p-1)\cdot1_p$ (rest).\\

 \emph{irreps}: $p=p\cdot1^2$.\\

 automorphisms: $Aut(\mathbb{Z}_p) = \mathbb{Z}_{p-1}$\\

2) $G$ =Direct product of two abelians, $G = A_1\times A_2$:\\

 $V =(\mathbb{Z}_2)^2$, $\mathbb{Z}_6 = \mathbb{Z}_2 \times \mathbb{Z}_3$, $\mathbb{Z}_2\times \mathbb{Z}_4$,
  $(\mathbb{Z}_3)^2$, $\mathbb{Z}_2\times \mathbb{Z}_5$, $\mathbb{Z}_3\times \mathbb{Z}_4$, $\mathbb{Z}_2\times \mathbb{Z}_7$
   and $\mathbb{Z}_3\times \mathbb{Z}_5 = \mathbb{Z}_{15}$ Class: $V$, $|V|= 4 = 1\cdot1_1(e) + 3\cdot1_2(a, b, ab)$;
   etc\ldots Burnside relation: $|G| = |G|\cdot1^2$\\

3) $G =\stackrel{{\rm\,o}}G$
  (rest of abelians): $\mathbb{Z}_4$, $\mathbb{Z}_8$, $\mathbb{Z}_9$, $\mathbb{Z}_2^3$, $V\times \mathbb{Z}_3$\\

4) Dihedrals: $D_3 = S_3$, $D_4$ ``octic'', $D_5$, $D_6 =
\mathbb{Z}_2\times S_3$, $D_7$\

 \quad \quad class, e.g. $D_7$: $14=1\cdot1_1(e)+1\cdot7_2(\alpha,\ldots)+3\cdot2_7(a,\ldots,a^6\ldots)$; \emph{irreps} e.g.
 \quad $D_7$: $14=2\cdot1^2+3\cdot2^2$\\

 5) Other, non-abelian. $Q = \mathbb{Z}_4\rtimes _{ /2} \mathbb{Z}_4$,
 Dicyclic. $Alt_4 = V\rtimes \mathbb{Z}_3$,\quad $Q_3 = \mathbb{Z}_3\rtimes\mathbb{Z}_4$ \\

 In total, for $|G| < 16$, there are 20 abelian groups + 8
 non-abelian. For an exhaustive study, consult [20].

\subsection{Characterization of groups.}

In this final Section of the review of the general theory of groups,
we include a couple of left-over topics and characterize one of
these small groups by several distinct properties.\\

Besides the stated Theorems of (1) Lagrange ($H\subset   G$ subgroup
$\Longrightarrow |G|: |H|$), (2) Cayley ($|G| = n\Longrightarrow
G\subset Sym_n$) and (3) Cauchy ($|G| = p^f\cdot m\Longrightarrow $
$\exists g$, $g^p=e$), the (4) \emph{Theorem(s) of Sylow} (1872)
extends
Cauhy's. \emph{Theorem of Sylow}: let $|G| = p^f\cdot m$, with $p$ and
$m$ coprimes. There are subgroups of order $p^f$, they are
conjugate, and their
number is $1+kp$, i.e. $\equiv 1$ mod $p$.\\

In particular, if $k=0$ the subgroup is \emph{normal}.\\

The proof, easy, is in any of the standard books (e.g. [34], p. 33).
Note this is a kind of reciprocal of Lagrange's, in the restricited
sense that there are not necessarily subgroups $H$ of any order
dividing $|G|$, but this is the case for pure power of prime
factors. To give a double example, $Alt_4$ (order 12) has subgroups
of order 2, 3 and 4, but not 6. And if $|G| = 21$ (two groups),
there are subgroups of order 3 and 7.\\

What is a measure of the non-simplicity of a finite group? Among the
normal subgroups, there are maximal ones: $H$ normal in $G$ is
\emph{maximal}, if there is no $H'$ in between: $H\subset H'\subset
G$, with $H'$ still normal in $G$; then $G/H$ is \emph{simple}
(trivial proof); repeating the process for $H$, we achieve a finite
decreasing chain, called a \emph{composition series}:\\

$G$;   $H_1$;   $H_2$;\ldots  ; $H_s$\ldots $I$;   $H_s$ simple, and
we have simple quotients $Q_1$, $Q_2$\ldots $H_s$; $G/H_1=Q_1$, etc.\\

Maximal (normal) subgroups might not be unique, but the quotients
are, up to reordering: this is the content of our next theorem:\\

\emph{Jordan-H\"{o}lder Theorem:}  The quotients  $\{Q_i\}$
of two composition series are the same, up to reordering.\\

Again, this is a (the fifth) classical theorem in finite groups,
proved in any textbook (again, we can quote ([34], p. 62). We shall
only exemplify the meaning of the result in several examples: 1) For
$Sym_4 = S_4$, we have: $S_4$; $Alt_4$; $V$; $\mathbb{Z}_2$; $I$,
with quotients $\mathbb{Z}_2$; $\mathbb{Z}_3$, $\mathbb{Z}_2$;
$\mathbb{Z}_2$.  2) For $Sym_5$: $S_5$; $Alt_5$, $I$; as $Alt_5$ is
already simple.- 3) For $G = Q$ (the quaternion group of order 8):
$Q$; $V$; $\mathbb{Z}_2$; $I$, with quotients $\mathbb{Z}_2$ three
times. Observe the obvious result $|G|=\prod_i|Q_i|$. \\

A group $G$ is \emph{solvable}, if the chain of any
Jordan-H\"{o}lder composition series has only as quotients abelian
(simple) groups of type $\mathbb{Z}_p$ ($p$ prime). When the chain
ends up in an nonabelian simple group, we speak of a composed group
in general; for example $Sym_6$ is \emph{composed}, as the
composition is $Alt_6$;
$I$.\\

We shall see in the next section that the only nonabelian simple
groups
of order less than 1000 are $Alt_5$ (60), $PSL_2(7)$ (168), $Alt_6$ (360), $SL_2(8)$ (504) and $PSL_2(11)$ (660).\\

We now take the case of $G = Sym_4$ as an excuse to show several
items one has to reckon with to understand completely any (finite)
group:\\

1) \emph{Definition}: $S_4=Sym_4$ is the permutation group in
four symbols; order
4! = 24\\

2) \emph{Equivalences}: Written also as $V\rtimes S_3  =
Hol(V)$.\\

3) \emph{Generators and relations}: If $(a,b,\alpha,\beta)$
generate $V$ and $S_3$, we have $a^2=b^2=(ab)^2=\alpha^3=\beta^2=
e$; $\alpha\cdot
a\cdot\alpha^{-1} =b$, $\alpha\cdot b\cdot \alpha^{-1} =ab$, etc.\\

4) As Coxeter group, it is $\circ\!\!\!-\!\!\!-\!\!\!\circ\!\!\!-\!\!\!-\!\!\!\circ$\\

5) \emph{Burnside relation}: $4! = 24 = 2\cdot1^2 + 2\cdot3^2 +
1\cdot2^2$; easily deduced from 3).\\

6) \emph{Class equation}: $1A(e)$, $2A[2]$, $2B[2^2]$, $3A[3]$,
$4A[4]$, or 24 = 1+6+3+8+6. (we write the classes [2], [$2^2$], etc. as partitions).\\

7) \emph{Subgroups and quotients}: Center $I$, Derived Subgroup
$Alt_4$,
Abelianized $\mathbb{Z}_2$, $Out= I$; $S_4$ is \emph{complete}.\\

8)  \emph{Lattice of subgroups}: too complex; see e.g. ([20],
Type 24/12)\\

9) \emph{The cross} (as centre is trivial, and there are no outer automorphisms\\

$\xymatrix{&  I\ar[d] & \\Alt_4\ar[r] &Sym_4\ar@{=}[d]\ar[r]&
\mathbb{Z}_2\\& Sym_4\ar@{=}[r]&Sym_4}$\\
\bigskip

10) \emph{Character Table}\ (see, e.g. [20]).
\begin{table}[h]
\begin{center}
\begin{tabular}{ r r r r r r }
 & 1A & 2A  & 2B & 3A & 4A \\
 \hline
$\chi^{(1}$ & 1 & 1 & 1 & 1 & 1 \\
$\chi^{(-1}$ & 1 & 1 & -1 & 1 & -1 \\
$D^{(2}$ & 2 & 2 & 0 &-1 & 0 \\
$D^{(3}$ & 3 & -1 & 1 &0 & -1 \\
$D^{(3'}$ & 3 & -1 & -1 &0 & 1 \\
\end{tabular}
\end{center}
\end{table}

\pagebreak

\section{Finite simple groups}

 \subsection{The search for finite simple groups: historical survey}

As stated in the Introduction, the search for all collections of
finite simple groups (\underline{FSG}) took well over a century,
from the simplest cases $\mathbb{Z}_p$ ($p$ prime) and the
alternating group ($Alt_{n>4}$), known since early 1830s, to the
Monster group $\mathbb{M}$, constructed around 1980, with $\approx
10^{54}$ elements. Most of the groups came in families, but other
are
isolated (``sporadic'').\\

Mathieu found (around 1860), rather by chance, the first set
of five ``sporadic'' FSG, that is, finite simple groups \emph{not}
in families. Mathieu was searching for groups more than
3-transitive, not simple groups: simplicity was proven later.\\

We shall see that the finite simple groups show up in several
\emph{families} (2+16, in fact) plus several (26, in fact)
\emph{sporadic} groups. The largest supply of groups is by groups of
matrices, that is subgroups or subquotients of
$GL_{\mathbb{K}}(V)\approx GL_n(\mathbb{K})$, meaning automorphisms
(invertible matrices) in a $n$-dimensional vector space $V$ over the
field $\mathbb{K}$; for \emph{finite} groups the field of numbers
$\mathbb{K}$ (as well as the dimension of $V$) has to be
\emph{finite}, obviously.\\

To repeat: The easier finite simple groups (FSG) are $\mathbb{Z}_p$
and $Alt_{n>4}$. That $\mathbb{Z}_p$ for $p$ prime is simple is
obvious, as it is abelian with no proper subgroups at all (Lagrange
theorem). As for simplicity of $Alt_{n}$ (Galois), we shall only
show the partition by classes and the subgroup structure of
$Alt_5$:\\

\begin{equation}
\begin{aligned}
&\textrm{Partitions in classes}\quad 60 = 1\cdot 1_1 + 1\cdot 15_2 + 1\cdot 20_3 + 2\cdot 12_5\\
&\textrm{Subgroup structure}\quad     \quad\quad\quad I(1)\quad
\mathbb{Z}_2(15)\quad \mathbb{Z}_3(10)\quad \mathbb{Z}_5(6)
\end{aligned}
\end{equation}

And no subgroup is normal, as they are conjugate within a class
(Sylow's theorem) or two. Then, if $Alt_5$ has no proper normal
subgroups, the same is true for $Alt_{n>5}$, (easy proof, by
induction, see e.g. [32]), so $Alt_n$ for $n > 4$ is simple
(Galois, 1832).\\

Galois also discovered some of the \emph{finite fields},
$\mathbb{F}_q$; as we shall see in detail later, for any prime
number $p$ and any natural number $f$, there is a finite field
$\mathbb{F}_q$, with $q = p^f$ elements, and this exhausts all the
finite fields; all of them are commutative (Wedderburn, 1908). In
the vector spaces $\approx\mathbb{F}_q^n$ there are plenty of
automorphism groups, as groups of invertible matrices $GL_n(q)$, or
subgroups; to extract the simple pieces of these groups (subgroups
or subquotients) is sort of mechanical.\\

It turned out that the classification, due to Cartan, of (infinite,
continuous) simple Lie groups, was to be repeated with matrix groups
over finite fields, but now the families are, in general,
\emph{biparametric}, depending of the field $\mathbb{F}_q$ ($q =
p^f$; $p$ prime, $f$ natural) and on dimension ($n$). Starting with
$GL$ the first \emph{biparametric} family is with the projective
quotient ($P$) of the unimodular ($SL$) restriction, namely\\

\begin{equation}
PSL_n(q)
\end{equation}
(here $P$ implies to divide $SL$ by its center). This forms a
\emph{doubly infinite family} of finite simple groups for any $n\geq
2$ and any $q = p^f$, with \emph{two} exceptions, to be recalled later.\\

Since 1955, Chevalley [55] (and others) completed the list of
Lie-type finite simple groups, also started by Dickson around 1900,
including the exceptional group $G_2$, by attacking the other
exceptional groups, $F_4$ and $E_6$ to $E_8$; here the families are
monoparametric, e.g. $F_4(q)$, etc.\\

Later, Steinberg showed (1959) [64] that the (continuous) Lie
algebras with \emph{outer} automorphisms originated more Lie-type
finite groups, namely for the four cases $A_n$, $n > 1$, $D_n$, ($n
> 4$), $D_4$ and $E_6$: all depend on $q$. Finally, Ree (Korea) and
Suzuki (Japan) completed (about 1960) the list of Lie-type families by
showing that the ``double/triple bond'' continuous Lie groups
also gave rise to more finite simple groups: this is the case for
$B_2$,
$G_2$ and $F_4$ (we shall see this in detail in section 4.5).\\

So in total there were (4 + 5 + 4 + 3 = 16) \emph{families} of
finite simple groups of Lie-type. Or (2 + 16 = 18) as the total
number
of \emph{families} of finite simple groups.\\

As mentioned above, FSG \emph{not} in these
families were first discovered by Mathieu (about 1860): the first five
sporadic groups. For more than a century, no more
sporadic FSG were discovered, until Janko discovered the next one,
$J_1$, in 1966 (order 175 560); after some frenzy activity, in the
decades of 1960s and next, the list was completed by the efforts of
a large community of mathematicians  (Leech, Conway, McKay,
Gorenstein, Fisher, Griess, Thomson, Aschbacher, etc.): there were
another two more related series of sporadics, the Leech-lattice set
(7 groups) and the Monster series (8 cases); to all these one must
add 6 totally unrelated (up to now) cases, the so-called
pariah groups, for a total of (5 +7 + 8; +6) = 26 sporadic groups. For the general history of FSG,
see the book [51].\\

We firmly believe (since around 1985) (it is really proven) that
the list of finite simple groups is now complete. From the extensive
literature, we extract [58], [47] and [57].

\subsection{Finite fields}

In physics we handle only the fields $\mathbb{R}$: the real numbers,
and $\mathbb{C}$: the complex numbers; however, there are also
fields with a finite number of elements, already discovered by
Galois. For any number $q = p^f$, where $p$ is prime and $f$
natural, there is a
finite field $\mathbb{F}_q$. Recall, for any field $K$, that $K^* := K\setminus\{0\}$ forms the multiplicative group.\\

We start with the simplest example:\\

In the set  $\{0, 1\}$, if we sum \emph{mod} 2, and multiply, we have the rules\\

\begin{equation}
0+0=0,\quad  0+1=1,\quad 1+1=0;\quad 0\cdot0=0\cdot1 = 0,\quad
1\cdot1 = 1
\end{equation}
which makes up a \underline{field of two elements}, called
$\mathbb{F}_2$; in this case $F_2^* =\{1\} = I$. As (0, 1) must
exist as \emph{different} in
any field, $\mathbb{F}_2$ is the smallest possible field of numbers.\\

The same construction works for \emph{any prime} $p$, namely:\\

Define a \emph{ring} structure in the set of $p$ elements as $\{0,
1, a, a^2,\ldots, a^{p-2}\}\equiv (0, 1, 2,\ldots, p-1)$ by sum
\emph{mod} $p$ and product $\neq0$ like in $\mathbb{Z}_{p-1}$: it is
trivial to show that both operations are commutative, and the
product is distributive with respect the sum: it is a \emph{field}
$\mathbb{F}_p$, as any element $\neq0$ has inverse for the product
(e.g. in $\mathbb{F}_5$ ($0 = e, 1, a, a^2, a^3$)
the multiplicative  inverse of $a$ is $a^3$, and $a^2$  is involutive). So

\begin{lem} For any prime number $p$, in the set of $p$
elements $(0, 1, a, a^2,\ldots, a^{p-2})$ with sum defined
\emph{mod} $p$ and product as in $\mathbb{Z}_{p-1}$, there is an
underlying \emph{field} structure, named $\mathbb{F}_p$. As
$\mathbb{F}_p^* = \mathbb{F}_p\backslash\{0\}$, it has $(p-1)$
elements,
and corresponds to the cyclic group $\mathbb{Z}_{p-1}$. The additive group is clearly $\approx\mathbb{Z}_p$.
\end{lem}

The minimal field is, as said $\mathbb{F}_2$, as $0\neq1$ always.
These finite fields have \emph{characteristic} $\chi\neq0$: for
$\mathbb{F}_p$, we have: $1+1+1+\ldots ^{(p}  +1 = 0$, so
$Char(\mathbb{F}_p) = p$.
One shows also that $Char(\mathbb{F}_q) = p$ if $q = p^f$.\\

 As a field, in $\mathbb{F}_p$ there are \emph{no} automorphisms $\neq Id$, i.e.
 we have $Aut(\mathbb{F}_p) = I$, as any automorphism $\alpha$ should verify  $\alpha(0+a) =\alpha(a)$
 and  $\alpha(1\cdot a) =\alpha(a)$:  $\alpha(0) = 0$ and  $\alpha(1) = 1$, hence
  e.g.  $\alpha(3) =\alpha  (1+1+1) = 3\cdot \alpha  (1) = 3$,
 etc.\\

There are \emph{more} finite fields. We just state the result (Moore, 1903); see [21]:

\begin{thm} For any power $f\in\mathbb{N}$ of a prime
number $p$, there
is a field $\mathbb{F}_q$ with $q = p^f$  elements, and any finite field is of this type.
\end{thm}

The field operations are:\\

The \emph{sum}, as in $(\mathbb{F}_p)^f=
\mathbb{F}_p\oplus\mathbb{F}_p\oplus\ldots\oplus^{(f}\mathbb{F}_p$
(i.e., as an \emph{elementary abelian group}).\\

The \emph{product}, as ``product'' in $\mathbb{Z}_{q-1}$, so
$|\mathbb{F}^*_q| = |\mathbb{Z}_{q-1}|$,
completed by $0\cdot(any) = 0$.\\

Notice the first law: in any $\mathbb{Z}_n$ there is a \emph{ring}
structure,
with sum and product \emph{mod} $n$, but only for $n = p^f$ can one deform the additive law to make up a \emph{field}.\\

We just check that the laws work for the simplest case, namely
$\mathbb{F}_4$. $\mathbb{F}_4$:  $\{0,1, 2, 3\}$. \emph{Sum}, like
$\mathbb{F}_2\oplus\mathbb{F}_2(=V)$, with e.g.  $\{e, a\} = \{0,
1\}$, with  $\{e, b\} =\{0, 2\}$, and $\{e, ab\} = \{0, 3\}$.
\emph{Product}, like 1, 2, 3 as in $\mathbb{Z}_{4-1}\approx(e,\omega
,\omega^2)$, $(\omega:= exp(2\pi i/3)$, namely $\mathbb{Z}_3$. This
works by the natural generalization for any power $f$ of any prime,
so $|\mathbb{F}_q| = p^f$, for any prime number $p$ and any natural
number $f$: $1, 2, 3,\ldots$ For example,

\begin{equation}
\mathbb{F}_2, \mathbb{F}_3, \mathbb{F}_4, \mathbb{F}_5,
\mathbb{F}_7, \mathbb{F}_8\quad \textrm{and}\quad
 \mathbb{F}_9\quad \textrm{are the fields with $\leq10$ elements}
\end{equation}

Finite fields $\mathbb{F}_q$ with $q = p^f$, $f >1$ have (field)
automorphisms, hence give rise to \emph{semilinear} applications
(see below). For example, the simplest case is $\mathbb{F}_4$: As
$Aut(V) = S_3$, and $Aut(\mathbb{Z}_3) = \mathbb{Z}_2$, where
$\mathbb{Z}_3 =\mathbb{F}_4^*$, and $V =
\mathbb{Z}_2\times\mathbb{Z}_2$,
the natural embedding $\mathbb{Z}_2\subset S_3$ is an automorphism for the \emph{sum} and for the \emph{product}!\\

\begin{equation}
Aut( \mathbb{F}_4: 0, 1, 2, 3) = \mathbb{Z}_2;\quad (2,
3)\Longrightarrow (3, 2);\quad \textrm{etc.}
\end{equation}

The same is true for all fields $\mathbb{F}_q$, $q = p^2$. We
refrain of showing $Aut(\mathbb{F}_q)$ for general $q = p^f$;
eventually we shall use $Aut(\mathbb{F}_9)$ (Sect. 5).
We just remark that $Aut(\mathbb{F}_{q^2})$ has always involutory automorphisms; see e.g. [52].\\

Summing up, all finite fields of numbers are known: for each prime
number $p$ and for each natural number $f$, there is a unique finite
field $\mathbb{F}_q$, where $q = p^f$, and this spans all finite
fields; these are true fields, that is, they are commutative. For $f
>1$, these fields have field automorphisms;
this means there are \emph{semilinear} maps and groups, as we shall see. Another good reference for finite fields
is in Bourbaki (Bourbaki [80]) or [30]. \\

\subsection{General series (PSL)}

For any field $\mathbb{K}$, the $n$-dimensional vector space over
$\mathbb{K}$ is unique, and is written as $\mathbb{K}^n$; so we have
in our finite-field cases

\begin{equation}
V = \mathbb{F}^n_{q}\quad \textrm{as a finite $\mathbb{F}$-vector
space, with $q^n$ points or elements}
\end{equation}

Notice the matrix group $Aut(V) := GL_n(q)$ has centre (diagonal
entries from $\mathbb{F}_q$), and the determinant map
$GL\longrightarrow \mathbb{F}^*_q$ has $SL$ as kernel, nearly by
definition; hence, $GL$ may serve to generate simple groups,
but itself is far from simple! (except $q=2$, $n >2$).\\

For later, we shall need also the notion of \emph{projective
spaces}: given a vector space $V$, $PV$ is by definition the set of
one-dimensional subspaces (lines or rays). In particular, if $dim$
$V = n$, as manifold $dim$ $PV = n-1$ by definition, equivalent to
remove the origin in $V$ and make vectors equivalent if parallel:
$PV\approx(V\backslash\{0\})/(\mathbb{K} \setminus\{0\})$,
 and for $\mathbb{K}$ finite, $V= \mathbb{K}^n$, so (as $\mathbb{K}=\mathbb{F}_q$)\\

\begin{equation}
|PV| = |(\mathbb{K}^n \backslash\{0\} / \mathbb{K}\backslash\{0\})|
= (q^n-1)/(q -1) = 1 + q + q^2 +\ldots + q^{n-1}
\end{equation}

The groups acting \emph{effectively} on the projective spaces are
$PGL_n(q) := GL_n(q)/\mathbb{F}^*$, similarly $PSL_n(q)$, and some
important subgroups. The following diagram clarifies the situation:

 \small{\begin{equation}\label{eq:3}
\begin{CD}
J @>>>SL_n(q)@>>>PSL_n(q)\\
@VVV @VVV @VVV\\
\mathbb{F}^*_q @>>>GL_n(q) @>>> PGL_n(q)\\
@VVV @VVV @VVV\\
\mathbb{F}^*_q/J@>>>  \mathbb{F}^*_q @>>>J
\end{CD}
\end{equation}
$SL$ is the kernel of the determinant map, as said; diagonal entries
$\mathbb{F}_q^*$ in $GL$ act like the centre, and the quotient is
\emph{defined} as $PGL$. Also, $J$ is the
intersection $\mathbb{F}^*_q\cap SL_n(q)$. We have now the very \emph{important} result (ca. 1900):\\

\begin{thm} (Dickson). $PSL_n(q)$ is SIMPLE for any $n\geq2$
and $q$,
except $n=2$ AND $q = 2, 3$.
\end{thm}

So the subquotient $PSL$ is the simple ``piece'' coming from $GL$.
To discuss the result, let us deal first with the exceptions:

\begin{equation}
\begin{aligned}
&PSL_2(2) = GL_2(2) = Sym_3 \equiv S_3 = Dih_3=\mathbb{Z}_3\rtimes \mathbb{Z}_2\quad \textrm{(order 3!=6)}\\
&PSL_2 (3) = Alt_4 = V\rtimes \mathbb{Z}_3\quad \textrm{(order 4!/2
= 12)}
\end{aligned}
\end{equation}

The first result comes from $| GL_2(2) | = 6$ and nonabelian. For the second, we have $| GL_2(3) | = 48$, etc.\\

So $PSL_n(q)$ is seen as constituting the first biparametric family
of finite simple
groups of Lie type. We shall NOT try to prove the Theorem (see Carter [21]; also [49]).\\

Next, we exhibit some matrix groups over the smallest fields: first we have\\

\begin{equation}
GL_1(q) =\mathbb{F}_q^*,\quad \textrm{abelian with $(q-1)$ elements}
\end{equation}

For $|\mathbb{F}_2|=2$, $GL_n(2) = SL_n(2) = PGL_n(2) = PSL_n(2)$, because $\mathbb{F}_2^* = I$. Some order values are\\

$|GL_2(2)| = (2^2-1)(2^2-2) = 6$, indeed $GL_2(2) = S_3$, see (73).\\

$|GL_3(2)| = (2^3-1)(2^3-2)(2^3-4) = 168$; indeed, one shows
$GL_3(2) = PSL_2(7)$, of
the same order: $(7^2-1)(7^2-7)/6/2 = 168$. It is the second smallest  nonabelian simple group; see also [54].\\

$|GL_4(2)| = 20160$, indeed $GL_4(2) = Alt_8$, of order $8\cdot7\cdot6\cdot5\cdot4\cdot3 = 20 160$. But notice\\

$|PSL_3(4)| = (4^3-1)(4^3-4)(4^3-4^2)/3/3 = 20 160$, but $PSL_3(4)\neq GL_4(2)$, [53].\\

For $p > 2$, we only record\\

$PSL_2(3) = Alt_4$, order 12.- $|PSL_3(3)| = (3^3-1)(3^3-3)(3^3-9)/2
= 5616.-$  $SL_2(4) = PSL_2(4)$,
order 60.- $PSL_2(5) = Alt_5 = SL_2(4)$. $PSL_2 (9) = Alt_6$, order 360.\\

The counting of the order in $PSL_n(q)$, employed already above, is understandable: First, we do for $GL_n(q)$:\\

\begin{equation}
|GL_n(q)| = (q^n-1)(q^n - q)(q^n - q^2)\ldots (q^n - q^{n-1})
\end{equation}
because, roughly speaking, the first row in $GL$ as a matrix can
have any $n$ numbers in $q$ except all zero, the second row has to
be independent of the first, so substract $q$ values, to guarantee
invertibility of the matrix, etc.  Now passing to $SL$ means to
divide the order by $(q-1)$ = $|\mathbb{F}^*_q|$, and the $``P''$
means to divide
by the order of the centre. We just exemplify for $PSL_4(3)$:\\

$|GL_4(3)| = (3^4-1)(3^4-3)(3^4-3^2)(3^4-3^3) = 24 261 120$\\

$|SL_4(3)| = | GL_4 (3)|/(3-1) = 12 130 560$, as $|\mathbb{F}_3^*| = 2$\\

$|PSL_4(3)|  = |SL_4(3)| /2$, as Centre $SL_4(3) = \mathbb{Z}_2$; so $|PSL_4(3)| = 6 065 280$.\\

It is easy to see the size of the smallest nonabelian finite simple group: it is\\

\begin{equation}
SL_2(4) = PSL_2(5) = Alt_5,\quad \textrm{order}\quad 5!/2 = 60 =
2^2\cdot3\cdot5
\end{equation}

So the general formula for the order is\\

\begin{equation}
|PSL_n(q)| = q^{n(n-1)/2} (q^2-1)(q^3-1)\ldots
(q^n-1)/\textrm{$|centre|$}
\end{equation}

There is a Theorem (of Burnside, see [2]) saying that the order of
any simple non-abelian group must have at least three different
prime factors, (one of which has to be 2, see later); it can be
checked in the above examples.
So the next two cases of finite simple groups, besides $Alt_5$, turn out to be\\

\begin{equation}
\begin{aligned}
&PSL_2(7) = GL_3(2),\quad \textrm{order}\quad 168 = 2^3\cdot3\cdot7\\
&Alt_6 = PSL_2(9),\quad \textrm{order}\quad 360 = 2^3\cdot3^2\cdot5
\end{aligned}
\end{equation}

For curiosities about both groups, see [54] and our section 5.\\

Notice that the group $PGL_2(q)$ is sharp 3-\emph{transitive} in the
projective line $\mathbb{F}P^1$, that is to say, any three points in
the line can be transformed in any three others, with no leftover
stabilizer ($\neq I$):  The line $\mathbb{F}P^1$ has $(q^2-1)/(q-1)
= (q+1)$ points (one adds the ``point at infinity''); the action of
$PGL_2$ is transitive with the affine group $Aff_1(q)$ (translations
and homotheties in the field) as little group (e.g. for the point
$\infty$). This affine group in turn acts still \emph{trans} in the
one-dimensional vector space $\approx \mathbb{K}$ (without
$\infty$), with stabilizer at zero $\mathbb{F}^*_q$: finally this
last group is still \emph{transitive} with \emph{trivial
stabilizer}, in $\mathbb{F}^*_q$, so the full action of $PGL_2(q)$
is sharp 3-\emph{transitive} in $\mathbb{F}P^1$ (compare Sect.
3.1); thus one deduces the size: $| PGL_2(q) | = (q+1)q(q-1)$, as it
is: = $(q^2-1)(q^2-q)/(q-1)$, for any $q$. So it
is always divisible by 6. If $q$ is odd, divisible by 24.\\

\underline{Semilinear groups.} Suppose that the field $\mathbb{K}$
has automorphisms; then we have a generalization of linear map in
$V\approx\mathbb{K}^n$ to a \underline{semilinear map} (I. Schur,
1903); $M: V\longrightarrow   V$ is \emph{semilinear}, if

\begin{equation}
\begin{aligned}
&M(x + y) = Mx + My\quad \textrm{(i.e., morphism for the sum in $V$)}\\
&M(\lambda x) =\lambda^{\alpha} M(x),\quad \textrm{where}\quad
\alpha:\lambda\longrightarrow\lambda^{\alpha}\quad \textrm{is an
\emph{automorphism} of the field $\mathbb{K}$}
\end{aligned}
\end{equation}

Correspondence: if  $\alpha = Id$ := trivial, we revert to
\emph{linear} maps. The reader should recall the case of the complex
field $\mathbb{C}$, with the conjugation as automorphism
$z\longrightarrow \bar{z}$ . The maps $M(\lambda x)
=\bar{\lambda}M(x)$  are called there \emph{antilinear}. Antilinear
maps are important in physics (Wigner), because in Quantum Mechanics
states are rays, not vectors, so the natural symmetry groups are the
projective ones, and they come from linear AND antilinear
maps, for $\mathbb{K} = \mathbb{C}$, which is the field case in physics. See [39].\\

By \emph{collineations} in a vector space we mean invertible
 semilinear maps [49]. Let us call $\Gamma L_n(q)$ the \emph{semilinear} $n$-dim. group, of
 invertible $n$-dim. linear $OR$ semilinear maps: $\mathbb{F}^n_q\longrightarrow   \mathbb{F}^n_q$. We have the
 following diagram [49], which generalizes (72): we suppose
  $\mathbb{K}$ is an arbitrary \emph{skew} field, in general non-commutative,
  with $Aut(\mathbb{K})$, $Int(\mathbb{K})$ and $Out(\mathbb{K}) = Aut/Int$ as group of automorphisms,
  internal and group of classes of \emph{autos}; the diagram is\\

  \small{\begin{equation}\label{eq:3}
\begin{CD}
\mathbb{Z}_n @>>>GL_n(\mathbb{K}) @>>> PGL_n(\mathbb{K})\\
@VVV @VVV @VVV\\
\mathbb{K^*}\approx H_n @>>>\Gamma L_n(\mathbb{K}) @>>> P\Gamma L_n(\mathbb{K})\\
@VVV @VVV @VVV\\
I = Int(\mathbb{K})@>>>   Aut(\mathbb{K})@>>>Out(\mathbb{K})
\end{CD}
\end{equation}

Here $H_n$ are the homotheties, or maps $x\longrightarrow x\lambda$
, with $\lambda$ in $\mathbb{K}$ (the vector space over
non-commutative $\mathbb{K}$ is supposed at the \emph{right}); in
the usual, field case (commutative $\mathbb{K}$, by the modern
definition) $Int(\mathbb{K}) = I$, so $\mathbb{K}^*\approx
\mathbb{Z}_n$, and $Out = Aut$. We shall use (80) rather seldom, in
cases of $\mathbb{F}_q$ with $q = p^2$ for example, and
$Aut(\mathbb{F}_q) = \mathbb{Z}_2$.

\subsection{Other regular series (O, Sp, U)}

The group $GL$ will provide us with most of the families
of finite simple groups, by considering its subgroups and subquotients.\\

Besides this first biparametric family of $FSG$, namely $PSL_n(q)$,
which in the Cartan classification corresponds to the $A_{n-1}$
series,
there are three more general ones for the case of simple Lie groups :\\

\begin{equation}
A_n, B_n, C_n\quad \textrm{and}\quad  D_n
\end{equation}

They also have meaning for the finite fields, and provide three
other biparametric families, as we shall expound now. $B$ and $D$
correspond to orthogonal groups, $C$ to the
\underline{symplectic} ones. One might ask the \emph{why} of
precisely these families, and why there are no more; we offer the
following short answers,
which hold for \emph{any} field:\\

Consider a finite vector space $V$ over a field $\mathbb{K}$ (so
$V\approx \mathbb{K}^n$). The rank-two tensors divide in two types:
endomorphisms, or $T^1_1$ tensors, and bilinear forms, say $T^0_2$
tensors $\approx T^2_0$; the latter split in either symmetric or
antisymmetric: under equivalences (by action of $GL(V)$) they keep
the symmetry character. Leaving the matrices or \emph{endomorphisms}
$T^1_1$, for the symmetric case the stabilizer are the orthogonal
groups by definition, and have dimension (as manifolds)\\

\begin{equation}
\textrm{dim}\quad GL_n -  \binom {n+1}{2} = n^2 - n(n+1)/2 =
n(n-1)/2
\end{equation}
which indeed is the dimension of the orthogonal $O(n)$ group (as Lie
group), stabilizer of a (regular or non-degenerate, definite or not)
bilinear symmetric form. For the antisymmetric case, the calculation
is

\begin{equation}
\textrm{dim}\quad GL_n -\binom {n}{2} = n^2 - n(n-1)/2 = n(n+1)/2
\end{equation}
that is, the dim of the symplectic group acting in $n$-dim space. As
we said, the orthogonal case covers the series $B_n$ (odd dimension,
$2n+1$) and $D_n$ (even, $2n$), whereas $C_n$ is the case for the
symplectic group, $Sp_n$: we add that, for the symplectic group, the
space
 dimension has to be \emph{even}, as $M$ regular antisymmetric
 matrix (det$\neq0$ and $M = - ^tM$) implies
  even dimension, as $\mathrm{det}(M) = (-1)^n \mathrm{det}(^tM)$. In the
  complex case ($K = \mathbb{C}$) the orthogonal group is
\emph{unique}, for a given dimension, but e.g. in the real case
$K=\mathbb{R}$ one
  should consider Sylvester's \emph{signature}, important in
physics as the Lorentz group is    $O(3, 1)$.\\

This ``explains'' the $O$ and $Sp$ groups (in arbitrary, perhaps
only even dimension, over any field). What about stabilizers of
other (higher) tensors? If dim $V$ = $n$, a \emph{generic} 3-tensors
will run with dimension like $n^3$, so $> n^2$ = dim$GL_n$; thus, in
general \underline{no} stabilizer: \emph{this is the precise reason
why there are only the O and Sp series of groups in arbitrary
dimension} (and, of course, we add the $SL$ group, corresponding to
the $A_{n-1}$ in Cartan's series, which leaves fixed an $n$-form or
volume form $\tau$, $|\tau | =|\mathbb{K}^*|$ (with dim = 1); in
this case, of course, dim
$SL_n$ = dim $GL_n$ - dim $\tau$   = $n^2 -1$.\\

This argument leaves open, of course, some special (non-generic)
cases; for example, one can argue that the $G_2$ or first
exceptional group in Cartan's series, corresponds to leaving fixed a
3-form $\omega$ in 7 real dimensions, and indeed $7^2 -\binom  {7}
{3}$ = 14 = dim $G_2$ [50]; besides, $G_2$ lies inside $SO(7)$. One
thus ``understands'' the group $G_2$ (and also $F_4$) over any
field; (the connection with \emph{octonions} (with base 1, $e_i$
(i:1 to 7)) is the following (see ``octonions'' at the end of this
section 4.): octonion product yields a $T^1_2$ tensor $\Gamma
(\Gamma : V\times V\longrightarrow V)$, which becomes a 3-form from
antisymmetry $e_ie_j = - e_je_i$ and the existence of a quadratic
form $Q$,
see [50].\\

We describe now briefly the orthogonal, unitary and symplectic
families over finite fields. The easier is the \emph{symplectic}: if
$V = q^{2n}$ is endowed with a regular (hence dim $2n$, even)
bilinear antisymmetric form $\omega$, the stabilizer is defined as
the \underline{symplectic group}, $Sp_n(q)$ (Many authors write it
as $Sp_{2n}(q)$). Given the field $\mathbb{K}$ (= $\mathbb{F}_q$ in
our case) and (even) dimension, the symplectic group is
\emph{unique} up to $GL$ equivalences. The group is unimodular, that
is $Sp_n\subset SL_{2n}$ (identity for $n=1$), because, if $Sp$
maintains a regular 2-form $\omega$  (= antisymmetric rank-two
tensor), it maintains its $n$-th power, which is a volume form:
$\omega^n =\tau$, fixed by $SL$ by definition, hence the group $Sp$
sits inside $SL$. One also shows the centre of $Sp$ is
$\mathbb{Z}_2$,
and let $PSp = Sp/\mathbb{Z}_2$. Now, the general result is\\

\begin{lem}
$PSp_n(q)$ is simple, for any $n \geq 1$ and $q$, with \emph{three} exceptions.
\end{lem}

Again, we quote only the exceptions: as $SL_2(\mathbb{K}) =
Sp_1(\mathbb{K})$ (identical definition, noted above, for any $K$:
in dim 2, the volume form $\tau$  is a 2-form $\omega$), the first
two exceptions are $Sp_1$($\mathbb{K}$ = $\mathbb{F}_2$ and
$\mathbb{F}_3$) = $GL_2(2)$ and $Alt_4$, respectively, which are not
simple, as we know already. The third exception is

\begin{equation}
Sp_2(2)\approx Sym_6\quad \textrm{ (order 720). The alternating
subgroup, $Alt_6$, of order 360, is simple.}
\end{equation}

The order of the finite symplectic groups turns out to be

\begin{equation}
| PSp_n(q) | = q^{n^2} (q^2-1)(q^4-1)(\ldots)
(q^{2n}-1)/(\textrm{$|centre|$})
\end{equation}
where $|centre|$=g.c.d.($n$, $q-1$).\\

The orthogonal case is more complicated, as the
characteristic of the field enters, and also the equivalent to
``signature'' in the real case $\mathbb{K} =\mathbb{R}$: for a fixed
dimension $n$ and field $\mathbb{K}$, there are may be more than one
symmetric bilinear
forms, mod $GL$ equivalence.\\

We just sum up the situation, for $\chi (\mathbb{K}):=
Char(\mathbb{K})\neq 2$: the $\chi (\mathbb{K})$ is important for
the following reason: a regular symmetric bilinear form $f: V\times
V\longrightarrow   \mathbb{K}$ defines a quadratic form $Q(x) :=
f(x, x)$; viceversa, $Q$ defines $f: f(x, y) = (Q(x+y) - Q(x)
-Q(y))/2$ \emph{provided} $\chi (\mathbb{K})\neq2$: this complicates
 affairs. So we shall consider only  $\chi (\mathbb{K})\neq2$.\\

The orthogonal group $O(n)$ admits, as in the continuous case, the
index-2 rotation subgroup $SO(n)$. Still, the group $PSO$ is
\emph{not} yet simple, in general: only the (in general smaller)
commutator normal subgroup $\Omega\subset SO\subset   O$ might be
projective-simple: $P\Omega_n(q)$ is simple and \emph{unique} for
$n$ odd = $2m+1$, (recall that $O/\Omega $ is abelian) (the
physicist reader might recall that the Lorenz group $O(3,1)$ admits
the $SO(3, 1)$ group
as normal but it is the ``orthochrone'' subgroup of the later, $SO_+^\dagger$, which is already simple).\\

\begin{lem}:  The groups $P\Omega_{2l+1}(q)$ is \emph{simple},
and unique,
for any $q$, $\chi(\mathbb{F}_q)\neq2$, and any $l\geq1$.
\end{lem}

There are NO exceptions (but some identities that we omit). The calculation of the order is straightforward:\\

\begin{equation}
| P\Omega_{ 2l+1}(q) | =
q^{l^2}(q^2-1)(q^4-1)(\ldots)(q^{2l}-1)/\textrm{$|centre|$}
\end{equation}
corresponding to the identity of dimensions in the continuous Lie
case, $|B_l| = |C_l|$, \quad so that $|O(2l+1)| =|Sp_l|$, although
of
course the groups are, in general, non-isomorphic.\\

But the even dimensional case $n = 2l$ gives rise to two families:
the difference is due to two inequivalent bilinear forms, and being
somehow subtle, we just
refer to the literature ([21], p.6): call $\Omega^{\pm}$ the two cases:\\

\begin{lem}
The groups $P\Omega^{\pm}_{2l}(q)$ are \emph{simple}, for any $q$ and any $l\geq1$
\end{lem}

The orders are

\begin{equation}
    | P\Omega^{+}_{2l}(q) | = q^{l(l-1)}(q^2-1)(q^4-1)(\ldots)(q^{2l-2} -1)(q^l-1)/(\textrm{$|centre|$})
\end{equation}

and

\begin{equation}
| P\Omega^{-}_{2l}(q) | = q^{l(l-1)}(q^2-1)(q^4-1)(\ldots)(q^{2l-2}
-1)(q^l+1)/(\textrm{$|centre|$})
\end{equation}

When $char(\mathbb{K}) = 2$, they do appear new simple orthogonal
groups, but their orders are among the given already; we do not quote them.\\

The \emph{unitary groups $U(n)$}. Unitary groups come up
because, as said, the finite fields $\mathbb{F}_q$ admit
automorphisms if, in $q = p^f$, $f >1$ (the reader should recall
that the usual unitary groups are \emph{complex}, and complex
conjugation, as automorphism of $\mathbb{C}$, plays a role); neither
the rational field $\mathbb{Q}$ nor the real $\mathbb{R}$ have
automorphisms (but the
trivial $Id$); it will be enough to consider only \emph{involutive} automorphisms.\\

So here we should use fields $\mathbb{F}_q$ with $q$ a power ($f
>1$) of a prime, and admitting an involutary automorphism; hence
$U_n(q=2^2)$, $U_n(q=3^2)$, etc., make sense. In the continuous
case, the unitary groups appear as compact forms in the $A_l$
series, but for finite fields obviously all groups are finite, hence
compact. In total, we consider $O$, $U$, $Sq$ (and $SL$) as
the originators of the four bi-parametric families of FSG.\\

The unimodular restriction $SU$ allows the projective quotient $PSU
= SU/(centre)$, which is generically simple. Indeed, writing $q^2$
for the $|field|$ to assure the involutary automorphism, we have the corresponding result (e.g. [21], 1.5):\\

\begin{lem}
The groups $PSU_n(q^2)$ are simple, except three cases. The order is\\

\begin{equation}
| PSU_n(q^2) |
=q^{n(n-1)/2}\cdot(q^2-1)\cdot(q^3+1)\cdot(q^4-1)\cdot(\ldots)\cdot(q^n
- (-1)^n)/(n, q+1)
\end{equation}
where $(n,q+1)$ is equivalent to the order of the centre.
\end{lem}

The exceptions are $PSU_2(4)$, $PSU_2(9)$ and $PSU_3(4)$.\\

In the \emph{continuum} case the three types $O$, $U$ and $Sp$ can
be related also to the reals $\mathbb{R}$, the complex $\mathbb{C}$
and the \emph{skew} field of the quaternions $\mathbb{H}$; this is
nice, because then the five exceptional
 Lie groups ($G_2$ to $E_8$) can be related to the octonions, and they make up
 \emph{no} families because lack of associativity in the division algebra $O$; see Sect. (4.6). In
 our finite case it is better to think of $SL$, $O$, $Sp$ and $U$ as unimodular, bilinear
 forms and semilinear maps, as we have done.\\

Besides the exceptions already mentioned, there are some identities
(similar somehow to the ``Cartan identities'' among simple
(continuous) Lie groups, e.g. $A_1$ = $B_1$ = $C_1$,
 or $SU(2)$ = $Spin(3)$ = $SpU$ ), which we omit.\\

\subsection{Exceptional series.}
The quoted important paper of Chevalley [55] extended the analysis
to the five exceptional groups of Killing-Cartan: as Lie groups, the
rank is the sub-index and in brackets we write the dimension:

\begin{equation}
G_2(14),\quad  F_4(52),\quad E_6(78),\quad E_7(133),\quad
\textrm{and}\quad   E_8(248)
\end{equation}

They give rise to \emph{five uniparametric} families of FSG over any
field $\mathbb{F}_q$; we just include the order, taken from  Griess [58]. The simple cases are of order:\\

\begin{equation}
|G_2(q)| = q^6(q^6-1)(q^2-1)
\end{equation}

$G_2(2)$, order 12 096, admits $K= PSU(9)$ as \emph{normal} subgroup of index two.\\

\begin{equation}
\begin{aligned}
&|F_4(q)| = q^{24}(q^{12}-1)(q^8-1)(q^6-1)(q^2-1)\\
\end{aligned}
\end{equation}

\begin{equation}
|E_6(q)| = q^{36}(q^{12}-1)(q^9-1)(q^8-1)(q^6-1)(q^5-1)(q^2-1)
\end{equation}

\begin{equation}
|E_7(q)| =
q^{63}(q^{18}-1)(q^{14}-1)(q^{12}-1)(q^{10}-1)(q^8-1)(q^6-1)(q^2-1)
\end{equation}

\begin{equation}
|E_8(q)| = q^{120}(q^{30}-1)(q^{24}-1
)(q^{20}-1)(q^{18}-1)(q^{14}-1)(q^{12}-1)(q^8-1)(q^2-1)
\end{equation}

Only $E_6(q)$  and $E_7(q)$ have centre (to divide by): $(3, q-1)$
and $(2, q-1)$ respectively. There are many relations with the
continuous case; for example, the centre of $E_{6,7}(\mathbb{R}$\ or
$\mathcal{C}$) is
$\mathbb{Z}_3$, $\mathbb{Z}_2$. The order $q^6$, $q^2$ in $G_2$ is related to the  $I_6$,  $I_2$ invariants, etc.\\

There are \emph{two} more families of finite simple groups of Lie
type. Recall the (continuous) simply-laced \emph{simple} (true) Lie
groups $A_n$, $D_n$ ($n > 4$), $D_4$ and $E_6$: these are the ones
exhibiting \emph{outer} automorphisms (type $\mathbb{Z}_2$, except
for $D_4$); in the Dynkin diagrams, that we generally omit, the
outer automorphisms are very clear: interchange and identification
of symmetric nodes (folding); in the case of the $A_n$ series with
the compact representative $SU(n+1)$, the outer automorphism can be
realized as identification of the similar nodes: for example,
$A_3\approx
\circ\!\!\!-\!\!\!-\!\!\!-\!\!\!\circ\!\!\!-\!\!\!-\!\!\!-\!\!\!\circ$
becomes, after \emph{folding}, $B_2 = C_2$:
$\bullet\!\!\!=\!\!\!=\!\!\!\circ$; the identified nodes correspond
to different node: so the Aut-stable subgroup of $SU(4)$ is $Sp(2)$
= $Spin(5)$, in our
notation ($B_2 = C_2$). In general $A_{2n+1}$, corresponding, as compact group, to $SU(2(n+1))$, becomes $Sp_{n+1}=C_{n+1}$.\\

For the $D_n$ series ($n > 4$), the continuous Lie group is $SO(2n)$
(or rather $Spin(2n)$), and the outer automorphism can be realized
as interchange of the two spinor representations; the Aut-stable
subgroup is $B_{n-1}=SO(2n-1)$. $D_4$ shows the maximal \emph{outer}
symmetry: it is Cartan's triality (permutation of the \emph{three}
external nodes: the outer symmetry group is $S_3$); the full folding
of $O(8)\approx D_4$ generates $G_2$. Finally, the primordial
representations of $E_6$ come up in conjugate pairs, mixed by the
outer automorphism, but of course some of them (like the adjoint,
dim 78) are real: the folding generates $F_4$. For all this see,
 e.g. (Jacobson [56]).\\

Steinberg and Tits, continuing the 1955 important work of Chevalley,
[55] came (in 1959) to the conclusion that these Lie algebras with
external \emph{autos} could generated \emph{more} FSG. Sometimes
these groups are called twisted groups, twisting being a
typical mathematical procedure when there are automorphisms. Notice,
first, the difference with the (usual) Lie theory treatment: the
points (subgroup) fixed by the outer automorphism in $A_{2n-1}$ give
rise, (as explained above), to the symplectic series $C_n$, etc;
similarly for the other three cases. We recall the results above.
$H\subset G$ means the subgroup $H$ fixed by the outer automorphism
of $G$:

\begin{equation}
Sp_n\subset   SL_{2n}.-\quad   O(2n-1)\subset   O(2n).-\quad
G_2\subset   O(8).-\quad   F_4\subset   E_6
\end{equation}

All these groups do exist also here over finite fields (indeed we
counted them already), but the Steinberg ``twist'' is different (we
do not elaborate). We refer to the references ([21] and [64]), and
only write the symbols of the new simple groups:

\begin{equation}
^2A_n\quad (n>1).-\quad ^2D_n.-\quad ^3D_4.-\quad\textrm{and}\quad
^2E_6
\end{equation}

Notice, in $^3D_4$, the twist is by the \emph{ternary} symmetry as
$Aut(D_4) = S_3$.
The orders of these ``Aut-twisted'' groups are [57]\\

\begin{equation}
\begin{aligned}
&\textrm{\underline{Group\quad\quad\quad\quad Order\quad\quad\quad\quad\quad\quad\quad\quad\quad\quad\quad\quad\quad\quad\quad\quad\quad\quad\quad Center}}\\
&^2A_n(q),n>1\quad      q^{n(n+1)/2}\prod(q^{i+1}  - (-1)^{i +1})\quad\quad\quad\quad\quad\quad\quad\quad\quad(n+1, q+1)\\
&^2D_n(q),n>3\quad     q^{n(n-1)}(q^n+1)\prod(q^{2i}  - 1) \quad\quad\quad\quad\quad\quad\quad\quad\quad\quad (4, q^n+1)\\
&^3D_4(q)\quad\quad\quad\quad     q^{12}(q^8 + q^4 +1)( q^6 -1)( q^2 -1)\quad\quad\quad\quad\quad\quad\quad\quad\quad\quad             1\\
&^2E_6(q)\quad\quad\quad\quad    q^{36}(q^{12}-1)(q^9 +1)(q^8 -1
)(q^6 -1 )( q^5 +1)( q^2 -1)\quad (3, q+1)
\end{aligned}
\end{equation}

Beware, as explained in detail in [57]: the two series $^2A_n$ and
$^2D_n$ are really already taken in account, as related to
unitary groups and even (-) orthogonal.\\

Finally there is another ``twist'' of the double-laced groups, $B_2
= C_2$, $G_2$ and $F_4$,    (notice these come already from
\emph{folding}, but only three of them) found subsequently by Ree
and Suzuki, this time unrelated to automophisms, but with
restrictions on the fields. Again we do \emph{not} elaborate,
limiting ourselves to
show the new groups: we take again the following table from (Griess [58], last page).\\

\begin{equation}
\begin{aligned}
&\textrm{\underline{Group\quad\quad Field\quad\quad\quad\quad Order}}\\
&^2B_2(q)\quad      q= 2^{2m+1}\quad           q^2(q^2+1)(q-1)\\
&^2G_2(q)\quad     q= 3^{2m+1}\quad           q^3(q^3+1)(q-1)\\
&^2F_4(q)\quad     q=2^{2m+1}\quad  q^{12}(q^6+1)(q^4-1)(q^3+1)(q-1)
\end{aligned}
\end{equation}

This terminates our description of the FSG in families.\\

There is here a resume of
the 18 = 2 + 4 + 5 + 4 + 3 families of FSG, with a minimum of details:\\

1) $\mathbb{Z}_p$ for $p$ prime: abelian, order any prime $p\geq2$. Smallest, $\mathbb{Z}_2$, order 2.\\

2) $Alt_n$ for any natural number $n > 4$: order $n!/2$. Smallest, $Alt_5$, order 60.\\

3) to 6): $PSL_n(q)$, $PSp_n(q)$, $PSUn(q)$, $P\Omega_{n\ odd}(q)$,
together with $P\Omega_{n\ even}(^\pm)(q)$: the classical four
\emph{bi}parametric $(n, q)$ families
 $SL$, $Sp$, $U$ and $O$. Restrictions in $n$, $q$ and exceptions mainly cleared up in the main text.\\

7) to 11): $G_2(q)$, $F_4(q)$, $E_6(q)$, $E_7(q)$ and $E_8(q)$ :
\emph{Uni}parametric
families, associated to the FIVE exceptional Lie groups.\\

12) to 15): Twisted by automorphisms: $^2A_n(q)$, $^2D_n(q)$,
$^3D_4(q)$, $^2E_6(q)$: the first
two still   biparametric families, related to $U$ and to $O(-,odd)$; the last two, uniparametric.\\

16) to 18): Double/triple bond twist: $^2B_2(q)$, $^2G_2(q)$  and
$^2F_4(q)$: three uniparametric families.\\

\pagebreak

The following table lists the FSG up to order 10000. We quote 20
groups.\\

\emph{Simple groups up to order 10000}\\

I) \emph{All} simple groups $G$, with $|G|<10$:\

$\mathbb{Z}_1=I,\quad \mathbb{Z}_2,\quad \mathbb{Z}_3,\quad
\mathbb{Z}_5,\quad \mathbb{Z}_7$\\

II) All \emph{nonabelian} simple groups up to order o: $10< o <100$\\

$Alt_5=SL_2(4)=PSL_2(5)$   \quad \quad order 60\\

III) All nonabelian simple groups up to order o: $100< o <1000$\\

$PSL_2(7)=GL_3(2)$     \quad \quad\quad \quad\quad order 168\

$Alt_6=PSL_2(9)$       \ \quad\quad \quad\quad \quad\quad order 360\

$SL_2(8)$              \ \quad \quad\quad\quad \quad\quad \quad\quad
\quad\quad order 504\

$PSL_2(11)$            \quad\quad\quad \quad\quad \quad\quad \quad\quad order 660\\

IV) All nonabelian simple groups up to order o: $1000< o <10000$\\

$PSL_2(13)$             \quad\quad\quad \quad\quad \quad\quad
\quad\quad order 1092\

$PSL_2(17)$             \quad\quad\quad \quad\quad \quad\quad
\quad\quad order 2448\

$Alt_7$                 \ \ \!\quad\quad\quad\quad\quad \quad\quad
\quad\quad \quad\quad order 2520\

$PSL_2(19)$             \quad\quad\quad \quad\quad \quad\quad
\quad\quad order 3420\

$SL_2(16)$              \!\quad\quad\quad\quad \quad\quad \quad\quad
\quad\quad order 4080\

$SL_3(3)$              \ \ \!\!\quad\quad\quad\quad \quad\quad
\quad\quad \quad\quad order 5616\

$PSL_2(23)$            \quad\quad\quad \quad\quad \quad\quad
\quad\quad order 6072\

$PSL_2(25)$            \quad\quad\quad \quad\quad \quad\quad
\quad\quad order 7800\

$M_{11}$               \ \ \!\quad\quad\quad\quad\quad \quad\quad
\quad\quad \quad\quad order 7920\

$PSL_2(27)$           \quad\quad\quad \quad\quad \quad\quad
\quad\quad order 9828\\

\subsection{Division algebras and octonions.}

We have referred to the complex numbers $\mathbb{C}$, quaternions
$\mathbb{H}$ and even octonions $\mathbb{O}$ several times in this
review. In this subsection, we collect several
results about the three ``division algebra'' extensions of the real
numbers $\mathbb{R}$.  All the three are used in physics in
different contexts.\\

If, in the vector space $\mathbb{R}^2$ we define in the second unit
$i = (0,1)$ the square as $i^2 = -1$, the pair of real numbers ($x$,
$y$) in the form $z = x + iy$ generate the algebra of complex
numbers $\mathbb{C}$: sum and product follow automatically, and both
operations are commutative, associative, and distributive with each
other; define \emph{conjugation} of $z$ as $\bar{z}:= x - iy$,
\emph{norm} as $\mathcal{N}(z) = \bar{z}z = x^2 + y^2$, real number
$\geq 0$ and \emph{inverse} ($z\neq0$) as $z^{-1}
=\frac{\bar{z}}{\mathcal{N}(z)}$. Then $\mathbb{C}$ is a
bidimensional division algebra over the reals
$\mathbb{R}$ ($\equiv$ in which any element $\neq 0$ has an
inverse); also $\mathbb{C}$ is a field, in the sense of
Sect. 1.2. As we already said, the field $\mathbb{R}$ has no proper
automorphisms, $Aut(\mathbb{R}) = I$, as any \emph{auto} should
verify $\alpha(0) = 0$, $\alpha(1) = 1$, so $\alpha(n/m) = n/m$,
even in the limit. But now in the field $\mathbb{C}$, consider
\emph{autos preserving} $\mathbb{R}$; then only conjugation
survives, and we have $Aut_{\mathbb{R}}(\mathbb{C}) = Z_2$.
Historically, the complex numbers appeared if one wanted to express
the zeros of an arbitrary polynomial, $P_n(x) = 0$, even with purely
real coefficients and quadratic. By the beginning of the 19th
century, the field of the complex numbers was well defined and
established, and very much used in mathematics (Gauss, Argand,
first; then Cauchy, Riemann, Weierstrass, etc.).\\

W. R. Hamilton was busy, in the 1830s, attempting fruitlessly, to
find an extension of the doublets $(x, y)$ for complexes
$\mathbb{C}$ to \emph{three} real numbers $(x, y, z)$ with the
division properties of the complex numbers: we know perfectly well
today why he did not succeed: we need a \emph{power of two} for the
total number of units; so Hamilton himself invented the quaternion
numbers $q = (u, x, y, z)$, (units $1+2(i, j) +1(ij = k)$ ) in
October, 1843, by extending to three \emph{new} (imaginary) units:
$i$, $j$ and $k:=ij$; but, in order to imitate the product,
conjugation, norm and inverse of $\mathbb{C}$, he had to suppose
anticommutativity: $ij = - ji$. As we described already the
quaternions $q$ in Sect. 2.4, we shall not elaborate, only to
remember that, writing $q = u + ix + jy + kz$ as $q = u +
\mathbf{x}$, for $\mathbf{x}$ a three-vector, conjugation is
$\bar{q} = u - \mathbf{x}$, norm is $\mathcal{N}(q) :=\bar{q}q = u^2
+ \mathbf{x}\cdot\mathbf{x}$ , real $\geq 0$ and inverse is $q^{-1}
= \frac{\bar{q}}{\mathcal{N}(q)}$. Quaternions $\mathbb{H}$ were
used firstly as 3-Dim rotations (as $SU(2)$ covers twice $SO(3)$),
but found not many other applications until Gibbs, Heaviside and
others at the last third of the 19th century used the imaginary part
($u = 0$) for the \emph{vector calculus}. The modern qualification
of $\mathbb{H}$ is as a \emph{skew field}. For an actual reference,
see [59].\\

It is easy to show that $Aut_{\mathbb{R}}(\mathbb{H}) = SO(3)$:
$e_{1,2,3}$ with the antisymmetric product acting as a 3-form in
3-space, so the invariance group is the unimodular subgroup $SL$;
but it has also to be an orthogonal transformation and  $SL\cap O
= SO$. Notice also conjugation in the quaternions is \emph{only}
antiautomorphism, as
$(qq')^- = \bar{q}'\bar{q}$.\\

Now with three \emph{independent} new units $e_1$, $e_2$  and $e_3$,
the total number of units is now $2^3= 8$ (1; $e_i$; $e_i e_j$;
($e_1$ $e_2$) $e_3$ with 1+3+3+1 = 8). To guarantee division, one
has to suppose, not only squares = - 1 (i.e. $e_i^2 = - 1$), and
anticommutativity, like in the quaternions, with $e_1 e_2 = - e_2
e_1$ etc., but also \emph{antiassociativity}, (called alternativity)
in the sense that $(e_1 e_2) e_3 = - e_1 (e_2 e_3)$. If, in full
analogy with the two previous cases of $\mathbb{C}$ and
$\mathbb{H}$, we define an octonion as $o = v + \boldsymbol{\xi}$
with $v$ real and a $\boldsymbol{\xi}$ vector in $\mathbb{R}^7$, we
can define again product, conjugate, norm and inverse as $o\cdot o'=
vv'-\boldsymbol{\xi}\cdot\boldsymbol{\xi}' + v\boldsymbol{\xi}'+
v'\boldsymbol{\xi} +\boldsymbol{\xi}\wedge\boldsymbol{\xi}'$ (the
vector product $\wedge$ implies a choice),  $\bar{o} = v
-\boldsymbol{\xi}$, $\mathcal{N}(o) = \bar{o}o\geq 0$, $o^{-1} =
\frac{\bar{o}}{\mathcal{N}(o)}$ ($o\neq0$): now the octonions
$\mathbb{O}$, as the reals $\mathbb{R}$, the complex $\mathbb{C}$
and the quaternions $\mathbb{H}$, are \emph{division algebras}, i.e.
a (real) vector space with a multiplication law, which allows
inverse for any number $\neq 0$. Octonions, for lack of a better
name, as called just a 8-dim \emph{division algebra} (over the
reals); see e.g. Baez [81].\\

Bott and Milnor proved in 1958 (see, e.g. [82]) that there are no
more real division algebras: one can generalize e.g. the octonions
to the sedenions, with a total of 16 units, but then
there are no universal inverses.\\

There is no problem now in considering vector spaces of any
dimension over $\mathbb{C}$ or over $\mathbb{H}$, as they were a
field (and a skew-field) respectively: the reader is already used to
$\mathbb{C}^n$. Noncommutativity of the quaternions forces one to
distinguish between $\mathbb{H}$-left vector spaces and
$\mathbb{H}$-right, according to which  $\lambda v$ or $v\lambda$ is
defined, with $\lambda$ in $\mathbb{H}$ (we used in the text the
\emph{right} case). However, non-associativity is easily seen to be
an obstacle to consider vector spaces over the octonions with more
than three dimensions; in particular, the projective \emph{plane}
$\mathbb{O}P^2$ exists, (it is called the ``Moufang plane''), but
not of higher dimensions; see [81].\\

\emph{Exceptional Lie groups}. The five exceptional Lie groups
of ([90], pag. 57) are all related to the octonions; we just want to
explain the relation of the first two:\\

We argued that $Aut(\mathbb{R}) = I$, $Aut_{\mathbb{R}}(\mathbb{C})
= Z_2$, $Aut_{\mathbb{R}}(\mathbb{H}) = SO(3)$. What about
$Aut(\mathbb{O})$? The three independent units (orthogonal ($\perp$
)) $e_1$, $e_2$ and $e_3$ have to move to three others; the first
can go to any point in the 6-dim sphere of norm-one imaginary
octonions, then the second ($\perp$) to the equator $\approx S^5$,
and the third is restricted to a $S^3$ by the images of $e_1$, $e_2$
and $e_1 e_2$: so the $Aut(Oct)$ group has 6 + 5 + 3 = 14
parameters, and it has to be orthogonal: it is called the group
$G_2$ (Cartan's name and classification), it has rank 2, with 14
parameters (dimensions) as Lie group, and it lies inside $SO(7)$ (in
particular it is connected and compact). The natural representation
is 7-dimensional, as $Aut(\mathbb{O})$, acting on the imaginary
octonions, and it has also the ``adjoint'' representation of dim 14.
For a ``dual'' interpretation of $G_2$ as stabilizer group of a
3-form, see [50]. So we have\\

\begin{equation}
Aut(\mathbb{O}) \equiv Aut(Oct) =G_2\quad  \textrm{(rank 2,
dimension 14)}
\end{equation}

It will take us some time to find the \emph{Aut} group for the
Moufang projective octonionic plane, so we just state the result
(see Baez [81] or Conway [59]):\\

\quad \quad \quad $Aut(\mathbb{O}P^2): =F_4$  (rank 4,
dimension 52)\\

As for the three other exceptional Lie groups, $E_{6,7,8}$, they are
related to some Jordan algebras over the octonions, but we omit a
complete description; see again Baez [81].\\

As a resume of the four division algebras, we write the following
Table:\\

\underline{Division Algebra\quad\quad\quad\quad dim (over $\mathbb{R}$)\quad\quad\quad Character\quad\quad\quad\quad\quad\quad Automorphism group}\\

\quad$\mathbb{R}$\quad\quad\quad\quad\quad \quad\quad\quad\quad\quad\quad \quad  1\quad \quad\quad \quad\quad \textrm{Comm.\& Associat.}\quad $Aut(\mathbb{R}) = I$\\

\quad C\quad or\quad$\mathbb{C}$\quad\quad\quad\quad\quad\quad\quad\quad\quad \!\!\!\!2\quad\quad\quad\quad\quad\quad yes \quad\quad yes\quad \quad \quad \quad \!\!$Aut_{\mathbb{R}}(\mathbb{C}) = Z_2$\\

\quad $\mathbb{H}$\quad\quad\quad\quad\quad \quad\quad\quad\quad\quad\quad \quad  4\quad\quad\quad\quad\quad\quad no \quad\quad yes\quad \quad \quad \quad \!\!$Aut_{\mathbb{R}}(\mathbb{H})=SO(3)$\\

\quad $\mathbb{O}$\quad or \quad
Oct\quad\quad\quad\quad\quad\quad\quad\quad\!\!\!\!\!8\quad\quad\quad\quad\quad\quad
no \quad\quad no\quad \quad \quad \
$Aut_{\mathbb{R}}(\mathbb{O})=G_2$

\pagebreak

\section{Sporadic groups}
\subsection{Introduction to sporadic groups.}

By definition, sporadic groups are finite simple groups (FSG)
(nonabelian, of course) not in the previous (2+16=18) families; the
name is due to W. Burnside [2], who attached it to the Mathieu
groups, five finite simple groups discovered by the French
mathematical
physicist \'{E}. Mathieu, starting back in 1861.\\

For over a century, no more sporadic groups were discovered (nor
much research went into that, for that matter!). Then, in the period
1960-1975 mathematicians all over the world completed them, starting
by Janko (1965) (group $J_1$); the list today consists altogether of
26 groups, ranging in size from Mathieu's $M_{11}$, of order
$11\cdot10\cdot9\cdot8 = 7920$, to the Monster group $\mathbb{M}$,
of order $\approx10^{54}$. There were several ways to consider the
problem: before Janko, people even thought the Mathieu's groups were
the only sporadic ones! On the hypothesis that all FSG were of even
order, that is, contained involutions, it was shown by Brauer (Cfr.
e.g. [83]) \emph{that the centralizers of involutions} $\{z| za =
az$, for $a$ the involution$\}$ would somehow select the type of
possible simple groups; indeed, that was an important tool to
discover many of the sporadic groups. Another clue was provided by
Fisher, in his search of higher-transpostition
groups (explained later).\\

Today we know that these sporadic groups gather together in 3+1
related series, the first three (generations) interconnected, with
respectively 5 + 7 + 8 (+ 6) = 26 sporadic groups; the isolated
(fourth) series is composed of the so-called 6 \emph{pariah} groups
(name due to Griess [58]). These related three generations contain
mostly subgroups or subquotients of the Monster group, which is in
this sense a sort of all-embracing
 group (but not quite): the Monster group $\mathbb{M}$, dealt with in detail below, is by
 far the biggest of the sporadic groups, with close to $10^{54}$ elements. These remaining,
 6 ``pariah'' groups, seem at the moment to be totally unrelated with anything else.\\

Let us stress here that these \emph{sporadic groups} are stranger
objects that, say, the exceptional Lie groups; for one thing, the
later give rise to \emph{families} of FSG (5 in fact, as we stated);
for another, these isolated Lie groups are no doubt connected with
the octonions, a well understood mathematical structure (for $G_2$,
see e.g. [50]), while the 26 sporadic groups do not depend, as far
as we can tell today, on any known clear mathematical structure: we
have to leave for the future a (full) understanding of these 26
sporadic groups. To be sure, they constitute well defined
mathematical structures, also the three generations seem to be
clearly inter-related, and for most of these groups a ``natural''
action in some sets is also known; all seem to ``depend'', somehow,
on the number 24, but we lack the level of understanding that we
have, for example, for the finite simple
groups of Lie type\ldots\\

Indeed, the three generations (called the ``happy family'' by Griess
[58]) do show some common skeleton, and they appear (as said) in
three neat sets, the first generation
fairly understood as a pair of isomorphic/nonisomorphic objects, as we shall just explain here:\\

The five Mathieu groups, $M_{11}$, $M_{12}$; $M_{22}$, $M_{23}$ and
$M_{24}$  originate in the equivalence (already quoted):

\begin{equation}
Alt_6\approx PSL_2(9)\quad\textrm{ (order 360)}.
\end{equation}\\

But the `extension$\cdot2$´ are different:

\begin{equation}
Sym_6\neq PGL_2(9)\quad\textrm{ (order 720)}.
\end{equation}\\

This connects with $M_{11}$ and $M_{12}$.\\

There is another relation for the other 3 Mathieu groups: According
a result from Artin [53], $|Alt_8|= 20160$ is the \emph{smallest}
order for which there are two \emph{nonisomorphic} finite simple
groups! We shall see later the relation of this with the other three
Mathieu groups, $M_{22,23,24}$.\\

For a recent review of FSG, see the monograph by Wilson (Wilson
[84]).\\

\subsection{The first generation: Mathieu groups.}
Mathieu was searching, back in 1861, groups more than 3-transitive,
and found \emph{five} new groups; it was shown later (Miller,
1900)
that they were also simple, see [61].\\

Recall (Sect. 3.1): a group $G$ operating in a space $\Omega$
($G\circ\!\!\!\!-\!\!\!\longrightarrow\Omega)$ acts
\emph{transitively}, if there is only one orbit, or equivalently any
point $P\in\Omega$ can be transformed into any other point $Q$ for
some $g\in G$, that is $g\cdot P = Q$. The action
$G\circ\!\!\!-\!\!\!\longrightarrow\Omega$ is (e.g.) three times
transitive (3-\emph{trans}) if any three different points $P$, $Q$,
$R$ can be transformed into three arbitrary different images, $P'=
g\cdot P$, $Q'= g\cdot Q$ and $R'= g\cdot R$. We also showed (Sect.
4.3) that $G := PGL_2(q)$, which is NOT simple, acting on the
projective line $\mathbb{F}_qP^1$ (of $q+1$ points) acts \emph{sharp
3-trans} (sharp: after the last action there are no left-over
stabilizer (but $I$)); it follows that $|G| = (q+1)\cdot
q\cdot(q-1))$.
The subgroup $PSL_2(q)$ is generally \underline{simple} (as noted), but it is only 2-transitive.\\

With reference mainly to the permutation group $S_n$, these notions of transitivity were already well developed by the 1860s.\\

Now we introduce the first two Mathieu groups, $M_{11}$ and
$M_{12}$. Recall the alternating group $Alt_n$ is simple for $n > 4$
(Galois). In particular, the smallest \emph{nonabelian} simple
groups, of order less than 2000,
are (we repeat, isomorphisms included)\\

\begin{equation}
\begin{aligned}
&\quad \quad Alt_5=PSL_2(5)=SL_2(4).-PSL_2(7)=GL_3(2).-Alt_6=PSL_2(9).\\
&\textrm{\underline{Order}}:\quad\quad 60\quad\quad\quad\quad\quad\quad\quad\quad\quad\quad  168\quad\quad \quad\quad\quad\quad\quad\quad\quad\quad 360\\
&\quad\quad\quad \quad SL_2(8).-PSL_2(11).-PSL_2(13)\\
&\textrm{\underline{Order}:}\quad\quad 504\quad\quad\quad\quad
660\quad\quad\quad\quad\quad 1092
\end{aligned}
\end{equation}

Focus on $Alt_6$: it has a natural extension to $Sym_6$, order 720;
but, as isomorphic to $PSL_2(9)$, it must have also another
extension$\cdot2$, to $PGL_2(9)$; one shows these last two groups
are \emph{not} isomorphic! The diagram clears this up:

\begin{equation}
\begin{CD}
Order\quad 360\quad PSL_2(9)@=Alt_6\\
\quad\quad \quad @VVV@VVV\\
Order\quad 720 \quad PGL_2(9)@. Sym_6\\
\quad \quad \quad @VVV@VVV\\
\quad\quad \quad \mathbb{Z}_2@=\mathbb{Z}_2\\
\end{CD}
\end{equation}\\

As we showed (Sect. 2.8 on $S_n$), the $Alt_n$ group admits always a
natural extension to $Alt_n\cdot2 = Sym_n$, due to an external
automorphism in $Alt_n$, mixing the two maximal cycles of equal
length. So now, $Alt_6$ should have another automormphism,
generating the other $Alt_6\cdot2$ extension, namely the quoted
$PGL_2(9)$.
Indeed it has, as was observed independently by Sylvester [60] in 1844, before Mathieu time!\\

That means: $Alt_6$ ($=PSL_2(9)$) has more than one \emph{outer}
automorphism: $\alpha$, say, to generate $Sym_6$, and $\beta,$ to
generate $PGL_2(9)$. Hence, as $\alpha$  and $\beta$ are involutive
and commute, $\alpha\beta$ must be a \emph{new} involutive external
automorphism, which gives rise to a \emph{third} (different)
extension!
 Call it $M_{10}$. The full group of classes of automorphism of $Alt_6$ is then $V = (\mathbb{Z}_2)^2$, and we have

 \begin{equation}
 Out(Alt_6) = V(e;\alpha, \beta, \alpha\beta) = \mathbb{Z}_2\times   \mathbb{Z}_2
 \end{equation}\\

 and the diagram

\begin{equation}
\xymatrix{ &Sym_6\ar[dr] &\\
Alt_6\approx PSL_2(9)\ar[ur]\ar[r]\ar[dr] &PGL_2(9)\ar[r] &P\Gamma
L_2(9)\\ &M_{10}\ar[ur] &\\
}
\end{equation}

ORDER \underline{\quad \quad \quad \quad 360 \ \quad \quad \quad \quad \quad \quad 720 \quad \quad \quad \quad \quad \quad \quad 1440}\\

As $9 = 3^2$, the field $\mathbb{F}_9$ has automorphisms (see Sect.
4.2), indeed an involutive one, which is instrumental in defining
the semilinear group $\Gamma L_n(9)$: extending these three
intermediate groups $Sym_6$, $PGL_2(9)$ and $M_{10}$ by the leftover
automorphism, we end up in the \emph{same} group, $P\Gamma L_2(9)$,
of order 1 440! By the way, $Alt_6$ is the \emph{only} $Alt_n$ group
with more than
one \emph{outo} $\neq e$ (Sylvester). As for $Sym_n$, only $Sym_6$ presents outer automorphisms.\\

Now recall $PGL_2(9)$ has to be sharp 3-transitive on the projective
line $\mathbb{F}_9P^1$, with 9+1=10 points, so order
=$10\cdot9\cdot8 = 720$ indeed, whereas $Sym_6$, of the same order,
is sharp 6-transitive on 6 symbols: $6! =
6\cdot5\ldots\cdot2\cdot1$. One shows now $M_{10}$ \emph{inherits}
this 3-sharp property of $PGL_2(9)$, but (and this is the
\emph{crucial point}) it admits also an \emph{augmentation} to a
certain $M_{11}$ group, which is \emph{sharp four-transitive}
 in 11 symbols, hence of order $11\cdot10\cdot9\cdot8 = 7920$, AND a second augmentation
 to a certain $M_{12}$ group, again \emph{sharp five-transitive} in 12, so of order $12\cdot11\cdot10\cdot9\cdot8 = 95 040$.
 This is not wholly understood, although a theory of augmentations can be read off in [43].\\

One shows then (Miller, see [61]) that both $M_{11}$ and $M_{12}$
(but clearly, not $M_{10}$) are \underline{simple groups}: the first
two sporadic groups! (Recall Sect. 3.4:
 extensions imply normal subgroup, but augmentations do not, by definition: in $K\longrightarrow  E\longrightarrow   Q$,
 $E$ \emph{extends} $K$, meaning $E/K\approx Q$, but if $H\subset   G$, we just say $G$ is an \emph{augmentation} of $H$).\\

As for the other three Mathieu groups, $M_{22,23,24}$ they are based
(as said) in another equivalence

\begin{equation}
GL_4(2)\approx Alt_8\quad \textrm{(order 20160)}
\end{equation}\\

\emph{and} non-equivalence:

\begin{equation}
|GL_4(2)| = | PSL_3(4) |,\quad \textrm{but}\quad GL_4(2)\neq
PSL_3(4)
\end{equation}\\

As $20 160 = 8!/2 = 21\cdot20\cdot48$, the action \emph{cannot be sharp} this time! One shows:\\

\quad \quad $Alt_8$ is (not sharp!) 2-transitive in 21 symbols, so call it also $M_{21}$.\\

And again, a new ``miraculous'' result comes up: $M_{21}$ has a
natural \emph{augmentation} to $M_{22}$, which is simple and
3-transitive in 22 symbols, with TWO more augmentations
 to $M_{23}$ (4-\emph{trans}) and to $M_{24}$ (5-\emph{trans}), none of them sharp but simple.
  We limit ourselves to state the groups again, and the order:\\

$M_{21}\equiv Alt_8$, \emph{simple}, order $8!/2 =
21\cdot20\cdot48$, 2-transitive in 21 symbols and 6-\emph{trans} in
8.

\begin{equation}
\begin{aligned}
&M_{22}:\quad \textrm{ by augmentation, order $22\cdot21\cdot20\cdot48$, 3-\emph{trans} in 22. \underline{Simple}}\\
&M_{23}:\quad \textrm{another augmentation, order $23\cdot22\cdot21\cdot20\cdot48$: 4-\emph{trans} in 23. \underline{Simple}}\\
&M_{24}:\quad \textrm{ another augmentation; order $24\cdot|M_{23}|$ : 5-\emph{trans} in 24 symbols. \underline{Simple}}\\
&\textrm{In particular,}\quad |M_{24}| = 244 823 040 =
2^{10}\cdot3^3\cdot5\cdot7\cdot11\cdot23.
\end{aligned}
\end{equation}

We only recall that the number 24 is a kind of ``Magic'' number,
related to the three generations of sporadic groups; here we see
other reasons why the same number 24 appears
 in different mathematical contexts.\\

It is remarkable than NO more 4- or 5-transitive groups (besides the
well-understood cases of $Sym_n$ and $Alt_n$)
 have been found since 1873; we lack any theoretical reason to support this; in fact, for a long
  time it was thought that the Mathieu's groups were the start of a whole series of groups \emph{more} than 3-transitive!.
  To-day, we have the theorem: there are no 6-transitive or higher, groups, except $S_n$ and $Alt_n$.\\

These five Mathieu groups are also related to some triplets called
``Steiner systems'' (see [61]),
 to some error-correcting codes, in particular the so-called ``Golay Code''
[61], [47] etc. The Golay code can be understood as a subspace of $\mathbb{F}^{24}_4$, whose automorphism group is $M_{24}$.\\

  The literature on the Mathieu groups is very extensive. We quote [61], [47], [60] among others.\\

Some of the Mathieu  groups, mainly $M_{24}$, have physical
applications; see Sect. 6.\\

\subsection{Second family of sporadic groups.}

They are associated to the Leech Lattice, and comprehend in total
seven groups;
 the Leech lattice was discovered (Leech, 1962) in coding theory (for a good story, see Ronan [62]).\\

Let us introduce a bit of lattice theory, e.g. [47]. In the plane
$\mathbb{R}^2$, there are \emph{three}
 ways to (periodic) tessellate regularly it (= to cover with regular polygons): triangles,
  squares and hexagons; the latter is the thightest covering. The second is an example of a plane \underline{lattice}. In
  general, a lattice in $\mathbb{R}^n$ is the $\mathbb{Z}$-span of a vector base, plus a quadratic form.\\

In higher dimensions, to find general (i.e., not necessarily
regular) tessellations is a standard (and difficult) problem in
mathematics (for example, a conjecture of Kepler (in 1611) was not
proven until 1998 (Hales; paper in 2005 [85])). It turns out that in
8 and in 24 dimensions, there are ``special'' lattices. The famous
$E_8$ lattice in 8 dimensions was discovered by Gosset in 1900; see
([47], pag 120). In particular, John Leech discovered the
\underline{Leech lattice} in dimension 24. It represents the best
packing of spheres in dimension 24, and Leech was using a device
discovered already by Witt in 1938 for the Mathieu's groups [86]. In
the plane, the best (hexagonal) packing means that a circle touches
six others; in 3-dim, best packing of spheres is with 12 (6+3+3)
contacts, but in 24 dimensions, the corresponding sphere touches
optimally 196 560 others.\\

The mathematician John H. Conway took the challenge (1967) of
calculating the automorphism
 group of such a lattice, say Aut(Leech). It turned out to be a giant finite group,
  (not simple), called today $Co_0$, with size of the order of the $10^{18}$ elements.
   With the help of Thomson, an expert on group theory, Conway established first \emph{three new simple groups},
    related to the Aut(Leech) non-simple group; the first was simply $Co_0/\mathbb{Z}_2$; here are the three of them by order:\\

\begin{equation}
\begin{aligned}
&\textrm{$Co_1$ group, order} = 2^{21}\cdot3^9\cdot5^4\cdot7^2\cdot11\cdot13\cdot23\approx4.16\times 10^{18}\\
&\textrm{$Co_2$ group, order} = 2^{18}\cdot3^6\cdot5^3\cdot7\cdot11\cdot23\approx4.23\times10^{13}\\
&\textrm{$Co_3$  group, order} =
2^{10}\cdot3^7\cdot5^3\cdot7\cdot11\cdot23\approx.5\times10^{12}
\end{aligned}
\end{equation}

It was Thomson who deduced, that $Aut(Leech)/\mathbb{Z}_2\approx
Co_1$. This is obtained as
 stabilizer of a point of the lattice, and also $Co_2$ and $Co_3$ are (particular)
  stabilizers of two and three lattice points (this is magnificently explained by Ronan in [62], see also [47]).
   Moreover, taking stabilizers of more points Thomson realized that new simple groups were appearing, although
some of them were already known: in total, \emph{seven} new finite
simple groups were linked to the 24-dimensional Leech lattice!
     The remaing four are:\\

\quad \quad Suzuki group \underline{Sz}.   Size:\quad    $2^{13}\cdot3^7\cdot5^2\cdot7\cdot11\cdot13 = 448 345 497 600$\\

\quad \quad McLaughling group   \underline{McL}.  Size:\quad  $2^7\cdot3^6\cdot5^3\cdot7\cdot11 = 898 128 000$\\

\quad \quad Higman-Sims group, \underline{HS}. Size:\quad
$2^9\cdot3^2\cdot5^3\cdot7\cdot11=44 352 000$; these three groups are related to ``5'' stabilizers\\

\quad \underline{HJ} or Janko $J_2$ group: $2^7\cdot3^3\cdot 5^2\cdot7  = 604 800$,   related to ``7'' stabilizers; discovered first by Janko; HJ stands for Hall-Janko.\\

For more information on these ``Leech family'' groups, see [47], also [58], [83], [84], etc.\\

\subsection{The Monster group.}

Let us tell now a bit of the origin of the third generation
 of sporadic groups, in particular the biggest of them all,
 the Monster group. We advance that the connection with the two previous
  generations was not immediate; in fact, it is related to the number
   196 560 of spheres touching a central one in dimension 24.\\

The biggest finite simple sporadic group, the Monster $\mathbb{M}$,
was discovered independently by B. Fischer and by R. L.
 Griess in 1973, and constructed by Griess in 1980. Griess claims the first
presentation of the Monster group $\mathbb{M}$ was on January 14, 1980.
 It is also called the ``Friendly Giant'',
  and named, sometimes, $F_1$. Its exact order is gigantic:\\

\begin{equation}
|\mathbb{M}|=2^{46}\cdot3^{20}\cdot5^9\cdot7^6\cdot11^2\cdot13^3\cdot17\cdot19
\cdot23\cdot29\cdot31\cdot41\cdot47\cdot59\cdot71\approx8\cdot10^{53}
\end{equation}
comparable to the number of protons in the Sun! It has 194 classes
of conjugate elements, so
 the same number of inequivalent \emph{irreps}. Notice
  the five missing primes: 37, 43, 53, 61 and 67,
   before 71: five out of 20. The \emph{lowest} dimensional \emph{irreps} are,
   of course, of dimensions factors of the order:\\

\begin{equation}
\begin{aligned}
&\textrm{Id},\quad 196 883\quad (= 47\cdot59\cdot71),\quad  21 296 876\quad (= 2^2\cdot31\cdot 41\cdot 59\cdot 71),\\
& 842 609 326\quad (= 2\cdot13^2\cdot29\cdot31\cdot47\cdot59)
\end{aligned}
\end{equation}

To see the way Fischer was led to the Monster, consider the dihedral
group $D_h$, of order $2h$ (Sect. 2.6), and the Coxeter diagram
$\circ\!\!\!-\!\!\!-\!\!\!-\!\!\!\circ\!\!\!-\!\!\!-\!\!\!-\!\!\!\circ$
; the nodes are involutions
 (transpositions); say $(a,b)$, with $(ab)^h=e$;
  Fischer concluded (1971) that for a FSG group to be generated by transpositions (i.e.,
  more involutions linked by different $h$'s), besides $Sym_n$ (a \emph{linear} chain, with $h=3$
   among neighbours, $h=2$ if not) and other known cases, there were \emph{three} new Sporadic FSG,
    somehow similar to second series of the Mathieu's groups (see our Sect. 5.1); they
     were eventually called $Fi_{22}$, $Fi_{23}$, and $Fi_{24}$: (the original group was not simple
and written $Fi_{24}'$; it has the simple group $Fi_{24}$ as
index-two subgroup) they are much bigger
      than the Mathieu groups (sizes below).  Later, it was conjectured the existence
       of a much bigger group, which would have up to 6-transpositions (i.e., its involution pairs $a_1a_2$
        would have order six at most). This is the path which took Fischer (and Griess also) to the
        Monster group (the Baby Monster, see next, did also appear).\\

For the moment, the only relation between $\mathbb{M}$ and
 the two previous generations of sporadic groups is that the first
  (non-Id) \emph{irrep} has dimension close to the number of touching
   spheres in the Leech lattice, namely 196 560: indeed Griess' first
   construction of $\mathbb{M}$ was as the automorphism group of a commutative non-associative
    algebra of dimension 196 884. But there are other constructions as well:
    see e.g. chap. 29 in [47]: in particular, both Tits and Conway gave soon
another two constructions of the Monster group; see e.g. [62].\\

 As for the five missing primes (37\ldots),
    the same ones had been already found by Ogg (1976) in relation to modular functions; see [11].\\

We leave for the next chapter the relation of the Monster group with
physics
 (which was also instrumental in the first constructions of $\mathbb{M}$), and remark here
 only another construction of the Monster as a kind of ``Coxeter group'', generated
  by involutions (see Sect. 2.5, Coxeter groups). In fact, any noncyclic finite simple
   group can be understood as quotient of a Coxeter group (perhaps of infinity order).
   Let $\mathcal{G}_{pqr}$, $p\geq q\geq r\geq2$ be a ``Dynkin'' graph with three
   legs of length $p+1$, $q+1$ and $r+1$ sharing a common endpoint (see Fig-1- in Gannon [11]);
   with $p=q=r=5$, all 16 points are involutions, and the order of products $(ab)$ is 3
    (if adjacent) or 2 (non adjacent); with a relation $\mathcal{R}$ (that we omit)
     the (quotient) group $\mathcal{G}_{pqr}/\mathcal{R}$, named $Y_{pqr}$ has
      order $2 |\mathbb{M}|^2$. From that one gets the monster $\mathbb{M}$.
      It has also 2, 3 and 4 classes of orders 2, 3 and 4, so the Character Table $\chi_d$ starts
       with $1A$, $2A$, $2B$, $3A$, $3B$, $3C$\ldots in horizontal and $\chi_1$, $\chi_{196883}$, $\chi_{21 296 876}$
         etc. in vertical. See the complete table in the Atlas [63].\\

\subsection{Other groups in the Monster family.}

A total of 8 FSG constitute the third generation of the ``Happy
Family'' of groups.
 As they have so far not found many applications in physics, we include just the list and the order.
  The second biggest, the so-called \underline{Baby Monster} $B$ (Conway) derives (today) easily from $\mathbb{M}$
   itself, and it is the second biggest sporadic group. Actually, it was suspected to exist before the Monster was!.\\

\begin{equation}
\textrm{Baby Monster,}\quad
|B|=2^{41}\cdot3^{13}\cdot5^6\cdot7^2\cdot11\cdot13\cdot17\cdot19\cdot23\cdot31\cdot47\approx4\cdot10^{33}
\end{equation}

Fisher discovered also, as said, three groups related to the second
set of Mathieu groups
$M_{22, 23, 24}$. They have symbols $Fi_{22,23, 24}$\\

$Fi_{22}$: order $2^{17}\cdot3^9\cdot5^2\cdot7\cdot11\cdot13$\\

$Fi_{23}$ : order $2^{18}\cdot3^{13}\cdot5^2\cdot7\cdot11\cdot13\cdot17\cdot23$\\

$Fi_{24}$ : order    $2^{21}\cdot3^{16}\cdot5^2\cdot7^3\cdot11\cdot13\cdot23\cdot29$\\

The last is sometimes written as $Fi_{24}'$, because the original
$Fi_{24}$ was \emph{not} simple.\\

The three other groups completing this third generation are (they
are very much related to the Monster $\mathbb{M}$ and to the
baby Monster $\mathbb{B}$)\\

\underline{HN} (for Harada-Norton): order \quad $2^{14}\cdot3^6\cdot5^6\cdot7\cdot11\cdot19$\\

\underline{Th} (for Thomson): order \quad $2^{15}\cdot3^{10}\cdot5^3\cdot7^2\cdot13\cdot19\cdot31$\\

\underline{He} (for Held): order \quad $2^{10}\cdot3^3\cdot5^2\cdot7^3\cdot17$\\

These groups are rather enigmatic. For example, $Fi_{22}$, HN and Th
have \emph{irreps} of dim 78, 133 and 248 respectively, i.e. as the
dimensions of the exceptional Lie groups $E_6$, $E_7$ y $E_8$.\\

\subsection{The ``Pariah'' groups.}

As we mentioned, Janko found the first sporadic group after
Mathieu's in 1965, of modest order, $|J_1| = 175560$; it does not
fit into the three generations of sporadic groups. But Janko also
discovered three more unconnected sporadic groups, $J_{2,3,4}$; the
second really belongs to the Leech lattice generation, as pointed
out by Hall. But $J_3$ and $J_4$ were genuine new isolated
``pariah'' groups (orders below).\\

Two of the remaining Pariah groups were related somehow to the
sporadic families: thus the Lyons group (1969) Ly covers the McL
group in the Leech family series, and the very same $J_4$ group is
also related to the $M_{24}$ group in the first family. The O'Nan
group (ON) and the Rudvalis group (Ru) complete the series of 6
genuinely new unrelated (``Pariah'') groups; a brief table follows
(mainly from
(Griess [58]), with orders and discoverers.\\

\quad \quad \quad Table. \underline{The SIX pariah groups}\\

\underline{Name\quad \quad \quad \quad Order\quad\quad\quad
\quad\quad \quad\quad \quad\quad \quad \quad \quad \quad
Discoverer\quad \quad \quad\quad \quad
\quad \quad Year}\\

Janko-1, $J_1$ \quad $2^3\cdot3\cdot5\cdot7\cdot11\cdot19$ \
\quad\quad\quad\quad\quad\quad\quad\quad\quad\quad\quad \  Janko\
\quad\quad\quad\quad\quad\quad\quad\quad
1965\\

Janko-3, $J_3$ \quad     $ 2^7\cdot3^5\cdot5\cdot7\cdot17\cdot19$\
\quad\quad\quad\quad\quad\quad\quad\quad\quad\quad\quad
Janko\quad\quad\quad\quad\quad\quad \quad \quad \  1968\\

Lyons, Ly\quad $
2^8\cdot3^7\cdot5^6\cdot7\cdot11\cdot31\cdot37\cdot67$\quad\quad\quad\quad\quad\quad\quad\quad\quad
Lyons\ \quad\quad\quad\quad \quad \quad \quad \quad 1969\\

Rudvalis, Ru\quad       $ 2^{14}\cdot3^3\cdot5^3\cdot7\cdot13\cdot29$\quad\quad\quad\quad\quad\quad\quad\quad\quad    Rudvalis \ \quad \  \ \quad\quad\quad\quad\quad \quad  1972\\

O'Nan, ON\quad       $2^9\cdot3^4\cdot5\cdot7^3\cdot11\cdot19\cdot31$ \ \quad\quad\quad\quad\quad\quad\quad\quad\quad        O´Nan\quad\quad\quad\quad\quad\quad\quad\quad 1973\\

Janko-4, $J_4$\quad $
2^{21}\cdot3^3\cdot5\cdot7\cdot11^3\cdot23\cdot29·\cdot31\cdot37\cdot43$\quad\quad
\quad\quad\quad\quad
Janko\quad\quad \quad\quad\quad\quad\quad \  \ \   1975\\

As a last comment, the order of a FSG is always divisible by 2:
today this is a \emph{theorem} (the Feit-Thomson theorem, 1963, see
[65]). The factor 3 is nearly always present, too.\\

We end up with a table of all (26) sporadic groups.

\pagebreak

\subsection{Table of sporadic groups by size.}

$M_{11}$ Mathieu
\quad\quad\quad\quad\quad\quad\quad\quad\quad\quad\quad\quad\quad\quad\quad\quad\quad\quad\quad\quad\quad\quad\quad\quad\quad\quad
7 920\

$M_{12}$ Mathieu \quad\quad \
\quad\quad\quad\quad\quad\quad\quad\quad\quad\quad\quad\quad\quad\quad\quad\quad\quad\quad\quad\quad\quad\quad\quad
95 040\

$J_1$ Janko
\quad\quad\quad\quad\quad\quad\quad\quad\quad\quad\quad\quad\quad\quad\quad\quad\quad\quad\quad\quad\quad\quad\quad\quad\quad\quad\quad
175 560\

$M_{22}$ Mathieu\quad \
\quad\quad\quad\quad\quad\quad\quad\quad\quad\quad\quad\quad\quad\quad\quad\quad\quad\quad\quad\quad\quad\quad\quad\quad
443 520\

$HJ$  Janko 2 ( Hall-Janko) \ \quad\quad \
\quad\quad\quad\quad\quad\quad\quad\quad\quad\quad\quad\quad\quad\quad\quad\quad\quad604
800\
\\

$M_{23}$ Mathieu
\quad\quad\quad\quad\quad\quad\quad\quad\quad\quad\quad\quad\quad\quad\quad\quad\quad\quad\quad\quad\quad\quad\quad\quad
10  200 960\

$HS$ (Higman-Sims) \
\quad\quad\quad\quad\quad\quad\quad\quad\quad\quad\quad\quad\quad\quad\quad\quad\quad\quad\quad\quad\quad
44 352 000\

$J_3$
Janko\quad\quad\quad\quad\quad\quad\quad\quad\quad\quad\quad\quad\quad\quad\quad\quad\quad\quad\quad\quad\quad\quad\quad\quad\quad\quad
50 232 960\

$M_{24}$ Mathieu \
\quad\quad\quad\quad\quad\quad\quad\quad\quad\quad\quad\quad\quad\quad\quad\quad\quad\quad\quad\quad\quad\quad\quad
244 823 040\

$McL$ MacLaughlin\quad \
 \  \ \quad\quad\quad\quad\quad\quad\quad\quad\quad\quad\quad\quad\quad\quad\quad\quad\quad\quad\quad
898 128 000\
\\

$He$  Held \quad \
\quad\quad\quad\quad\quad\quad\quad\quad\quad\quad\quad\quad\quad\quad\quad\quad\quad\quad\quad\quad\quad\quad\quad
4 030 387 200\

$Ru$
Rudvalis\quad\quad\quad\quad\quad\quad\quad\quad\quad\quad\quad\quad\quad\quad\quad\quad\quad\quad\quad\quad\quad\quad
145 926 144 000\

$Sz$
Suzuki\quad\quad\quad\quad\quad\quad\quad\quad\quad\quad\quad\quad\quad\quad\quad\quad\quad\quad\quad\quad\quad\quad\quad
448 345 497 600\

$ON$ O'Nan \quad\quad \
\quad\quad\quad\quad\quad\quad\quad\quad\quad\quad\quad\quad\quad\quad\quad\quad\quad\quad\quad\quad
460 815  505 920\

$Co_3$
Conway\quad\quad\quad\quad\quad\quad\quad\quad\quad\quad\quad\quad\quad\quad\quad\quad\quad\quad\quad\quad\quad\quad
495 766 656 000\
\\

$Co_2$
Conway\quad\quad\quad\quad\quad\quad\quad\quad\quad\quad\quad\quad\quad\quad\quad\quad\quad\quad\quad\quad\quad42
305 421  312 000\

$Fi_{22}$
Fischer\quad\quad\quad\quad\quad\quad\quad\quad\quad\quad\quad\quad\quad\quad\quad\quad\quad\quad\quad\quad\quad
64  561  751  654 400\

$HN$
(Harada-Norton)\quad\quad\quad\quad\quad\quad\quad\quad\quad\quad\quad\quad\quad\quad\quad\quad
273  030 912 000 000\

$Ly$
Lyons\quad\quad\quad\quad\quad\quad\quad\quad\quad\quad\quad\quad\quad\quad\quad\quad\quad\quad\quad\quad
51 765 179  004 000 000\

$Th$
Thomson\quad\quad\quad\quad\quad\quad\quad\quad\quad\quad\quad\quad\quad\quad\quad\quad\quad\quad\quad
90 745 943  887  872 000\
\\

$Fi_{23}$
Fischer\quad\quad\quad\quad\quad\quad\quad\quad\quad\quad\quad\quad\quad\quad\quad\quad\quad\quad
4 089 470 473 293  004 800\

$Co_1$ Conway
\quad\quad\quad\quad\quad\quad\quad\quad\quad\quad\quad\quad\quad\quad\quad\quad\quad
4 157 776 806 543 360 000\

$J_4$
Janko\quad\quad\quad\quad\quad\quad\quad\quad\quad\quad\quad\quad\quad\quad\quad\quad\quad\quad\quad
86 775 571  046  077  562 880\

$Fi_{24}$
Fischer\quad\quad\quad\quad\quad\quad\quad\quad\quad\quad\quad\quad\quad\quad
1 255 205 709 190 661  721 292 800\

$B$   (Baby Monster)\quad\quad\quad\quad\quad 4 154 7681 481 226 426
191 177 580 544 000 000\

$\mathbb{M}$ Monster\quad\quad\quad\quad\quad\quad
$2^{46}\cdot3^{20}\cdot5^9\cdot7^6\cdot11^2\cdot13^3\cdot17\cdot19\cdot23\cdot29\cdot31\cdot41\cdot47\cdot59\cdot71$\

\quad\quad\quad\quad\quad\quad\quad\quad\quad\quad\quad\quad\quad\quad\quad\quad\quad\quad$(|\mathbb{M})\sim
8.04\times10^{53})$.

\pagebreak

\section{Physical applications}

\subsection{Rotations and permutations.}

In this final Section we shall apply finite groups to
Bose$\backslash$Fermi particles (Sect. 6.1), explain the relation of
the Monster group with string theory, a physical construct,
(Sects. 6.2, 6.3) and comment very briefly on the group $M_{24}$ in relation to the $K3$ compactification problem (Sect. 6.4).\\

\underline{Symmetries} occur frequently in Physics; that is, a
particular physical system is invariant under some (usually
geometric) transformations, like translations, rotations,
dilatations, etc. There are also ``internal'' symmetries, even more
important. For example, ``all directions are equivalent in 3-space''
amounts to invariance under the 3-dimensional rotations group
$SO(3)$, etc. In classical mechanics this leads (through Noether's
Theorem) to conservation of the angular momentum vector
$\mathbf{J}$, to plane (if unperturbed) planetary orbits, etc.
Interactions are described today by ``gauge forces''; these also
signal some gauge groups, like $U(1)$ for electromagnetism, $SU(3)$
color for the strong force, etc.\\

In Quantum Mechanics (Q.M.), the symmetries one wants to contemplate
have to be implemented as \emph{projective} representations of the
respective group $\mathcal{G}$. This is because the physical states
are (as said), instead of points (like in phase space), rays in
Hilbert space $\mathcal{H}$, and the projective unitary group
$PU(\mathcal{H})$ is the pertinent object, preserving rays and
unitarity (probability); so, if $\mathcal{G}$ is the (classical)
symmetry group one wants to implement, one should seek
representations $\mathcal{G}\longrightarrow PU(\mathcal{H})$. That
was very clear from the instauration of modern Q.M. in 1925. Von
Neumann and Wigner worked out the first cases; three books existed
from the very
beginning, [5], [6] and [7].\\

Projective representations of a group $G$ are usually obtained from
the linear ones of a bigger group $\hat{G}$ (See Sect. 3.1); it is
remarkable that some of these higher groups were already found much
earlier in crystallography, where they were called e.g. ``binary
tetrahedral'' group(s), for the case of the ordinary regular
tetrahedron $T_3$. (This was possible because $SU(2)$, as covering
of $SO(3)$ was found early (ca. 1840) in connection with the
\emph{quaternions}). The following (repeated) diagram specifies the
situation (The Alternative group $Alt_4$, with 12 elements, rotates
the 4 vertices of the tetrahedron $T_3$):

\begin{equation}
\begin{aligned}
&Z_2\longrightarrow 2\cdot Alt_{4}-\!\!\!-\!\!\!\longrightarrow Alt_{4}\\
&\parallel\quad \quad \quad \cap\quad \quad \quad \quad \quad \cap\\
&Z_2\longrightarrow SU(2)\longrightarrow SO(3)\\
\end{aligned}
\end{equation}

Here $2\cdot Alt_4$ lies inside $SU(2)$, as $Alt_4$ lies inside
$SO(3)$. All the projective \emph{irreps} of $SO(3)$ come from the
linear ones of the ``covering group'' $SU(2)$ (Topologically, $SU(2)
= Spin(3)$ is the universal covering group of the $3d$ rotations
group, as $Spin(n)$ is for $SO(n)$). Now, as we mentioned, the
\emph{irreps} of $SU(2)$ are conventionally named $D_j$, (where $j$
= 0, 1/2, 1, 3/2, \ldots), are complex (real if $j$ integer), and of
dimensions $2j+1$; as the \emph{irreps} of $SO(3)$ are with $j$
integer: this is the \underline{very reason} why half-integer
angular momentum appears in Q.M. The affair is not innocuous, as the
Spin-Statistics \emph{theorem} (W. Pauli, 1940) is equivalent to:
half-integer spin particles obey the exclusion principle. Now it is
very clear to anybody understanding chemistry that this principle is
the true \underline{differentiating principle} in Nature, the fact
that inspite of (most) stable matter being built up with only three
components (electrons, protons and neutrons), it offers such
agreable distinction of composites and forms!\\

We would like to re-state this assert in the form: objects are
different in Nature because the fundamental symmetry group (namely,
$SO(3)$) is not simply-connected! (so it has a double covering, SU(2)).\\

We are tacitly using another symmetry, which is permutation
symmetry. The main advantage of Democritus versus Aristotle in the
times of the Greeks (25-23 centuries before present) is that atoms
(or today rather, elementary particles) by definition, are specified
once a \emph{finite} number of properties are known (independently
of space-time position); for instance: mass, electric charge and
spin specify perfectly the electron (in the atomistic perspective).
Now, an assembly with $N$ electrons in interaction should be
invariant under the $S_N$ permutation symmetry (of order $N!$),
since, being identical, all experience the same forces. But it turns
out that nature does not use all the \emph{irreps} of this group
$S_N$, but only the simplest, the one-dimensional ones: as
$Sym_N/Alt_N = \mathbb{Z}_2$, there are precisely two (as we know)
one-dimensional \emph{irreps}; so the quantum state $|\Psi
>$ of $N$ identical particles has only two possibilities under
exchange:

\begin{equation}
|\Psi (1, 2,\ldots i,\ldots, j,\ldots N)
>=\pm|\Psi (1, 2,\ldots j,\ldots, i,\ldots N)
>
\end{equation}
which go with the names of Bose-Einstein (BE, +; 1924) and
Fermi-Dirac statistics (FD, $\!-\!-$; 1925/6), the first alternative
is fulfilled by \emph{bosons}, by definition, the other by
\emph{fermions}. Normally contituents of matter are fermions, like
electrons and quarks, whereas carriers of forces are bosons, like
photons, gluons etc. There have been several attempts to generalize
the BE/FD statistics to \emph{parastatistcs},
without too much success.\\

This spin-statistics connection is an universal rule, for which no
exceptions have been found. BE statistics is instrumental in forming
\underline{coherent} states of matter (e.g. in the laser), while FD
is the guarantee of chemical
valence, hence of all shapes and forms in Nature, as said.\\

We state all this as a triumph of science facing philosophy: never,
in their wildest dreams, have ever philosophers thought of a
property of matter guaranteeing the formation of shapes and forms,
inspite of the (very simple and identical) atomic constituents\ldots
This is an advertisement to people, philosophically  minded, who
enter into modern science from the other side (first philosophy):
they would hardly ``grasp''
the lessons of quantum mechanics\ldots\\

There are several books devoted to representation theory of groups
as related to quantum mechanics; besides the already quoted ones, we
might add [66] and [67].

\subsection{Monstrous Moonshine.}

In November 1978, J. McKay in Montreal remarked that

\begin{equation}
                196 884 = 1 + 196 883
\end{equation}

This is more than a joke: the left-hand side refers to the expansion
of $j(\tau )$, a modular function, whereas the right side counts the
first two dimensions of the \emph{irreps} of the Monster group!
Hardly two branches of mathematics were more apart: the theory of
\emph{modular forms} is an outgrowth of the theory of \emph{elliptic
functions}, in its turn a development of complex (analytic)
functions, while the Monster, as we said (Sect. 5) is the biggest
sporadic FSG. J. Conway and S.P. Norton [68] coined the expression
``Monstrous Moonshine'' to label
this phenomenon.\\

Summing up for the ``explanation'' (and following closely [11]):
today (since around 2000) we say: There is a \emph{vertex operator
algebra} (a construct from Physics: string theory; see below),
called the Moonshine module $V^{\sharp}$, which interpolates in
(116): its automorphism group is the Monster group $\mathbb{M}$, and
their graded dimensions are the coefficients of the modular
$j$-function. As we know already about $\mathbb{M}$, let us
``introduce'' the
$j$-function.\\

The upper half plane $H\subset \mathbb{R}^2 = \mathbb{C}$, is
defined by $H:= \{\tau\in\mathbb{C} |Im(\tau )>0\}$: it admits the
$SL_2(\mathbb{R})$ group as isometries: this group acts in the whole
complex plane $\approx \mathbb{R}^2$, and the real axis $\mathbb{R}$
$(Im\  z = 0)$ is invariant, as it is
$\mathbb{R}\cup\{\infty\}\approx \mathbb{R}P^1$. The action is by
homographies: if $ad-bc=1\neq0$,

\begin{equation}
\{a, b; c, d\}:\tau\longrightarrow\frac{a\tau+b}{c\tau+d}
\end{equation}

The action is \emph{ineffective} (Sect. 3.1), with kernel diag
$\{\pm1, \pm1\} = \mathbb{Z}_2$. The effective group is
$PSL_2(\mathbb{R})\approx SO_+^\dag (2,1) $. It is well-known that
$H$ is like the hyperbolic plane: a simply connected (non-compact)
surface with constant (negative) curvature; in fact
$SL_2(\mathbb{R})\approx Spin(2, 1)$, is the double cover
(\emph{not} universal) of $SO_+^\dag (2,1)$: this is called,
sometimes, the \emph{split} form of the $B_1$ Lie algebra ($=A_1$)).
An important subgroup of $G = SL_2(\mathbb{R})$ is $S :=
SL_2(\mathbb{Z})$ (inspite of
$\mathbb{Z}$ being not a field, $S$ makes sense, as the inverses belong to it).\\

Roughly speaking, any \emph{discrete} subgroup $G$ of
$SL_2(\mathbb{R})$ forms a kind of \emph{lattice}, and therefore
generates, as an orbit in $H$, a compact surface; a $G$-modular
function is a meromorphic function $f:
\hat{H}\longrightarrow\mathbb{C}$ invariant under $G$ ($\hat{H}$ is
a ``completion'' of $H$ adding some points ``at infinity''): for
$\left(
                                                                   \begin{array}{cc}
                                                                     a & b \\
                                                                     c & d \\
                                                                   \end{array}
                                                                 \right)\in
G$

\begin{equation}
f(\frac{a\tau+b}{c\tau+d})=f(\tau)
\end{equation}

Now $\mathbb{Z}\subset   \mathbb{R}$, and
$\mathbb{R}/\mathbb{Z}\approx S^1$ is compact: functions on compact
spaces can be parameterized by angles, and are like periodic
functions in general spaces, hence they do admit a Fourier series
expansion. The alluded $j(\tau)$ function is meromorphic, it is a
function of this type, and admits the series (with $q(\tau ) :=
e^{2\pi i\tau} = q(\tau+1)$ periodic)

\begin{equation}
    j(\tau ) = 1/q + 744 + 196 884q + 21 493 760q^2 + 864 299 970q^3 +\ldots
\end{equation}

The $1/q$ ``singularity'' is not really there, because (as said) $H$
is ``completed'' with $\infty$; the constant term $744(=24\times31)$
is irrelevant. Now the \emph{Moonshine} phenomenon (Conway and
Norton, 1979) [68] is the equivalence of numbers in (119) with the
dimensions of the \emph{irreps} of $M$: besides (116) we have more
equivalences, namely

\begin{equation}
\begin{aligned}
&21 493 760 =1+196 883+ 21 296 876;\\
& 864 299 970 =1+1+196 883+196 883 + 21 296 876 + 842 609 326
\end{aligned}
\end{equation}

Later, other ``Moonshine'' cases were found (Thomson, Kac\ldots) for
other groups; for example, the exceptional group $E_8$ is
related to $j(\tau )^{1/3}$, see below.\\

The central structure to understand the equivalences (116) and (119)
is an infinite-dimensional graded $\mathbb{M}$-module

\begin{equation}
V = V_0\oplus   V_1 \oplus  V_2\oplus   V_3\ldots\oplus   \ldots
\end{equation}
where each vector space $V_i$ undergoes an $\mathbb{M}$-action
through one or several \emph{irreps}. The equivalence can be written
in the form (take $j = J + 744$)

\begin{equation}
q J(\tau ) =\sum^{\infty}_{n=0} q^n\textrm{dim}(V_n) = 1 + 196
884q^2 + 21 493 760q^3 +\ldots
\end{equation}

The ``proof'' that this approach works, is that a similar process
can be written down for other ``Moonshine'' groups (McKay,
Kac,\ldots); for example (McKay)

\begin{equation}
j(\tau )^{1/3} = q^{-1/3}(1 + 248q + 4124q^2 + 34752q^3 + \dots)
\end{equation}
where 248, 3875 = 4 124 -248 -1 and 30 380 = 34 752 - 3875 -
$2\cdot248 -1$) are the dimensions of the \emph{irreps} of the $E_8$
group (as a Lie group); and again, $j^{1/3}$ is another modular
function\ldots There are many other examples, mainly developed by
McKay, some of which involve the so-called Kac-Moody algebras, an
affine extensions of ordinary, Lie algebras (that we do not
explain. See the reprint book [72]); however, each tier contains a finite-dimensional representation of the (ordinary, not extended) Lie algebra.\\

Before going on, let us recall a curious relation between the Leech
lattice (Sect. 5.3) and the Monster $\mathbb{M}$: we already
mentioned the number 196 590: in the Leech lattice, it is the number
of norm-4 vectors, and in fact the ``$\theta$-series'' (for any
lattice in space one can construct its elliptic $\theta$-series, see
e.g. (Lang [22])) is our Monster function $J(\tau ) + 24$ times the
standard $\eta$ function to the power 24. This is another example of
the Moonshine phenomenon. For a general overview, see ([69], [70]
and [71]).

\subsection{String Theory.}

In 1988 Frenkel \emph{et al}. wrote the book ``Vertex Operator
Algebras and the Monster'' [9] about a physical construction of the
Monster group starting from string theory, and in the same year R.
Borcherds [10] completed the work, enlarging the concept of Lie
algebras beyond the natural extension, Kac-Moody
algebras; see [72].\\

A few words on string theory. Around 1974 Schwarz and Scherk, two
outstanding physicists, proposed that a unified theory of the whole
world of elementary particles and forces (including gravitation)
could be attempted starting, not with (a) particle moving in
spacetime (describing a curve, say) and quantizing it, but as a
piece of a \emph{string} (either closed or open), moving like a
surface (called ``worldsheet'', with metric (1, 1)) in a higher
space and with quantizable excitations; these excitations could be
computed, and included particles like the gravitons and the photons.
(String theory really started earlier (ca. 1968) as a putative
theory of \emph{hadrons}). Theory developed consistently for the
next years, and since about 1985 presented itself in \emph{five}
forms, with the following characteristics (see
e.g. the two standard references by Green \emph{et al.}, [73], and Polchinski [74]).\\

The strings were all \emph{supersymmetric}, that is, their
excitations included fermions as well as partner bosons; the five
types lived in \emph{ten} dimensions (this is fixed by the absence
of a dilatation \emph{anomaly} only in this dimension; for the
purely Bose string this dimension was 26. Notice 26 = 24 + (1, 1)
and 10 = 8 + (1, 1), as well as 24 = 3·8: this numerology still has
a hidden meaning!). There might be one or two fundamental
supersymmetries (Type I and Type II superstrings); there might be
also internal symmetry (``gauge'') groups, but also constrained by
absence of anomalies: the only possible gauge groups were $O(32)$ or
$E_8\times E_8$, both of rank 16 and dimension 496, which is the
third perfect number (after 6 and 28); again, this numerology is not
yet
understood.\\

As the world around us has ostensibly only 4 = (3, 1) dimensions,
some reason must exist for not observing the extra (six) dimensions:
this is the \emph{compactification problem}, far from being solved,
even
today (spring-2013).\\

String theory has many other problems, but a big explosion occurred
in 1995, when E. Witten showed [87] that all five existing viable
superstring theories were different aspects, of a unique scheme
(called ``M-theory''), but the posterior progress in $M$-theory has
been very
scarce.\\

To describe string interactions is not an easy task either. In
1988/89 Borcherds constructed, as said, an extension of Lie algebras
beyond the Kac-Moody level: already around 1965 both Kac and Moody
had extended Lie algebras to some infinite-dimensional (affine)
algebras, one for each (finite) Lie algebras. Borcherds went a step
further, and constructed a ``Monster Lie algebra'' (see, e.g. [75]
and Chaps. 29 and 30 of [47]); it uses a lattice, enlargement of the
Leech lattice, in (25,1) dimensions, using the numerical
``coincidence'' for light-like vectors

\begin{equation}
\sum^{24}_{i=0}n_i^{2}=(70)^2
\end{equation}
(to understand this recall [88] $\sum^{N}_{i=1} n_i^2 =
N(N+1)(2N+1)/6)$.\\

With that construct,  Borcherds proved in 1990 that indeed the [9]
vertex algebras satisfy the Conway-Norton conjectures (including,
but generalizing, the first McKay observation 196883+1=196 884).
Borcherds got the Fields Medal in Mathematics in 1998 for this work
[76].\\

Unfortunately, we do not find Borcherds's achievements easy to
expose; we believe sometime will be needed to make this theory
accesible. So we leave it at that.\\

\subsection{K3 and $M_{24}$}

As the last topic in this review paper we want to mention a recent
connection between a \emph{complex} surface, $K3$, and the Mathieu
group $M_{24}$, work due to the Japanese school [13]. First, the
actors:\\

If $(\mathcal{V}, g)$ is a $n$-dimensional riemannian manifold, the
\emph{holonomy group} $Hol=Hol(g)$ is the collection of orthogonal
transformations $g\in O(n)$ of a frame (orthobase) $\varepsilon$
 moved along a closed loop $\gamma$ (for better understanding, see e.g.
[27]), that is $\varepsilon' = g\cdot\varepsilon$. For a general
manifold, $Hol(g)$ is $O(n)$. For a K\"{a}hler $n$-manifold $Hol$
lies inside $U(n)$, and if the first Chern class $c_1=0$, it is
inside $SU(n)$. Now the $K3$ manifold ($K3$ stands for Kummer,
K\"{a}hler and Kodaira; the name is due to A. Weil, 1952) is the
topologically unique \emph{complex} surface (four real dimensions)
with $SU(2)$ holonomy. K3 is instrumental in many models of
(partial) compactification. In 2010, Eguchi \emph{et al} [13]
established a remarkable relation between some properties of this
$K3$ surface (called ``elliptic genera'') and the \emph{irreps} of
the largest Mathieu group, $M_{24}$. Again, we do not elaborate.\\

Further studies on the $K3-M_{24}$ relation can be seen in [79] and [89].\\

Another relation between the Monster group $\mathbb{M}$ and physics
was found by E. Witten [77]; namely, in three-dimensional gravity
there is a famous ''BTZ'' black hole [78] and Witten found a
relation between the number of quantum states in this black hole and
the dimensions of the \emph{irreps} of the Monster.\\

It is too early to attribute any special meaning to this
coincidence\ldots.\\

\section{   ACKNOWLEDGMENTS}

This Report originated in a Seminar delivered by the author in the
``Centro de Física Pedro Pascual'' in Benasque, Spain, in the summer
of 2010. Prof. Carl BENDER (St. Louis) invited me to produce a full
review for publication; after him, Prof. Joshua FEINBERG (Tel Aviv,
Israel) insisted. The author apologizes for the long delay; lately
it was Alexandra HAYWOOD, from the IOP, who acepted the several new
deadlines. An incomplete Report has appeard in [90].\\

For help with the TeX version I thank my young collaborator, C.
Rivera; himself and several colleagues here in Zaragoza University
cleared up some points. This work has been supported by the Spanish
CICYT (grant FPA 2006-02315) and the aragonese DGIID-DGA (grant
2007-E242).

\pagebreak

\section{L I T E R A T U R E}

\quad \quad\quad \quad \quad\quad \underline{$\mathbf{Sect. I}$}\\

[1].-  F. Klein, ``Erlangenprogram''. Math. Ann. \underline{43}
(1893), 63.- \emph{Le programme d'Erlangen}. Gauthier-Villars 1974.\

[2].- W. Burnside: \emph{The Theory of Groups}. Cambridge U.P.
1897.\

[3].-  G. A. Miller, H. F. Blichfeldt and L. E. Dickson.
\emph{Theory and Applications of Finite Groups}. J. Wiley, New York
1916.\

[4].-  A. Speiser: \emph{Die Theorie der Gruppen von endlicher
Ordnung}. Springer, Berlin 1937.\

[5].- H. Weyl: \emph{The Theory of Groups and Quantum Mechanics}.
Dover 1960.\

 [6].- B.L. van der
Waerden: \emph{Group Theory and Quantum Mechanics}. Springer,
Berlin, 1932.\

 [7].-  E.P. Wigner: \emph{Group Theory and
Applications to Quantum Mechanics}. Academic Press, 1959.\

 [8].-  E.U. Condon and G. Shortley. \emph{The Theory of Atomic
Spectra}. Cambridge U.P., 1935.\

 [9].-  I. Frenkel, J.
Lepowski and A. Meurman: \emph{Vertex Operator Algebras and the
Monster}. Academic Press, San Diego 1988.\

 [10].-
R.E. Borcherds: ``What is Moonshine? Vertex Algebras, Kac- Moody
Algebras, and the Monster.'' Proc. Nat. Acad. Sci. (PNAS) USA
\underline{83} (1986), 3068-3071. Proc. Int. Cong. Maths.(ICM).
Berlin 1998.\

 [11].-  T. Gannon: ``Monstrous Moonshine: the
first 25 years''. Bull. London Math. Soc. \underline{38} (2006),
1-33.\

 [12].-  T. Gannon: \emph{Moonshine beyond the Monster},
Cambridge U.P. 2006.\

 [13].-  T. Eguchi \emph{et al}., ``Notes on K3
Surface and Mathieu Group $M_{24}$.'' Exp. Math. \underline{20} \
(2011), 91-96.\

[14].-  R. L. Griess, ``The Friendly Giant'', Inv. Math.
\underline{69} (1987), 1-102.\

[15].- D. Gorenstein: \emph{Finite Simple Groups}. Plenum Press, New
York 1982.\

[16].-  H. Weyl, \emph{The Classical Groups}. Princeton U.P. 1939.\

[17].-  S. McLane, ``Concepts and Categories in Perspective'', in
\emph{A Century of Mathematics in America}, A.M.S. Vol. 1, 323-366
(1988).\

[18].-  W. Lederman, \emph{Theory of Finite groups}. Oliver \& Boyd,
1953.\

[19].-  R. D. Carmichael, \emph{Groups of Finite Order}, Dover
1956.\

[20].-  A. D. Thomas and G. V. Wood, \emph{Group Tables}, Shiva
Pub., 1980.\

[21].-  R. W. Carter, \emph{Simple Groups of Lie type}. J. Wiley,
New York, 1972.\

[22].-  S. Lang: \emph{Algebra}. Addison-Wesley, 1965.\

[23].-  G. Birkhoff and S. McLane. \emph{Modern Algebra}. McMillan,
1941.\

[24].-  F. Klein, \emph{Elementary Mathematics from Higher
Standpoint}. Dover   2004.\

[25] .- \emph{In Wikipedia}: Sets, Relations and Groups.\

[26].-  J. G. Hocking and G. S. Young, \emph{Topology}.
Addison-Wesley 1964\

[27].-  S. Kobayashi and K. Nomizu: \emph{Foundations of
Differential
                  Geometry}, 2 Vols., J. Wiley 1963, 1969.\

[28].-  L.J. Boya and C. Rivera, ``Grupos Abelianos Finitos''. La
Gaceta,          R.S.M.E., \underline{13}(2, 2010),
    229-244.\

[29].-  B. L. van der Waerden, \emph{Moderne Algebra}. Springer,
Berlin 1931.\

[30].-  R. B. Ash, \emph{Basic Abstract Algebra}. Dover (2000).\

[31].-  L. Tondeur, \emph{Lie Groups}: Lect. Notes Math.
$\underline{7}$. Springer 1965.\\

\quad \quad\quad \quad \quad\quad \underline{$\mathbf{Sect. II}$}\\

[32].-  E. Artin, \emph{Lectures on Galois theory}. Dover 1998\

[33].-  D.J.S. Robinson, \emph{A Course in the Theory of Groups}.
Springer 1996.\

[34].-  B. Huppert, \emph{Endlicher Gruppen, I und II}. (Springer
1967).\

[35].- H. U. Besche, B. Eick and E. A. O´Brien: "A Millenium
Project: constructing         small groups". Int. J. Alg. \& Comp.
\underline{12}, 5 (2002), 623-644.\

[36].-  L.J. Boya and M. Byrd, ``Clifford periodicity from Finite
Groups''. J.   Phys. $\mathbf{A}$ \underline{32}, (1999),
L201-L205.\

 [37].-  A. G. Kurosh, \emph{Theory of Groups}, 2 Vols. Chelsea, 1956.\

[38].-  L. J. Boya in ``Hyderad 2010 Int. Cong. Math.'', unpublished.\\

\quad \quad\quad \quad \quad\quad \underline{$\mathbf{Sect. III}$}\\

[39].-  E. P. Wigner, Ann. Math. \underline{40} (1939), 149.\

[40].-
  C.W. Curtis, \emph{Pioneers of Representation Theory}. Am.
Math. Soc. 1999.\

[41].-  H. S. M. Coxeter and W. Moser, \emph{Generators and
Relations for Discrete  Groups}.- Springer, Berlin 1980.\

 [42].- Th. Kahan, \emph{Theorie des groups en Physique Classique et
Quantique}.- Dunod, Paris, 1960.\

[43].- N.L. Biggs and A.T. White, \emph{Permutation groups}.-
Cambridge U.P. 1979.\

 [44].-  S. McLane, \emph{Homology} .- Springer 1967.\

 [45].-  L. Pontriagin, \emph{Topological Groups},
Princeton U. P. 1946.\

 [46].-  Y. Berkovich, \emph{Groups of
Prime Power Order}.- W. de Gruyter, Berlin;            Vol. I,
2008.\

 [47].-  J. H. Conway and N. J. A. Sloane, \emph{Sphere
packings, Lattices and             Groups}.- Springer 1988.\

 [48].- B. Simon, \emph{Representations of Finite and Compact
Groups.} Am. Math. Soc., Providence, R.I. 1996, p.-43-.\

[49].-  J. Dieudonné, \emph{La geometrie des Groupes classiques}.-
Springer 1955.\

 [50].-  L.J. Boya and  R. Campoamor ``Composition
Algebras and the two faces of $G_2$'', Int. J. Geom. Meth. Mod.
Phys.
$\mathbf{7}$(3) (2010), 367-378\\

\quad \quad\quad \quad \quad\quad \underline{$\mathbf{Sect. IV}$}\\

[51].-  J. Stillwell, \emph{Mathematics and Its History}.- Springer
2010.\

 [52].- Jean P. Serre, \emph{A Course in
Arithmetic}.- Springer 1973.\

 [53].- E. Artin, `` The Orders
of The Classical Simple Groups'', Comm. Pur.          Appl. Math.,
$\mathbf{8}$ (1955), 455 - 472.\

 [54].- J. Gray, ``From the
History of a Simple Group'', Math. Intell. $\mathbf{4}$(1982),
59-67.\

 [55].- Claude Chevalley, `` Sur certains groupes
simples'', T\^{o}hoku Math. J. (2),           $\mathbf{7}$(1955),
14-66.\

[56].-  N. Jacobson, \emph{Lie Algebras}.- Dover (1962).\

[57].- M. Aschbacher, \emph{Finite Group Theory}. Cambridge U.P.
2000.\\

\quad \quad\quad \quad \quad\quad \underline{$\mathbf{Ch. V}$}\\

[58].- R. L. Griess, `\emph{Twelve Sporadic Groups}. Springer 1998.\

[59].- J.H. Conway and R. Smith \emph{On quaternions and octonions}.
A.K. Peters 2003.\

[60].-  H. S. M. Coxeter, \emph{The Beauty of Geometry} (essay 7).
Dover (1968).\\

 [61].- P. J. Greenberg, \emph{Mathieu Groups}. New York
University, 1973.\

[62].- M. Ronan, \emph{Symmetry and the Monster}.- Oxford U.P.
2006.\

[63].- (Atlas): Conway \emph{et al}.: \emph{Atlas of Finite Groups}.
Oxford U.P. 1985.\

[64].- R. Steinberg ``Variations on a theme of Chevalley'', Pac. J.
Math. $\mathbf{9}$ (1959), 875-891.\\

\quad \quad\quad \quad \quad\quad \underline{$\mathbf{Ch. VI}$}\\

[65].- W. Feit and J. G. Thomson, ``Solvability of groups of odd
order'', Pac. J. Math. $\mathbf{13}$(1963), 755-1029.\

[66].-  H. Bacry,  \emph{Lecons sur la Theorie des Groupes.}
Dunod/Gordon and Breach, N.Y. 1967\

[67].-  R. Gilmore, \emph{Lie Groups and Applications.} J. Wiley
1974\

[68].-  J. Conway and S. Norton, Bull. L. M. S. $\mathbf{11}$
(1979), 308-339.\

[69].- K. Harada, ``Moonshine'' of Finite Groups. Eur. Math. Soc.,
Z\"{u}rich 2010.\

[70].-  J. Lepowski and J. McKay eds.: ``Moonshine. The First
Quarter Century and beyond''. Cambridge U.P. 2010.\

[71].- A. A. Ivanov, \emph{The Monster Group and Majorana
Involutions}. Cambridge U.P. 2009.\

[72].-  P. Goddard and D. Olive eds. \emph{Kac-Moody and Virasoro
Algebras}. World Scientific 1988.\

[73].-  M. Green, J. Schwarz and E. Witten, \emph{SuperString
Theory}. Cambridge U.P. 1985.\

[74].- J. Polchinski, \emph{String Theory}. Cambridge U.P. 1998.\

[75].- R. Borcherds in European Congress of Mathematicians, Vol. I.:
``Sporadic groups and String Theory''. Birkh\"{a}user, Basel 1994.\

[76].- P. Goddard, ICM Berlin (1998).\

[77].- E. Witten, ICM Madrid 2006.\

[78].- M. Bañados. C. Teitelboim and T. Zanelli.- arXiv hep-th 92
04099 (1992).\

[79].- C. N. Cheng, "Umbral Moonshine".- arXiv math.RT 1204.2779,
12-IV-2012.\

[80].- N. Bourbaki, \emph{Elements de Mathematique}, Livre II,
ch.5.- Hermann, Paris 1964.\

[81].- J. Baez, "The octonions", Bull. Am. Math. Soc. $\mathbf{39}$
(2002), 145-205.\

[82].- J. Milnor and J. Stasheff, \emph{Characteristic classes},
Princeton U.P. 1974.\

[83].- R. Brauer and C. H. Sha, editors :\emph{Theory of Finite
Groups.} Benjamin, N.Y. 1969.\

[84].- R. A. Wilson, \emph{The Finite Simple Groups}. Springer,
Berlin 2007.\

[85].- T. C. Hales ``A proof of the Kepler conjecture''. Ann. Math.
$\mathbf{162}$(3) (2005), 1065-1185.\

[86].- E. Witt, ``\"{U}ber Steinersche Systeme'', Abh. Math.
Seminar. Univ. Hamb. $\mathbf{12}$(1938), 265-275. Also Ibid.
256-264.\

[87].- E. Witten ``Strong theory dynamics in various dimensions''
Nucl. Phys. B $\mathbf{443}$ (1995), 85-126; arXiv: hep-th/ 95 03
124.\

[88].- C. J. Moreno and S. S. Wagstaff, \emph{Sums of Squares of
Integers.} Chapman \& Hall, N.Y. 2006.\

[89].- C. N. Cheng, ``K3 Surfaces, $\mathcal{N}$=4 Dyons, and the
Mathieu Group $M_{24}$'' (arXiv: 1005.5415, [hep-th], 28-5-2010).\

[90].- L. J. Boya, arXiv math-phys (2011), 1105-3055, ``Sporadic
Groups'': Bull. Cal. Math. Soc. $\mathbf{103}$ (2011), 59-70.

\end{document}